\documentclass[5p]{elsarticle}

\makeatletter
\def\ps@pprintTitle{%
 \let\@oddhead\@empty
 \let\@evenhead\@empty
 \def\@oddfoot{}%
 \let\@evenfoot\@oddfoot}
\makeatother

\usepackage{graphicx}
\usepackage{amsmath}
\usepackage{amssymb}
\usepackage{amsthm}
\usepackage{subfig}
\usepackage{url}
\usepackage[utf8x]{inputenc}
\usepackage{caption}
\usepackage{booktabs}
\usepackage{paralist}
\usepackage[numbers]{natbib}
\usepackage{csquotes}
\usepackage{pdflscape}
\usepackage{tikz}
\usepackage[inline]{enumitem}

\tikzset{font={\fontsize{10pt}{12}\selectfont}}

\newcommand{\bd}{\boldsymbol{d}}
\newcommand{\be}{\boldsymbol{e}}
\newcommand{\bff}{\boldsymbol{f}}
\newcommand{\bg}{\boldsymbol{g}}
\newcommand{\bh}{\boldsymbol{h}}
\newcommand{\bs}{\boldsymbol{s}}

\newcommand{\bp}{\boldsymbol{p}}
\newcommand{\bw}{\boldsymbol{w}}
\newcommand{\bN}{\boldsymbol{N}}
\newcommand{\balpha}{\boldsymbol{\alpha}}
\newcommand{\bbeta}{\boldsymbol{\beta}}
\newcommand{\blambda}{\boldsymbol{\lambda}}
\newcommand{\bpi}{\boldsymbol{\pi}}
\newcommand{\bsig}{\boldsymbol{\sigma}}
\newcommand{\bvarrho}{\boldsymbol{\varrho}}
\newcommand{\bxi}{\boldsymbol{\xi}}

\newcommand{\PP}{\mathbb{P}}
\newcommand{\diag}{\operatorname{diag}}

\theoremstyle{definition}
\newtheorem{definition}{Definition}
\newtheorem{Thm}{Theorem}

\newtheorem{Pro}{Proposition}

\usetikzlibrary{shapes,decorations,arrows,calc,arrows.meta,fit,positioning}
\tikzset{
    -Latex,auto,node distance =1 cm and 1 cm,semithick,
    state/.style ={ellipse, draw, minimum width = 0.7 cm},
    point/.style = {circle, draw, inner sep=0.04cm,fill,node contents={}},
    bidirected/.style={Latex-Latex,dashed},
    el/.style = {inner sep=2pt, align=left, sloped}
}

\begin{document}

\begin{frontmatter}

\title{Large-scale Analysis and Simulation of Traffic Flow using Markov Models}

\author{Ren\'at\'o Besenczi\corref{cor1}}
\ead{besenczi.renato@inf.unideb.hu}

\author{Norbert B\'atfai\corref{}}
\ead{batfai.norbert@inf.unideb.hu}

\author{P\'eter Jeszenszky\corref{}}
\ead{jeszenszky.peter@inf.unideb.hu}

\author{Roland Major\corref{}}
\ead{roland.major@inf.unideb.hu}

\author{Fanny Monori\corref{}}
\ead{fannymonori@gmail.com}

\author{M\'arton Isp\'any\corref{}}
\ead{ispany.marton@inf.unideb.hu}

\address{Department of Information Technology, \\
Faculty of Informatics, University of Debrecen\\4002 Debrecen, PO Box 400, Hungary}
\cortext[cor1]{Corresponding author.}


\begin{abstract}
Modeling and simulating movement of vehicles in established transportation infrastructures, especially in large urban road networks is an important task. It helps with understanding and handling traffic problems, optimizing traffic regulations and adapting the traffic management in real time for unexpected disaster events. A mathematically rigorous stochastic model that can be used for traffic analysis was proposed earlier by other researchers which is based on an interplay between graph and Markov chain theories. This model provides a transition probability matrix which describes the traffic's dynamic with its unique stationary distribution of the vehicles on the road network. In this paper, a new parametrization is presented for this model by introducing the concept of two-dimensional stationary distribution which can handle the traffic's dynamic together with the vehicles' distribution. In addition, the weighted least squares estimation method is applied for estimating this new parameter matrix using trajectory data. In a case study, we apply our method on the Taxi Trajectory Prediction dataset and road network data from the OpenStreetMap project, both available publicly. To test our approach, we have implemented the proposed model in software. We have run simulations in medium and large scales and both the model and estimation procedure, based on artificial and real datasets, have been proved satisfactory. In a real application, we have unfolded a stationary distribution on the map graph of Porto, based on the dataset. The approach described here combines techniques whose use together to analyze traffic on large road networks has not previously been reported.
\end{abstract}


\end{frontmatter}

\section{Introduction}

In the past decade, research and development of Smart City applications have become an active topic. These applications provide services to inhabitants which can make everyday life easier \cite{ismagilova2019smart}. In addition, these services contain solutions such as intelligent city planning, crowd-sourcing, as well as crisis and disaster management \cite{Zheng2014}. These applications will also both generate and make us of big data analytics which will arise from the wide availability of cloud computing. In addition, complex sensor systems are being implemented, called the Internet of Things (IoT) that could aid the operation of these applications. By the year 2050, 70\% of Earth's population is expected to live in urban areas \cite{bocquier2005world}. City infrastructures will face new challenges from many factors; one such factor is urban traffic.

In the recent years, the research and development of smart city applications focusing on urban transportation and traffic have become a vivid topic. Since more and more people live in urban areas, a solution for the problem of dense traffic is highly demanding. Moreover, in the past few years, many developments have occurred in the automobile industry that fundamentally changed the way we think about personal mobility. Autonomous or driverless cars are being introduced by several tech giants and car companies, some of them are on the market already. From driving-aid system (e.g. pedestrian observers, lane support systems), which have become ubiquitous by the end of the 2010s, we can envision the widespread use of driverless cars in the 2020's. In addition, pure electric cars have been on the market for a few years and will certainly continue to expand their market share.

In a previous paper \cite{OOCWC1}, we introduced a smart city application that can support the operation of self driving and electric cars in urban environments and may provide a common research platform for investigating the connection between autonomous cars and smart cities. Our main question was: what can a city controlled IT solution offer to these cars to allow them to be operated more efficiently? The application was a prototype and some of its features were not defined or developed satisfactorily. One such feature was the traffic simulation algorithms included in the application. In a previous paper \cite{OOCWC2}, we raised the question whether we can develop such an algorithm to control the simulated cars that can hold the real distribution of cars. An additional question is how can we fit this algorithm by estimating its parameters based on real data, e.g., which has the form of trajectories. Our results presented in this paper are an answer to these questions.

In \cite{Crisostomietal2011}, see also \cite{Faizraetal2015} and \cite{Faizrahnemoon2016}, a stochastic model is proposed which can handle the traffic in an urban network by using a mathematically rigorous method. This model is based on discrete time Markov chain on the road graph which plays the role of the state space. In the traffic interpretation, the transition probability matrix describes the dynamic of the traffic while its unique stationary distribution  corresponds to the traffic equilibrium (or steady) state on the road network. In this steady-state, the distribution of vehicles remains invariant locally in time under the transport dynamic. Thus, this stationary distribution of the Markov chain can be interpreted as the momentary true distribution of the vehicles on the road network. 

Our contributions in this paper are the following. Based on \cite{Crisostomietal2011} we introduce the concepts of a Markov random walk, which describes the motion of an individual vehicle, and Markov traffic, which describes the entire traffic on the road network, respectively. We determine the stationary distribution of the Markov traffic as a multinomial distribution, see formula \eqref{multinomial}. We present how the ergodic theory of finite Markov chains implies that a complex traffic event can be approximated well by the stationary distribution of a Markov chain on the road network; see Theorem \ref{Markov_traffic_ergodic}. This theorem also yields the theoretical ground of our simulation algorithm. We reparametrize the model by introducing the concept of two-dimensional stationary distribution which possesses equi-distributed marginals that are the unique stationary distribution of the transition probability matrix, respectively. To estimate this parameter matrix the weighted least squares (WLS) estimation as a kind of composite (quasi-) likelihood methods is applied, see \cite{Hjort2008Varin}. In Theorem \ref{main_thm}, we show that the WLS estimator of the two-dimensional stationary distribution can be expressed explicitly. Moreover, this estimation method provides a computationally effective technique in a large scale since the MapReduce paradigm can be easily applied for it. Finally, we present how a city-controlled IT solution can be developed which is able to simulate the traffic on a road network that fits to the real data.

Note that the joint application of Markov chains and large graphs to analyze the behavior of complex systems is well known in several fields, e.g., distributed systems \cite{dabrowski2011hunt}, geophysics \cite{Cavers2015Vasudevan} and biology \cite{Lesne2006}. 

This paper is structured as follows: the remainder of this chapter provides a short introduction to our software system and shows a brief outlook for other simulation tools and approaches. In section \ref{modeling-traffic-flows}, we describe our proposed traffic simulation model in detail. In section \ref{results}, we discuss our findings, and in section \ref{conclusion}, we conclude the paper. The Appendix provides the proofs of theoretical results and a toy example to demonstrate the proposed estimation method.

\subsection{Previous and related works}

This research follows our development of a traffic simulation platform initiative called rObOCar World Championship (or OOCWC for short) \cite{OOCWC1}, \cite{OOCWC2}, \cite{OOCWC3}, \cite{OOCWC4}, \cite{dssv-talk}. The OOCWC is a multiagent-oriented environment for creating urban traffic simulations. The goal is to offer a research and educational platform to investigate urban traffic control algorithms and the relationship between smart cities and self-driving cars. In our vision, a software system that is controlled by a city administration can support these cars well, because a city possesses all the necessary information (congestions, accidents, detours, etc.) for effective route planning. The traffic simulations are performed by one of its components called {\textit{Robocar City Emulator}} (RCE). The first rapid prototype of RCE was called Justine, which is an open source project released under the GNU GPL v3 and can be downloaded from GitHub.\footnote{\url{https://github.com/nbatfai/robocar-emulator}} It has 3 main components, the RCE itself, and two visualization tools, the rcwin and the rclog. This prototype uses the OpenStreetMap (OSM) database and processes it with the Osmium Library. The result of this processing is a routing map graph and a Boost Graph Library graph. The routing map graph is then placed into a shared memory segment. This prototype was also called \enquote{Police Edition}, because it could simulate police pursuit (macroscopic, with multiple police agents). In addition, it was somewhat capable of simulating traffic, but only for a technical demonstration, because every traffic unit moved randomly. In the following, we only focus on the traffic simulation algorithm of the RCE. The traffic simulation model of the RCE is based on the Nagel-Schreckenberg (NaSch) model \cite{nasch}, because it uses a cell based approach, as well. Moreover, it can be considered as a standalone multi-agent system. The simulation takes place on a rectangular part of the OSM. We slice all edges for parts 3 meters long. Based on NaSch terminology, these parts are called cells. Therefore, the cell length is equal to 3 meters. In contrast with the NaSch model, where a cell may contain many cars, edges can contain a given number of cars only (calculated as the edge length divided by the length of the cell, or part, i.e.~3 meters), since in our simulation, one cell can contain only one car. In the original implementation, the simulation algorithm moves the cars by random walk. So, when a car arrives to a graph vertex (i.e.~intersection), it selects the next edge (i.e.~next street) according to uniform distribution. For a detailed description of the operation of RCE, see paper \cite{OOCWC1}.

A simulation can be initialized based on measured data or some prescribed distribution, e.g., uniform one, on the road network. During the simulation, we can observe how the distribution of the cars change. One of the reasonable requirements of any simulation algorithm is to keep the distribution of the vehicles on the road network invariant in a short time period. One statistics that can represent this distribution is the order of the streets based on the number of cars on them. An important aspect is that the order of the street should remain the same during the simulation when it is already in a steady state: this is the case when we call the distribution stationary. In paper \cite{OOCWC2}, we showed that in the original edition of the OOCWC the order of the streets changes almost randomly even when the simulation has been running for a long time, so the requirement of stationary distribution does not hold. In this paper, a model is proposed to answer this problem.

There exist several traffic simulation platforms. Multi-Agent Transport Simulation \cite{horni2016multi} is an open-source framework for implementing agent-based simulations and transport planning models in a large-scale. Its primary simulation algorithm is queue-based \cite{Charypara} and aims to reach an \enquote{equilibrium state} by co-evolutionary algorithms \cite{popovici2012coevolutionary}. Simulation of Urban Mobility \cite{SUMO2012} is a portable package that supports microscopic traffic simulations on large road networks. Aimsun\footnote{\url{https://www.aimsun.com/}} is a modeling software environment that supports macroscopic, and microscopic traffic simulations. PTV Vissim\footnote{\url{http://vision-traffic.ptvgroup.com/en-us/products/ptv-vissim/}} is a microscopic, multi-modal traffic simulation package. It supports simulation of traffic patterns and can display all road users and their interactions. Although the above mentioned applications are widely used in traffic analysis and planning, the main focus of their simulation algorithms is on microscopic traffic events. In contrast, our software system focuses only on the traffic flow of the whole city, or, to be more precise, the traffic graph.

Several approaches exist for short-term traffic flow prediction. These models are based on for example Box-Jenkins time-series analyses with ARIMA model \cite{lee1999application}, \cite{stathopoulos2003multivariate}, \cite{ghosh2009multivariate}, \cite{xue2008short}, Kalman filter theory \cite{wang2005real}, \cite{ngoduy2011low}, non-parametric methods (k-NN, kernel, local regression) \cite{davis1991nonparametric}, \cite{smith2002comparison}, \cite{turochy2004relating}, \cite{smith1997traffic}, exponential smoothing \cite{messer1993advanced}, \cite{castro2009online}, spectral analysis \cite{nicholson1974prediction} or wavelets \cite{jiang2005dynamic}, \cite{xie2006wavelet}, \cite{cheng2007mining}. In addition, several approaches use machine learning and data mining techniques, such as support vector regression \cite{jeong2013supervised}, artificial neural networks \cite{chan2012neural}, \cite{park1998short}, \cite{dia2001object}, Bayesian networks \cite{sun2006bayesian} or deep learning \cite{lv2015traffic}. Some applications can be found based on computational intelligence techniques, e.g., linear genetic programming \cite{brameier2007basic} or fuzzy logic \cite{iokibe1993traffic}, \cite{li2006type}, \cite{zhang2008short}, but seldom can we find approaches based on Markov models \cite{necula2014dynamic} and \cite{Crisostomietal2011} mentioned previously. 

\section{Modeling traffic flow by Markov chains on graphs}
\label{modeling-traffic-flows}

In this section, we overview a traffic simulation model that uses tools of graph theory and Markov chains. First, in sections \ref{model-basics} and \ref{model-detailed} we outline the basic concepts in the fields of graph theory and finite Markov chains, respectively. Then, in \ref{Markov_traffic}, we describe the proposed model called \enquote{Markov traffic} in detail. As a case study, we use a publicly available trajectory dataset, namely, the Taxi Trajectory Prediction dataset\footnote{\url{kaggle.com/c/pkdd-15-predict-taxi-service-trajectory-i}}; in section \ref{public-dataset}, we outline the key points of how we selected and processed the trajectory data. In section \ref{OSM-graph}, we describe how we process OSM data and build a traffic graph from this data. Section \ref{inference} is devoted to the statistical inference for Markov traffic.

\subsection{Basic concepts and notation}
\label{model-basics}

In this section, we outline the concepts of graph theory that are necessary for modeling traffic flow. A standard textbook on graph theory is \cite{JensenGutin2007}.

\smallskip

\textbf{Road network, line digraph and degree distributions.} Let $G = (V,E)$ be a directed graph (digraph) where $V$ and $E$ denote the set of vertices or nodes and the set of directed edges or arcs or arrows of the graph, respectively. In the sequel, vertices are denoted by $u, v, w,$ edges are denoted by $e, f, g$. For a directed edge $e = (v,w) \in E$ we also use the notation $v \rightarrow w$. We suppose that $G$ is a simple digraph in the sense that it does not contain multiple arrows, i.e., two or more edges that connect the same two vertices in the same direction. The digraph $G$, called road network in this paper, represents the road system of a city. More precisely, we have the following definition, see \cite{pan2013crowd}, 
\begin{definition}
A \textbf{road network} $G$ is a simple directed graph, $G = (V,E)$, where $V$ is a set of nodes representing the terminal points of road segments, and $E$ is a set of directed edges denoting road segments.

A \textbf{road segment} $e = (v,w) \in E$ is a directed edge in the road network graphs, with two terminal points $v$ and $w$. The vehicle flow on this edge is from $v$ to $w$.
\end{definition}

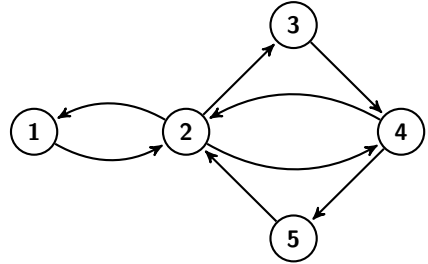
\begin{figure}
\centering
\begin{tikzpicture}[->,>=stealth',shorten >=1pt,auto,node distance=2cm,
                    thick,main node/.style={circle,draw,font=\sffamily\small\bfseries}]
 
  \node[main node] (3)  {3};
  \node[main node] (2) [below left of=3] {2};
  \node[main node] (1) [left of=2] {1};
  \node[main node] (4) [below right of=3] {4};
  \node[main node] (5) [below right of=2] {5};  
  \path[every node/.style={font=\sffamily\small}]
    (1) edge [bend right] node {} (2)
    (2) edge node {} (3) 
        edge [bend right] node {} (1)
        edge [bend right] node {} (4)        
    (3) edge node  {} (4)
    (4) edge [bend right] node {} (2)
         edge node {} (5)
    (5) edge node {} (2);    
\end{tikzpicture}
\caption{A simple road network.}
\label{graph-example-1}
\end{figure}
Fig.~\ref{graph-example-1} presents a simple example for road network.

Let $S$ denote the diagonal set of $V$, i.e., $S := \{(v, v) | v \in V \}$. If $v \rightarrow v$, i.e., $(v, v) \in E$, for a $v \in V$, then it is called a loop which connects the vertex $v$ to itself. If there would be a loop in a road network then there would be a physical road segment $v \rightarrow v$ ($v \in V$) in the road network. From a practical point of view, we lock out this possibility because in this case a vehicle can move in an infinite cycle. Thus, we suppose that $E \cap S = \emptyset$. Later however, when we define the stochastic traffic on a road network, we allow loops in order that we ensure that a vehicle may remain at the same node or edge of the road network after a time step. For $v \in V$ define $v^{−} := \{e \in E\, |\, \exists u \in V : e = (u, v)\}$ and $v^{+} := \{e \in E \,|\, \exists w \in V : e = (v, w)\}$, i.e., $v^{−}$ and $v^{+}$ are the sets of arrows in and out the node $v$, respectively. Note that $deg^{−}(v) = |v^{−}|$ and $deg^{+}(v) = |v^{+}|$ is the indegree and outdegree of $v$, respectively, where $|\cdot|$ denotes the cardinality of a set.

For a digraph $G = (V, E)$ another digraph can be associated by the following way. Let the set $V'$ of vertices of this new digraph be the set of directed edges $E$ of $G$ and let the set $E'$ of its directed edges consist of the ordered pair $(e, f)$ where $e, f \in E$ such that there exist $u, v, w \in V$ that $e = (u, v)$ and $f = (v, w)$, i.e., $u \rightarrow v \rightarrow w$ is a path (or dipath) in $G$ of length 2. This associated digraph is called a directed line graph, see Section 4.5 in \cite{JensenGutin2007}, shortly line digraph, or line road network (network line graph, see \cite{1209207}), and it is denoted by $\text{L}(G) = (V', E')$. The elements of $E'$ can be described by triplets $(u, v, w)$, where $u, v, w \in V, (u, v), (v, w) \in E$, and for a directed edge in $\text{L}(G)$ we use the notation $(u, v) \rightarrow (v, w)$ too. 

The basic difference between the digraph and line digraph views of a road network is that the former assigns the vehicles moving in a city to the vertices while the latter to the edges. One can refer the former as first-order (primal) network while the latter as second-order (dual) network, see \cite{Wuetal2018} (\cite{Portaetal2006}). These two kinds of graphs are both useful because in a road network, certain measurements are associated with the crossings (vertices), and certain measurements are associated with the road segments (directed edges). When we are concerned with comparing measurements associated with crossings, then we will be concerned with the adjacency relationships of crossings, and so with the road network. However, when we are concerned with measurements associated with road segments we will be concerned with the adjacency relationships of road segments, and so our analyses will involve the line road network.

\begin{figure*}[!t]
    \centering
    \subfloat[Indegree distribution (vertices).]{\includegraphics[width=0.4\textwidth]{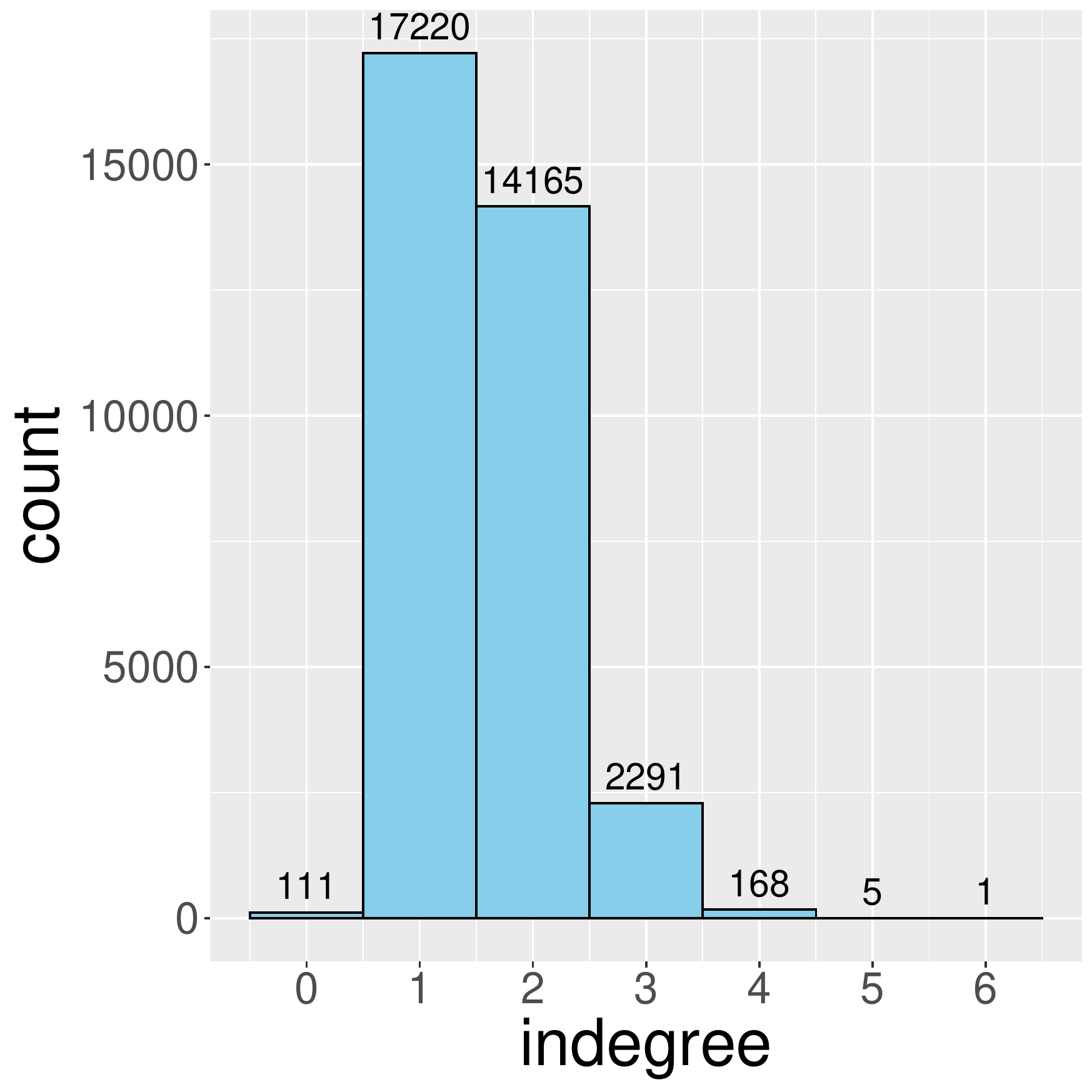}\label{ind_v}}
    \hspace{0em}
    \subfloat[Outdegree distribution (vertices).]{\includegraphics[width=0.4\textwidth]{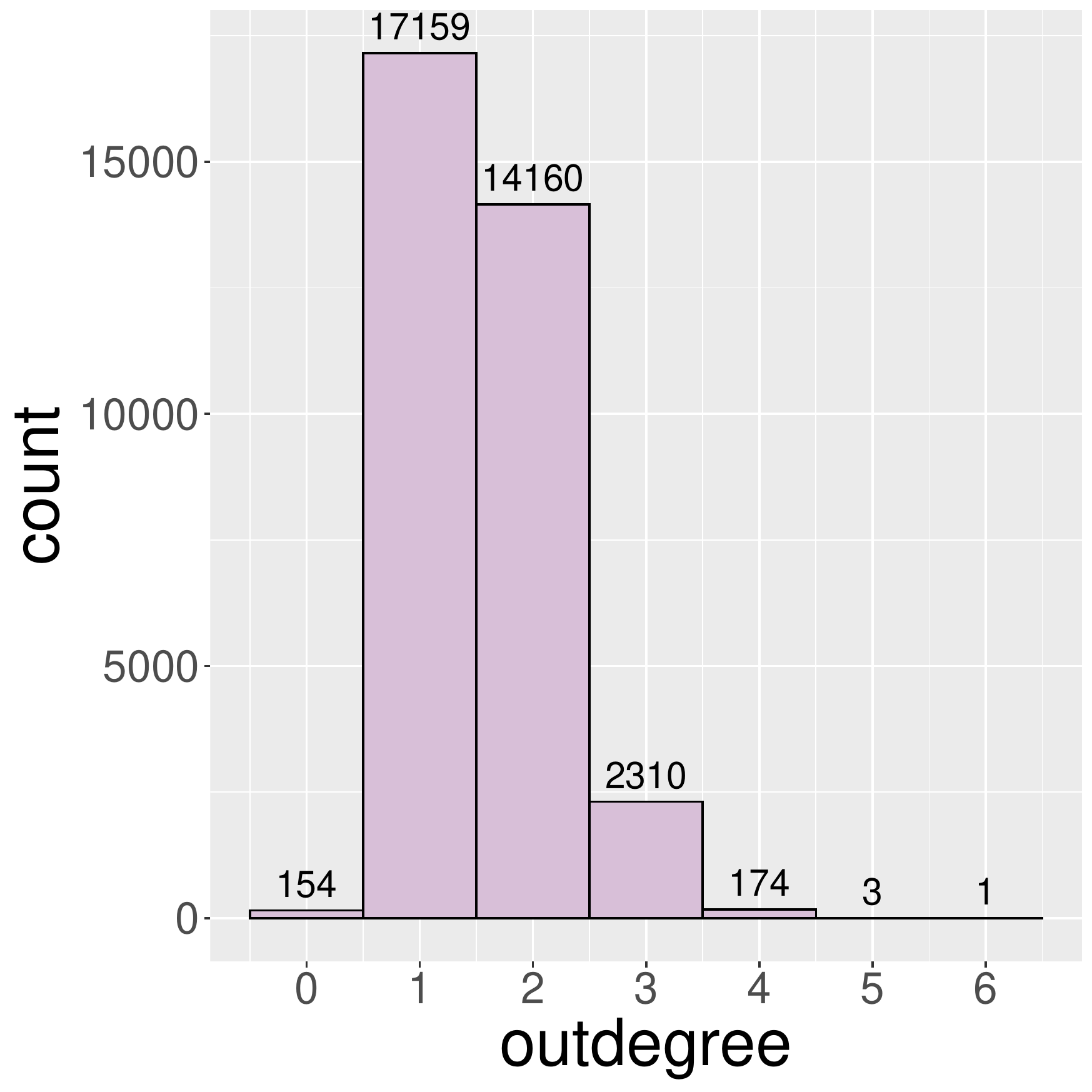}\label{outd_v}}
    \\
    \subfloat[Indegree distribution (edges).]{\includegraphics[width=0.4\textwidth]{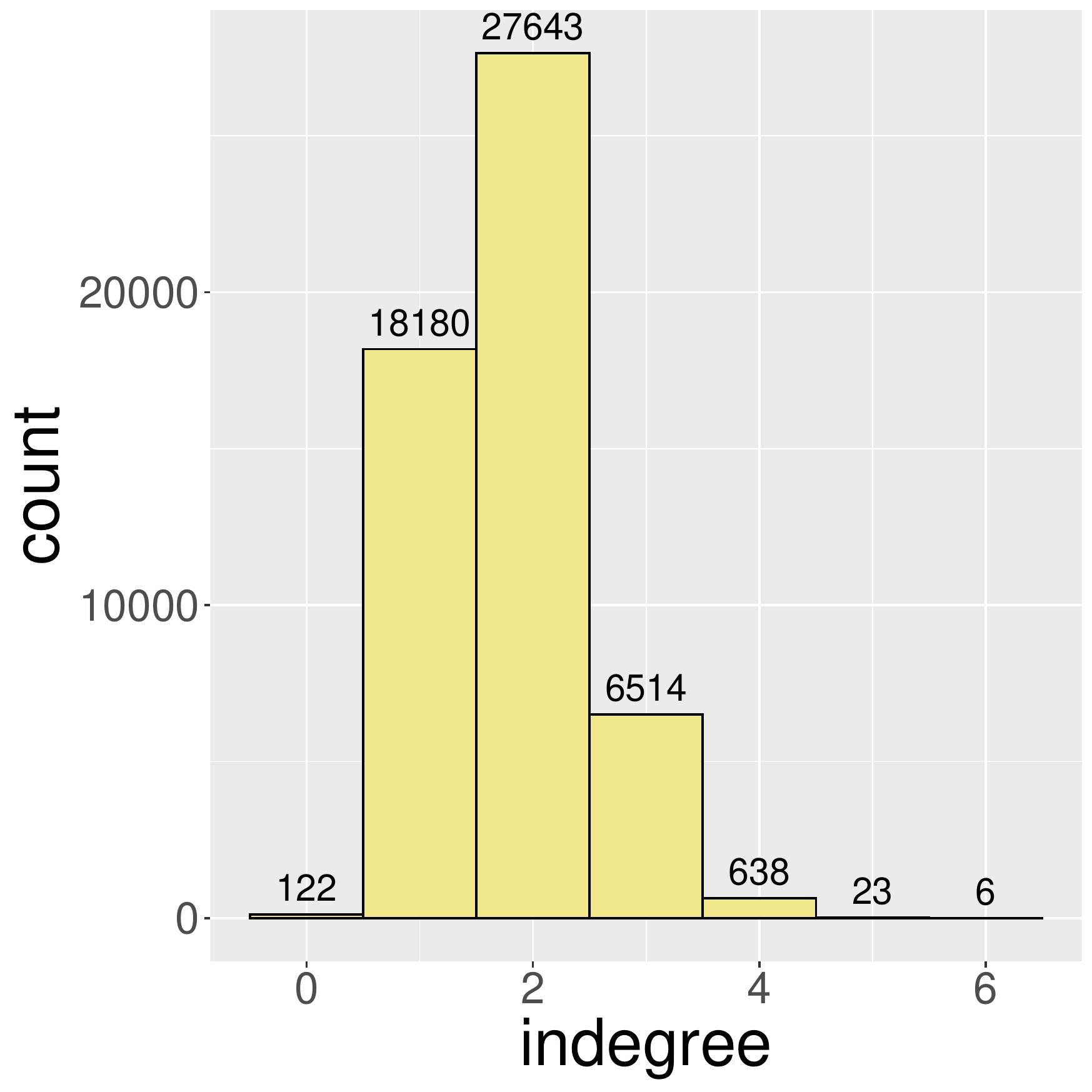}\label{ind_e}}
    \hspace{0em}
    \subfloat[Outdegree distribution (edges)]{\includegraphics[width=0.4\textwidth]{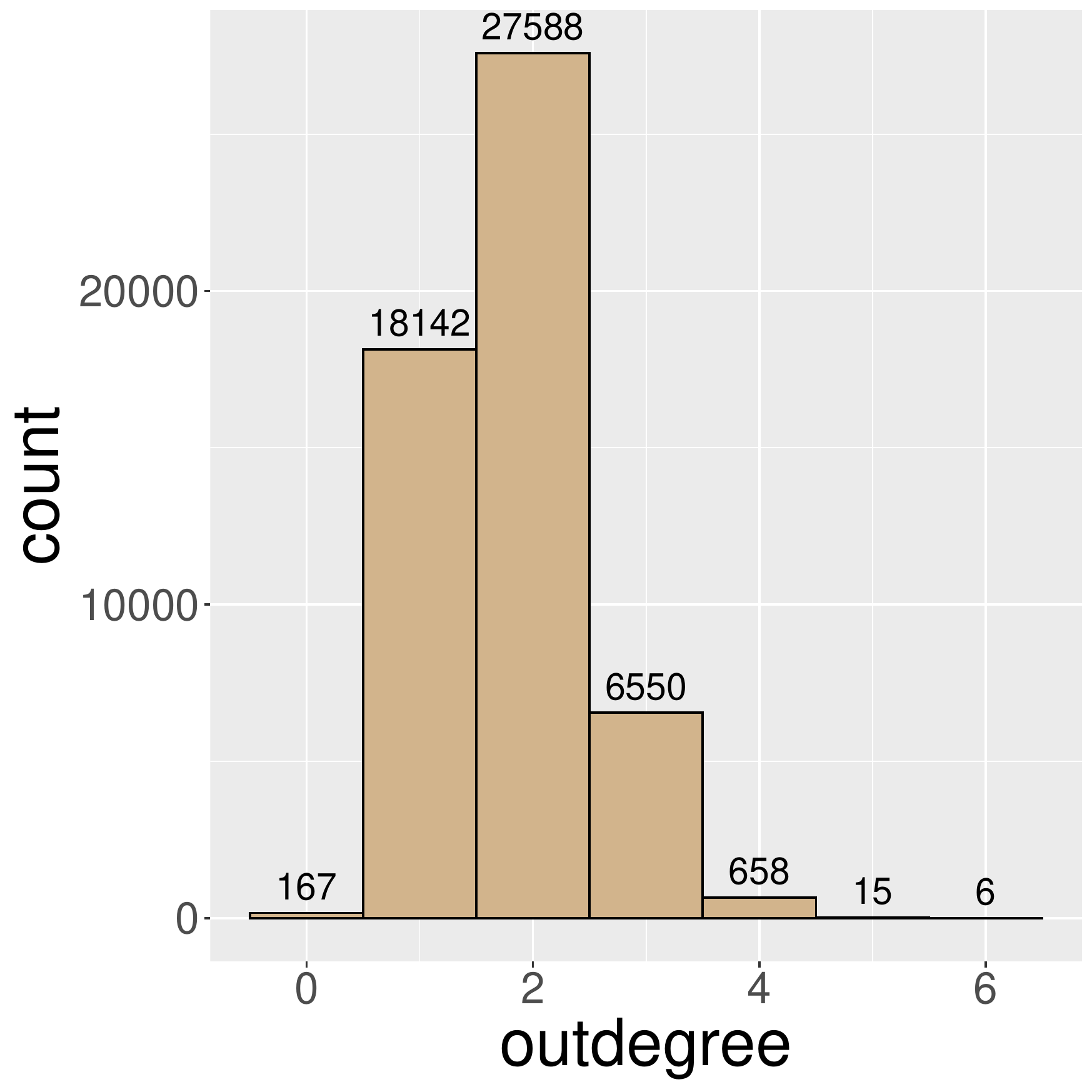}\label{outd_e}}
    \hspace{0em}
    \caption{The degree distribution histograms of the Porto map traffic graph.}
    \label{degree_dist}
\end{figure*}

The degree distributions of the digraphs $G$ and $\text{L}(G)$, respectively, are given in the following way. For all $i = 0, 1, 2, \dots$ define $n^{+}_i := |\{v \in V\, |\, deg^{+}(v) = i\}|$. Then, the pairs $(i, n^{+}_i), i = 0, 1, 2, \dots$, form the frequency histogram for the outdegree distribution of $G$. The indegree frequency histogram is defined similarly as $(i, n^{−}_i), i = 0, 1, 2, \dots$, where $n^{−}_i := |\{v \in V\, |\, deg^{−} (v) = i\}|$. On the other hand, for all $i = 0, 1, 2, \dots$, define $m^{+}_ i :=\\ \sum_{v \in G^{+}_i} deg^{−} (v)$ where $G^{+}_i := \{v \in V\,|\, deg^{+} (v) = i\}$. Then, the pairs $(i, m^{+}_i), i = 0, 1, 2, \dots$, form the frequency histogram for the outdegree distribution of $\text{L}(G)$. Note that $n^{+}_i = |G^{+}_i|$ for all $i$. Similarly, the pairs $(i, m^{−}_i), i = 0, 1, 2, \dots$, form the frequency histogram for the indegree distribution of $\text{L}(G)$ where $m^{−}_i := \sum_{v \in G^{−}_i} deg^{+}(v)$ and $G^{−}_i := \{v \in V\, |\, deg^{−}(v) = i\}$. (Note that $n^{−}_i = |G^{−}_i|$ for all $i$.) One can easily see that the supports of the two indegree (outdegree) histograms are the same. For the Porto example (described later in this paper), the above mentioned degree distributions can be seen in Fig.~\ref{degree_dist}. These histograms corroborates the fact that the Porto's road network is a sparse graph since there is no node with higher in- and outdegree than 6 and the ratio of the number of edges and the number of nodes is less than 2, see Fig.~\ref{porto_map}.

\smallskip

\textbf{Connectedness, periodicity and adjacency matrix.} Recall that a sequence $v_1, \dots,v_\ell \in V$, $\ell\in \mathbb{N}$ ($\mathbb{N}:=\{1,2,\ldots\}$), is called walk of length $\ell$ if $v_1 \rightarrow v_2 \rightarrow \dots \rightarrow v_\ell$. A walk is called path if its elements are different vertices. For a pair $u, v \in V, u \neq v$, it is said that $v$ is reachable from $u$ if there exists a walk $v_1, v_2, \dots, v_\ell$ such that $u = v_1$ and $v = v_\ell$. Clearly, if $v$ is reachable from $u$ then there is a path from $u$ to $v$. A digraph $G$ is said to be strongly connected (diconnected) if every vertex is reachable from every other vertex.

We can define vectors (functions) and matrices (operators or kernels) on $V$ in the following way. Let $\balpha: V\to \mathbb{R}$ be a real function on $V$ and let $\mathcal{F}(V,\mathbb{R})$ denote the set of real functions on $V$. We shall also use the notations $\balpha(v)=\alpha_v$ for all $v\in V$ and $\balpha=(\alpha_v)_{v\in V}$. Then, $\mathcal{F}(V,\mathbb{R})$ is a finite dimensional vector space with the usual inner product 
\begin{equation*}
        \langle \balpha, \bbeta\rangle = \balpha^\top \bbeta := \sum_{v\in V} \alpha_v \beta_v ,
\end{equation*}
$\balpha,\bbeta\in \mathcal{F}(V,\mathbb{R})$. Let $T:V\times V\to \mathbb{R}$ be a real function. We shall also use the notations $T(u,v)=t_{uv}$, $u,v\in V$ and $T=(t_{uv})_{u,v\in V}$. Then $T$ is called matrix, operator or kernel on $V$ and induces a linear operator on $\mathcal{F}(V,\mathbb{R})$ in the following way: for each $\balpha\in \mathcal{F}(V,\mathbb{R})$, $T(\balpha) \in \mathcal{F}(V,\mathbb{R})$ is defined as $T(\alpha)_u := \sum_{v\in V} t_{uv} \alpha_v$, $u\in V$. For the sake of simplicity, we write $T(\balpha)= T\balpha$ as a matrix-vector product. If the support of $T$ (the set $\{(u,v)\,|\, u,v\in V: t_{uv}\neq 0\}$ in $V\times V$) is a subset of $E$ ($E\cup S$) then $T$ is called $G$-subordinated in strong (weak) sense.

Let $A = (a_{uv})_{u,v\in V} $ denote the adjacency matrix of the digraph $G$, i.e., $A$ is a matrix on $V$ and $a_{uv} = 1$ if and only if $(u,v) \in E$ and $0$ otherwise. Clearly, the support of $A$ is $E$, i.e., $A$ is a $G$-subordinated matrix in strong sense ($a_{vv} = 0$ for all $v \in V$). Note that $A$ is not necessarily symmetric. The number of directed walks from vertex $u$ to vertex $v$ of length $k$ is the entry on $u$-th row and $v$-th column of the matrix $A^k$. For example, in Fig.~\ref{graph-example-1}, the number of directed walks of length 4 from vertex 2 to vertex 4 is 4, see \ref{example}. One can easily check that $G$ is strongly connected if and only if there is a positive integer $k$ such that the matrix $I + A + \dots + A^k$ is positive, i.e., all the entries of this matrix are positive. The indegree and outdegree of a vertex $v$ can be expressed by the adjacency matrix as $deg^{−}(v) = \sum_{u \in V} a_{uv}$ and $deg^{+}(v) = \sum_{u\in V} a_{vu}$. Introduce the vectors $\bd^{−} := (deg^{−}(v))_{v\in V}$ and $\bd^{+} := (deg^{+}(v))_{v \in V}$. Then, we have $\bd^{−} = A^T\textbf{1}$ and $\bd^{+} = A\textbf{1}$ where $\textbf{1} := (1)_{v \in V}$ is the constant unit function. 

Clearly, the line digraph of a strongly connected digraph is also strongly connected. Namely, if $e=(u,v)\in V^\prime (=E)$ and $f=(w,z)\in V^\prime (=E)$ are arbitrary such that $e\neq f$, then, since $G$ is strongly connected, there exists a walk (or a path) of length $\ell$ in $G$ such that $v=v_1\rightarrow v_2\rightarrow \ldots\rightarrow v_\ell=w$, where $v_1,\ldots,v_\ell\in V$ ($\ell\in\mathbb{N}$), and thus we have $e=(u,v)\rightarrow (v_1,v_2)\rightarrow\ldots\rightarrow (v_{\ell-1},v_\ell)\rightarrow (w,z)=f$, i.e., there exists a walk (or a path) of length $\ell$ in $\text{L}(G)$ between the vertices $e,f\in V^\prime$. If $u\rightarrow v\rightarrow u$ for a pair $u,v\in V$ then we have $(u,v)\rightarrow (v,u)\rightarrow (u,v)$ in the line digraph, i.e., vehicles can turn back at vertex $u$ into $v$. Sometimes the traffic regulations do not allow this kind of reversal, i.e., the edge set $E^\prime$ in $\text{L}(G)$ must not contain some triplet $(u,v,u)$ while some of these triplets are needed that $\text{L}(G)$ be strongly connected. By deleting all of the unnecessary triplet $(u,v,u)$, $u,v\in V$, such that the remaining line digraph be still strongly connected we get the minimal strongly connected line digraph of $G$. This line digraph is denoted by $\text{ML} (G)$. For example, the vertices of $\text{ML} (G)$ for $G$ defined in Fig.~\ref{graph-example-1} are given in Table~\ref{markov-kernel-r}.

Recall that a cycle $C\subset V$ in digraph $G$ is a path $v_1\rightarrow v_2\rightarrow \ldots\rightarrow v_\ell\rightarrow v_1$. Here $\ell(C)=\ell$ is called the length of $C$. A digraph $G$ is said to be aperiodic if the greatest common divisor of the lengths of its cycles is one. Formally, the period of $G$ is defined as $per(G) :=\gcd\{\ell >0: \exists C\subset V \textrm{ cycle such that } \ell(C)=\ell \}$. Then, $G$ is called aperiodic if $per(G)=1$. Clearly, if a digraph $G$ is aperiodic then its line digraph $\text{L}(G)$ is also aperiodic. This statement follows from the following fact: if $v_1\rightarrow v_2\rightarrow \ldots\rightarrow  v_\ell\rightarrow v_1$ is a cycle then $(v_1,v_2)\rightarrow(v_2,v_3) \rightarrow\ldots\rightarrow(v_\ell,v_1)\rightarrow (v_1,v_2)$ is a cycle in $\text{L}(G)$. Thus, if $\ell >0$ and there exists a cycle $C\subset V$ such that $\ell(C)=\ell$ then there exists a cycle $C^\prime \subset  V^\prime$ such that  $\ell(C^\prime)=\ell $.
 
It is well known that the adjacency matrix $A$ of an aperiodic, strongly connected graph $G$ is primitive, i.e., irreducible and has only one eigenvalue of maximum modulus. Primitivity is equivalent to the following quasi-positivity: there exists $k\in\mathbb{N}$ such that the matrix $A^k>0$, see Section 8.5 in \cite{HornJohnson2013}.

\smallskip

\textbf{Open road networks and their closures.} In general, there are vehicles which leave or enter the city. To model these two possibilities $V$ is augmented by a new ideal vertex 0 which denotes the world outside of the city. This approach is similar to that is applied for public transport in \cite{Faizrahetal2013}. Let $\overline{V} := V \cup \{0\}$. Then, additional directed edges which contains vertex 0 are also added to $E$. In this case, $(v, 0)$ denotes that the vehicles can leave the city at vertex $v$, and $(0, v)$ denotes that new vehicles can enter the city at vertex $v$, where $v \in V$. Let $\overline{E}$ denote the augmentation of $E$ by directed edges for getting into and out of the city. Note that, for $\overline{E}$, it is not allowed to contain the loop $(0, 0)$. The augmentation of $G$ is denoted by $\overline{G} = (\overline{V}, \overline{E})$ and it is called the closure of a road network $G$. For $e = (v, w) \in \overline{E}$ we also use the notation $v \rightarrow w$. In what follows, we suppose that there exist $u, v \in V$ such that $u \rightarrow 0$ and $0 \rightarrow v$.

In the sequel, each definition (strong connectedness, periodicity, line digraph) given for $G$ can be extended for $\overline{G}$ in a natural way. Note that in the augmented line digraph $\text{L}(\overline{G}) = (\overline{V}', \overline{E}')$ the elements of the edge set $\overline{E}'$ can be described by triplet $(u, v, w)$, where $u, v, w \in \overline{V}$ and if $v = 0$ then $u, w \neq 0$ and if $u$ or $w$ is 0 then $v \neq 0$ because triplets $(0,0,v)$, $(v,0,0)$ and $(0,0,0)$ are excluded from $\overline{E}'$. One can easily see that if $G$ is strongly connected then its closure $\overline{G}$ is also strongly connected. Moreover, the strongly connected components of $G$, if there exist more than 1, can be connected through the ideal vertex 0, resulting in a strongly connected $\overline{G}$. Thus, in this case, the augmented line digraph will also be strongly connected. Clearly, if G is aperiodic then $\overline{G}$ is aperiodic too.

In the rest of this paper, it is assumed that the road network is closed, i.e., the vehicles can not get into and out of the road system of the city augmented with the ideal vertex 0.

\subsection{Probability distributions and Markov kernels on road networks}
\label{model-detailed}

In this section, we summarize some basic concepts and results of the theory of finite Markov chains with their interpretations and consequences for traffic flow modeling. Some textbooks on this field are \cite{Asmussen2003} and \cite{Bremaud1999}.

We can define two kinds of probability distributions on a road network $G$ considering the set $V$ or $E$ as the state space, respectively. In fact, we can handle both cases altogether by defining probability distributions on the set of vertices of the digraph $G$ or the line digraph $\text{L}(G)$, accordingly. However, we should note that the Markov kernels defined on the line digraph must be defined with particular attention.

A probability distribution (p.d.) on $V$ is the vector $\bpi := (\pi_v)_{v \in V}$ where $\pi_v \geq 0$ for all $v \in V$ and $\sum_{v \in V} \pi_v = 1$. That is a p.d. $\bpi$ is a normalized $V \rightarrow \mathbb{R}_+$ function. We can think of $\pi_v$ as the proportion of the number of vehicles which drive through the crossing $v$ with respect to the whole number of vehicles in the city at a fixed time period. A Markov kernel or transition probability matrix on $V$ is defined as a real kernel $P := (p_{uv})_{u,v \in V}$ such that $p_{uv} \geq 0$ for all $u, v \in V$ and $\sum_{v \in V} p_{uv} = 1$ for all $u \in V$, i.e., $\bp_u := (p_{uv})_{v \in V}$ is a p.d. on $V$ for all $u \in V$. The quantity $p_{uv} \in [0, 1]$ is called the transition probability from vertex $u$ to vertex $v$. The kernel $P$ is said to be $G$-subordinated if $p_{uv} > 0$ for a pair $u, v \in V$ implies $(u, v) \in E$ or $u = v$, i.e., $P$ as a matrix on $V$ is $G$-subordinated in the weak sense. It is well known, see \cite{jarvis1999graph}, that for a Markov kernel $P$ on $V$, an associated graph $G_P = (V, E_P)$ can be introduced in the following way: for a pair $u, v \in V$ (where the case $u = v$ is also allowed) $(u, v) \in E_P$ if and only if $p_{uv} > 0$. Thus, $P$ is $G$-subordinated if and only if $E_P \subseteq E \cup S$, i.e., $G_P$ is the subgraph of the graph $G$ extended with its diagonal $S$. In this case, we also use the notation $P = (p_e)_{e \in E \cup S}$. In other words, a $G$-subordinated Markov kernel $P$ is a stochastic matrix on $V$ with support $E \cup S$. Then, the sum condition for Markov kernel $P$ can be rewritten as:
\begin{equation}\label{Psum}
\sum_{w:v \rightarrow w} p_{vw} + p_{vv} = 1, \quad v \in V.
\end{equation}

\begin{figure}
\centering

\begin{tikzpicture}[->,>=stealth',shorten >=1pt,auto,node distance=2.3cm,
                    thick,main node/.style={circle,draw,font=\sffamily\footnotesize\bfseries}]
 
  \node[main node] (3)  {3: 1/7};
  \node[main node] (2) [below left of=3] {2: 2/7};
  \node[main node] (1) [left of=2] {1: 1/7};
  \node[main node] (4) [below right of=3] {4: 2/7};
  \node[main node] (5) [below right of=2] {5: 1/7};  

  \path[every node/.style={font=\sffamily\scriptsize}]
    (1) edge [bend right] node [below] {1/2} (2)
          edge [loop left] node {1/2} (1)
    (2) edge node [left] {1/4} (3) 
        edge [bend right] node[above] {1/4} (1)
        edge [bend right] node[below] {1/4} (4)        
         edge [loop above] node {1/4} (2)
    (3) edge node [right] {1/2} (4)
          edge [loop above] node {1/2} (3)
    (4) edge [bend right] node [above] {1/4} (2)
         edge node [right] {1/4} (5)
          edge [loop right] node {1/2} (4)
    (5) edge node [left] {1/2} (2)
          edge [loop below] node {1/2} (5);    
\end{tikzpicture}
\caption{A Markov kernel (on edges) with its stationary distribution (on vertices) on the road network in Fig.~\ref{graph-example-1}.}
\label{pi_graph}
\end{figure}
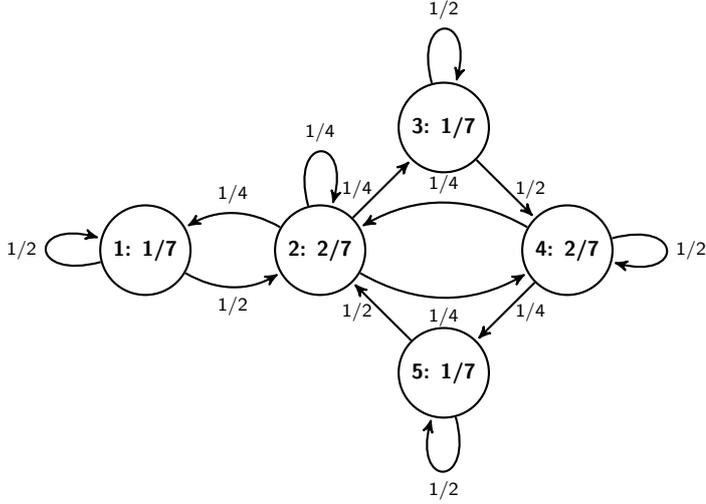

A p.d. $\bpi$ on $V$ is a stationary distribution (s.d.) of the kernel $P$ if $\sum_{u \in V} \pi_u p_{uv} = \pi_v$ for all $v \in V$. For a $G$-subordinated Markov kernel $P$ this formula, the so-called global balance equation, can be expressed as:
\begin{equation}\label{pi_eq}
\sum_{u:u \rightarrow v} \pi_u p_{uv} + \pi_v p_{vv} = \pi_v, \quad v \in V.
\end{equation}
Fig.~\ref{pi_graph} presents a Markov kernel with its s.d. on the road network in Fig.~\ref{graph-example-1}.

Since the vehicles are moving along the road segments of the road network $G$ it is natural to choose the set $E$ for the role of the state space. In this case, we can define probability distributions on the set of vertices again, however, we have to consider the line digraph $\text{L}(G)$ (or $\text{ML}(G)$). Formally, a probability distribution (p.d.) on $\text{L}(G)$ is the vector $\bpi^\prime:=(\pi^\prime_e)_{e\in E}$ where $\pi^\prime_e\ge 0$ for all $e\in E$ and $\sum_{e\in E} \pi^\prime_e =1$. Similarly to p.d.'s on $V$, the p.d. $\bpi^\prime$ can also be considered as a normalized $E\to \mathbb{R}_+$ function or a normalized $V^\prime\to \mathbb{R}_+$ function. If we want to emphasize the vertices of the original road network $G$ instead of the edges then the notation $\pi^\prime_e = \pi^\prime_{uv}$ is also used where $e=(u,v)\in E$. We can think of $\pi^\prime_e$ as the proportion of the number of vehicles at the road segment $e$ with respect to the whole number of vehicles in the city at a fixed time point. Note that $G$ endowed with $\bpi^\prime$ is a weighted digraph which is often called a network in itself as well.

\begin{table*}[ht!]
\center
\begin{tabular}{c|cccccccc}
&(1,2)&(2,3)&(3,4)&(4,2)&(2,1) &(2,4)&(4,5)&(5,2)\\ \hline 
(1,2)&1/2&1/4&0&0&0 &1/4&0&0\\ (2,3)&0&1/2&1/2&0&0 &0&0&0\\                 (3,4)&0&0&1/2&1/4&0 &0&1/4&0\\              (4,2)&0&1/4&0&1/2&1/4 &0&0&0\\ (2,1)&1/2&0&0&0&1/2 &0&0&0 \\                (2,4)&0&0&0&0&0&1/2&1/2&0\\                 (4,5)&0&0&0&0&0&0&1/2&1/2 \\(5,2)&0&1/4&0&0&1/8&1/8&0&1/2
\end{tabular}
\caption{An example for a Markov kernel on the minimal line digraph of the road network in Fig.~\ref{graph-example-1}.}
\label{markov-kernel-r}
\end{table*}

A transition probability matrix (or Markov kernel) on $E$, i.e., on the line digraph $\text{L}(G)$, can be defined as a real kernel $P^\prime:=(p^\prime_{ef})_{e,f\in E}$ such that $p^\prime_{ef}\ge 0$ for all $e,f\in E$ and $\sum_{f\in E}p^\prime_{ef} = 1$ for all $e\in E$. A p.d. $\bpi^\prime$ on $E$ is a s.d. of the kernel $P^\prime$ if $\sum_{e\in E} \pi^\prime_e p^\prime_{ef} = \pi^\prime_f$ for all $f\in E$. Since $G$ represents a road system we may suppose that if $e\neq f$ then $p^\prime_{ef}>0$ implies that $(e,f)\in E^\prime$, i.e., there exist $u,v,w\in V$ such that $e=(u,v)$ and $f=(v,w)$, and hence, $u\rightarrow v\rightarrow w$ is a walk of length 2. In other words, we may suppose that the Markov kernel $P^\prime$ is $\text{L}(G)$-subordinated. Similarly to Markov kernels on $V$, $P^\prime$ also induces an associated graph $G_{P^\prime}=(V^\prime,E^\prime_{P^\prime})$. Thus, $P^\prime$ is $\text{L}(G)$-subordinated if and only if $E^\prime_{P^\prime}\subseteq E^\prime\cup S^\prime$ where $S^\prime$ denotes here the diagonal $\{(e,e)\,|\, e\in E\}$ in $L(G)$. In this case, we use the notation $p^\prime_{ef}=p^\prime_{uvw}$ as well. In fact, $p^\prime_{uvw}$ denotes the probability that a vehicle on the road segment $(u,v)$ will go further to the road segment $(v,w)$ in the next time point. Moreover, in the case of $e=f=(u,v)$, let $p^\prime_{ee}=p^\prime_{uv}$ be the probability that a vehicle remains on the same road segment  in the next time point which can be non-zero as well. Thus, since $P^\prime$ is a Markov kernel, we have that, for all $u\rightarrow v$,
\begin{equation}\label{Rsum}
\sum_{w:v\rightarrow w} p^\prime_{uvw} + p^\prime_{uv} =1
\end{equation}
and the global balance equation is given as:
\begin{equation}\label{rho_eq}
            \sum_{u:u\rightarrow v} \pi^\prime_{uv} p^\prime_{uvw} + \pi^\prime_{vw} p^\prime_{vw} = \pi^\prime_{vw}
\end{equation}
for all $v\rightarrow w$.

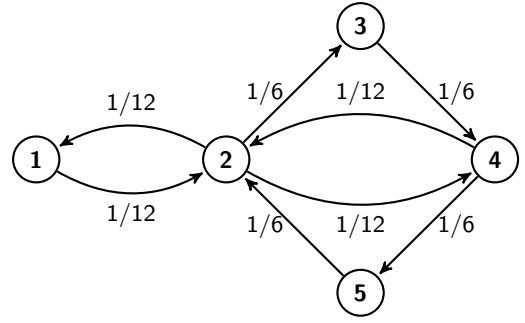
\begin{figure}
\centering
\begin{tikzpicture}[->,>=stealth',shorten >=1pt,auto,node distance=2.5cm,
                    thick,main node/.style={circle,draw,font=\sffamily\small\bfseries}]
 
  \node[main node] (3)  {3};
  \node[main node] (2) [below left of=3] {2};
  \node[main node] (1) [left of=2] {1};
  \node[main node] (4) [below right of=3] {4};
  \node[main node] (5) [below right of=2] {5};  

  \path[every node/.style={font=\sffamily\small}]
    (1) edge [bend right] node [below] {1/12} (2)
    (2) edge node [left] {1/6} (3) 
        edge [bend right] node[above] {1/12} (1)
        edge [bend right] node[below] {1/12} (4)        
    (3) edge node [right] {1/6} (4)
    (4) edge [bend right] node [above] {1/12} (2)
         edge node [right] {1/6} (5)
    (5) edge node [left] {1/6} (2);    
\end{tikzpicture}
\caption{The stationary distribution of the Markov kernel in Table~\ref{markov-kernel-r}.}
\label{rho_graph}
\end{figure}

An example for the Markov kernel $P^\prime$ on the minimal line digraph $\text{ML}(G)$ of the road network $G$ in Fig.~\ref{graph-example-1} is shown in Table \ref{markov-kernel-r}. Fig.~\ref{rho_graph} shows the unique stationary distribution $\bpi^\prime$ of the Markov kernel $P^\prime$.

\smallskip

\textbf{Probability distributions and Markov kernels on open road networks.} Probability distributions and Markov kernels on the closure $\overline{G}$ of an open road network $G$ can be defined similarly by considering the set $\overline{V}$ or $\overline{E}$ as the state space, respectively. A p.d. on $\overline{V}$ is the vector $\bpi:=(\pi_v)_{v\in \overline{V}}$ where $\pi_v\ge 0$ for all $v\in \overline{V}$ and $\sum_{v\in \overline{V}} \pi_v =1$. Note that $\pi_0$ denotes the proportion of the number of vehicles which drive in or out of the city's roads at a time point. A Markov kernel $P$ on $\overline{V}$ is said to be $\overline{G}$-subordinated if $p_{uv}>0$ for a pair $u,v\in \overline{V}$ implies $(u,v)\in \overline{E}$ or $u=v\in V$. Note that $\overline{G}$-subordination implies that $p_{00}=0$, i.e., the vehicles cannot move from 0 to 0, thus they either enter to the road network or leave the road network. The equations \eqref{Psum} and \eqref{pi_eq} remain hold too. Equation \eqref{Psum} can be rewritten as
\begin{equation*}
\begin{split}
            &\sum_{w\in V:v\rightarrow w}p_{vw}+p_{v0}+  p_{vv}=1, \quad  v\in V, \\
            &\sum_{w\in V :0\rightarrow w}p_{0w} = 1.
\end{split}
\end{equation*}
The global balance equation \eqref{pi_eq} for the s.d. can be rewritten as
\begin{equation*}
\begin{split}
           \sum_{u\in V:u\rightarrow v}  & \pi_u p_{uv} +\pi_0 p_{0v} + \pi_v p_{vv} = \pi_v, \, 
           v\in V, \ 0\rightarrow v, \\
            \sum_{u\in V:u\rightarrow v}  & \pi_u p_{uv} + \pi_v p_{vv} = \pi_v, \quad  v\in V, \ 0\nrightarrow v, \\
            \sum_{u\in V:u\rightarrow 0} &\pi_u p_{u0} = \pi_0.  
\end{split}
\end{equation*}

We can define Markov kernels on the line digraph $\text{L}(\overline{G})$ of the augmented road network $\overline{G}$, and thus on the augmented edge set $\overline{E}$ similarly to the case of $\text{L}(G)$. A Markov kernel $P^\prime$ on $\text{L}(\overline{G})$ is called $\text{L}(\overline{G})$-subordinated if $p^\prime_{ef}>0$ for $e,f\in \overline{E}$ implies $(e,f)\in \overline{E}^\prime$ or $e=f$. Note that $(e,f)\in \overline{E}^\prime$ implies that $e=(u,v)$ and $f=(v,w)$ where $u,v,w\in\overline{V}$ excluding the triplets $(0,0,v)$, $(v,0,0)$ and $(0,0,0)$. We shall also use the notation $p^\prime_{uvw}=p^\prime_{ef}$ if $e=(u,v)$ and $f=(v,w)$ and $p^\prime_{uv}=p^\prime_{ee}$ if $e=(u,v)$. However, three additional conditions should be added. The first one is that $p^\prime_{u0u}=0$ for all $u\in V$ such that $u\rightarrow 0\rightarrow u$. This means that if a vehicle is on the edge $(u,0)$, i.e., it leaves the city at vertex $u$ then it cannot be on the edge $(0,u)$ at the next time point, i.e., it cannot enter at vertex $u$ in the road network again, immediately. The second one is that $p^\prime_{0v0}=0$ for all $v\in V$ such that $0\rightarrow v\rightarrow 0$, i.e., vehicles can enter and leave the city at node $v$. This means that if a vehicle enters the city then it cannot leave the city at the next time. Finally, the third one is that $p^\prime_{u0}= p^\prime_{0v}=0$ for all $u,v\in V$ such that $u\rightarrow 0$ and $0\rightarrow v$. That is a vehicle cannot remain on the road network at the edge $(u,0)$ after two consecutive time points and if a vehicle enters into the road network at the edge $(0,v)$ (or at the vertex $v$) the first time then it does not remain on this edge after the next time point and it goes further immediately in the road network. Under these conditions the equations \eqref{Rsum} and \eqref{rho_eq} remain hold. Equation \eqref{Rsum} can be rewritten as:
\begin{equation*}
\begin{split}
\sum_{w\in V:v\rightarrow w} &p^\prime_{uvw} + p^\prime_{uv0} + p^\prime_{uv} =1, 
\quad u,v\in V,\ u\rightarrow v, \\
\sum_{w\in V:v\rightarrow w} & p^\prime_{0vw} =1, \quad v\in V,\ 0\rightarrow v,  \\
\sum_{v\in V\setminus\{u\}:0\rightarrow v} & p^\prime_{u0v} =1, \qquad u\in V,\ u\rightarrow 0. 
\end{split}
\end{equation*}
Equation \eqref{rho_eq} can be rewritten as:
\begin{equation*}
\begin{split}
\sum_{u\in V:u\rightarrow v} &\pi^\prime_{uv} p^\prime_{uvw} + \pi^\prime_{0v}p^\prime_{0vw}+ 
\pi^\prime_{vw} p^\prime_{vw} = \pi^\prime_{vw}, 
\\ & \qquad v,w\in V,\ v\rightarrow w, \\
\sum_{u\in V:u\rightarrow v} &\pi^\prime_{uv} p^\prime_{uv0} + \pi^\prime_{v0} p^\prime_{v0} = \pi^\prime_{v0}, 
\, v\in V,\ v\rightarrow 0,  \\
\sum_{u\in V\setminus\{w\}:u\rightarrow 0} &\pi^\prime_{uv} p^\prime_{u0w} +  \pi^\prime_{0w} p^\prime_{0w} = \pi^\prime_{0w}, \,
 w\in V, \ 0\rightarrow w.
\end{split}
\end{equation*}

The stationary distribution in all cases, i.e., for Markov kernels on road networks, line road networks and their closures, can be derived by solving the above appropriate linear equations numerically. Since the state space (the road network) is finite there exists at least one stationary distribution. 
However, in some cases, the stationary distribution is not uniquely defined by these equations. 

\smallskip

\textbf{Uniqueness of stationary distribution.}
We show that there is a direct connection between the existence of s.d. of the Markov kernels $P$ and $P^\prime$ and the strongly connected property of the physical road network $G$ if the Markov and graph structures are compatible with each other. The Markov kernel $P$ on $V$ is called $G$-compatible if, for any $u,v\in V$ such that $u\neq v$, $p_{uv}>0$ if and only if $(u,v)\in E$. Note that the $G$-compatibility implies the weak $G$-subordination for a Markov kernel $P$, however the converse is not true. Similarly, the Markov kernel $P^\prime$ on $E$ is called $G$-compatible if it is $\text{L}(G)$-compatible Markov kernel on $\text{L}(G)$, i.e., for any $e,f\in E$ such that $e\neq f$, $p^\prime_{ef}>0$ if and only if $(e,f) \in E^\prime$. This is equivalent to the statement that $p^\prime_{uvw}>0$, $u,v,w\in V$, if and only if $(u,v),(v,w)\in E$. Since $(e,f)\in E^\prime$ if and only if there exist $u,v,w\in V$ such that $e=(u,v)$ and $f=(v,w)$ we can define the $G$-compatibility of a Markov kernel $P^\prime$ as, for any $e,f\in E$ such that $e\neq f$, $p^\prime_{ef}>0$ if and only if there exist $u,v,w\in V$ such that $e=(u,v)$ and $f=(v,w)$. 

Clearly, if $P$ is $G$-compatible then the strong connectivity of $G$ implies that the associated graph $G_P$ to the Markov kernel $P$ is also strongly connected. In this case, the Markov kernel (the transition matrix) $P$ is called irreducible. Thus, by Theorem 1 in \cite{jarvis1999graph}, see also Theorem 3.1 and 3.3 in Chapter 3 of \cite{Bremaud1999} the following theorem holds.

\begin{Thm}\label{statsol}
If a road network $G$ is strongly connected then there is a unique stationary distribution $\bpi$ ($\bpi^\prime$) to any $G$-compatible Markov kernel $P$ ($P^\prime$). Moreover, this distribution satisfies $\pi_v>0$ for all $v\in V$ ($\bpi^\prime_{uv} >0$ for all $(u,v)\in E$).
\end{Thm}

The main consequence of this theorem is that, in case of any physical road network augmented by the ideal vertex 0, all of the Markov kernels defined on the road network that has positive transition probability on all roads have unique stationary distribution. Thus, it is reasonable to suppose that a real traffic which follows a Markovian dynamic has a local unique stationary distribution in a short time period that can be explored by observing the traffic. 

\subsection{Markov random walk and Markov traffic on road networks} 
\label{Markov_traffic}

Let $(\Omega,\mathcal{A},\PP)$ be a probability space. Then a $V$-valued ($E$-valued) random variable (r.v.) is a $X:\Omega\to V$ ($Y:\Omega\to E$) measurable function, i.e., $X^{-1}(v)\in\mathcal{A}$ for all $v\in V$ ($Y^{-1}(e)\in\mathcal{A}$ for all $e\in E$). In this case, $X$ ($Y$) is a random function on the set $V$ of vertices (on the set $E$ of edges). For example, $X$ ($Y$) can be the random position of a vehicle on the road network $G$, where the position refers to the actual vertex (edge) which the vehicle belongs to. Then, $\PP(X^{-1}(v))=\PP(X=v)$ ($\PP(Y^{-1}(e))=\PP(Y=e)$) denotes the probability that a vehicle is at the vertex $v\in V$ (at the edge $e\in E$). Clearly, by $\bpi_X(v):= \PP(X=v)$, $v\in V$, a r.v. $X$ induces a p.d. $\bpi_X$ on $V$. Similarly, by $\bpi^\prime_Y(e):= \PP(Y=e)$, $e\in E$, a r.v. $Y$ induces a p.d. $\bpi^\prime_Y$ on $E$. 

A sequence $\{X_t\}_{t\in\mathbb{Z}_+}$ of $V$-valued r.v.'s is a Markov chain on the state space $V$ if the Markov property holds:
\begin{equation*}
\begin{aligned}
         \PP(X_t=v_t | X_{t-1}=v_{t-1},\ldots, X_0=v_0) \\ =  \PP(X_t=v_t | X_{t-1}=v_{t-1})
\end{aligned}
\end{equation*} 
for all $t\in\mathbb{N}$, $v_0,\ldots,v_t\in V$. If $X,X^\prime$ are $V$-valued r.v.'s then for the conditional distribution $P=(p_{v v^\prime})_{v,v^\prime\in V}$, $p_{v v^\prime}:=\PP(X=v | X^\prime=v^\prime)$, $v,v^\prime\in V$, we shall also use the notation $X|X^\prime$. Clearly, $X|X^\prime$ is a Markov kernel on $V$. Similarly, a Markov chain $\{Y_t\}_{t\in\mathbb{Z}_+}$ of $E$-valued r.v.'s can also be defined through the Markov kernel $Y|Y^\prime$ on the state space $E$.

The main concepts of this paper are the Markov random walk and the Markov traffic  defined in the following way.

\begin{definition}
Let the road network $G$ be strongly connected and let $P$ be a $G$-compatible Markov kernel on $V$ with unique s.d. $\bpi$. Moreover, let $\{X_t\}_{t\in\mathbb{Z}_+}$ be a Markov chain on $V$ such that $\bpi_{X_0}=\bpi$ and $X_t|X_{t-1} \sim P$ for all $t\in\mathbb{N}$. 

Then, $\{X_t\}_{t\in\mathbb{Z}_+}$ is called \textbf{Markov random walk} on the road network $G$ with Markov kernel $P$.

The set of $k$ ($k\in\mathbb{N}$) mutually independent Markov random walks on $G$ with Markov kernel $P$ is called \textbf{Markov traffic} of size $k$ and it is denoted by the quadruple $(G,P,\bpi,k)$.
\end{definition}

Similarly, $\{Y_t\}_{t\in\mathbb{Z}_+}$ is a Markov random walk on the line road network if it is a Markov chain on the state space $E$ such that $\bpi^\prime_{Y_0}= \bpi^\prime$ and $Y_t|Y_{t-1}\sim P^\prime$  for all $t\in\mathbb{N}$.

Note that if $P$ is the uniform Markov kernel on $G$ then we obtain the standard random walk of the graph theory, see the survey \cite{Lovasz1996}.

A Markov random walk is an individual Markov traffic with $k=1$ in the sense that it describes the movement of a random vehicle which follows the stochastic rules defined by the Markov kernel. For a pair $u,v\in V$ the notation $u\Rightarrow v$ will mean that $(u,v)\in E\cup S$, i.e., either $u\rightarrow v$ or $u=v$. If $X_1,X_2$ are random functions on $V$ then $X_1\Rightarrow X_2$ means that the two-dimensional distribution of $(X_1,X_2)$ is concentrated on $E\cup S$. Clearly, for any Markov random walk $\{X_t\}_{t\in\mathbb{Z}_+}$ we have $X_t \Rightarrow X_{t+1}\Rightarrow\ldots \Rightarrow X_{t+n}$ for all $t$ and $n\in\mathbb{N}$. One can also call $\{X_t\}_{t\in\mathbb{Z}_+}$ as a first-order random walk on the road network where a vehicle moves from vertex $u$ to vertex $v$ with probability $p_{uv}$. On the other hand, $\{Y_t\}_{t\in\mathbb{Z}_+}$ can be referred as a second-order random walk where the vehicles move from edge to edge, i.e., we have to consider where the vehicle came from, the vertex visited before the current vertex. The second-order random walk has also been considered in graph analysis, see \cite{Wuetal2018}.

The state space of a first-order Markov traffic can be described by the function space $\mathcal{F}(V,\mathbb{N}_0)$ where $\mathbb{N}_0 :=\{0,1,2,\ldots\}$. A function $\bff=(f_v)_{v\in V}$ in $\mathcal{F} (V,\mathbb{N}_0)$ is called traffic configuration or counting function and $f_v$ measures the number of vehicles at vertex $v\in V$. Let $|\bff|$ denote the size of the traffic configuration $\bff$ defined by $|\bff|:=\sum_{v\in V}f_v$. The size of a traffic configuration counts the number of vehicles on the road network at a time. Let $\mathcal{F}_k$ ($k\in\mathbb{N}$) denote the subset of traffic configurations of size $k$. A p.d. $\bvarrho$ on $\mathcal{F} (V,\mathbb{N}_0)$ is a function $\bvarrho : \mathcal{F} (V,\mathbb{N}_0) \to [0,1]$ such that $\sum_{\bff} \bvarrho(\bff)=1$. For a p.d. $\bpi$ on the road network $G$ define the induced p.d. $\bvarrho$ on $\mathcal{F}_k$ as a multinomial distribution with parameters $k$ and $\bpi$, see Chapter 35 in \cite{Johnson1997Kotz}, given by the formula
\begin{equation}
\label{multinomial}
    \bvarrho(\bff):= k! \prod_{v\in V}
    \frac{\pi_v^{f_v}}{f_v!}
\end{equation}
for all $\bff\in \mathcal{F}_k$. By this formula the probabilities of complex events of the traffic can be computed.

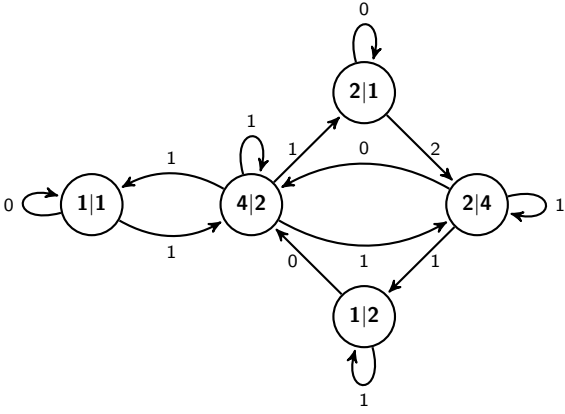
\begin{figure}
\centering

\begin{tikzpicture}[->,>=stealth',shorten >=1pt,auto,node distance=2.1cm,
                    thick,main node/.style={circle,draw,font=\sffamily\footnotesize\bfseries}]
 
  \node[main node] (3)  {2\textbar1};
  \node[main node] (2) [below left of=3] {4\textbar2};
  \node[main node] (1) [left of=2] {1\textbar1};
  \node[main node] (4) [below right of=3] {2\textbar4};
  \node[main node] (5) [below right of=2] {1\textbar2};  

  \path[every node/.style={font=\sffamily\scriptsize}]
    (1) edge [bend right] node [below] {1} (2)
          edge [loop left] node {0} (0)
    (2) edge node [left] {1} (3) 
        edge [bend right] node[above] {1} (1)
        edge [bend right] node[below] {1} (4)        
         edge [loop above] node {1} (2)
    (3) edge node [right] {2} (4)
          edge [loop above] node {0} (3)
    (4) edge [bend right] node [above] {0} (2)
         edge node [right] {1} (5)
          edge [loop right] node {1} (4)
    (5) edge node [left] {0} (2)
          edge [loop below] node {1} (5);    
\end{tikzpicture}
\caption{A transport matrix (on edges) on the road network in Fig.~\ref{graph-example-1} from configuration $\bff=(1,4,2,2,1)$ (left in vertices) to configuration $\bg=(1,2,1,4,2)$ (right in vertices) with $k=10$.}
\label{transport_matrix}
\end{figure}

Then, we define the concept of Markov kernel on $\mathcal{F}_k$. A Markov kernel $R$ on $\mathcal{F}_k$ is a function $\mathcal{F}_k \times \mathcal{F}_k \to [0,1]$ such that $\sum_{\bg\in\mathcal{F}_k} R(\bff,\bg)=1$ for all $\bff\in\mathcal{F}_k$. We demonstrate that every Markov kernel $P$ induces a natural Markov kernel on $\mathcal{F}_k$. The matrix $K=(k_{uv})_{u,v\in V}$ is called transport matrix from traffic configuration $\bff$ to $\bg$ on the road network $G$ if $K:V\times V\to \mathbb{N}_0$ is weakly $G$-subordinated, $\sum_{v\in V} k_{uv}=f_u$ for all $u\in V$, and $\sum_{u\in V} k_{uv}=g_v$ for all $v\in V$. In fact, $K$ has row and column marginals $\bff$ and $\bg$, respectively and, heuristically, $K$ defines a way for transporting the vehicles from configuration $\bff$ into $\bg$ on the road network. An example for transport matrix can be seen in Fig~\ref{transport_matrix}. For a pair $\bff,\bg\in \mathcal{F}(V,\mathbb{N}_0)$ let $\mathcal{M}(\bff,\bg)$ denote the set of all transport matrix from $\bff$ to $\bg$. Define the Markov kernel $R$ on $\mathcal{F}_k$ in the following way 
\begin{equation}
\label{multinomial_kernel}
      R(\bff,\bg) := \prod_{u\in V}
      f_u! \sum_{K\in \mathcal{M}(\bff,\bg)}
      \prod_{u,v:u \Rightarrow v}
      \frac{p_{uv}^{k_{uv}}}{k_{uv}!}
\end{equation}
where $\bff,\bg\in \mathcal{F}_k$. Then, $R$ maps a p.d. $\bvarrho$ into the p.d. $R\bvarrho$ on the state space $\mathcal{F}_k$ in the following way:
\begin{equation}
\label{Rkernel_trans}
     (R\bvarrho)(\bg):= \sum_{\bff\in\mathcal{F}_k} \bvarrho (\bff) R(\bff,\bg)
\end{equation}
for all $\bg\in\mathcal{F}_k$. To check that $R$ is a Markov kernel indeed we note that, by the multinomial theorem, 
\begin{equation}
\label{multi_kernel_sum}
\begin{split}
    \sum_{\bg\in\mathcal{F}_k} 
     R(\bff,\bg)= &\prod_{u\in V} f_u!
     \sum_{\sum\limits_{v\in V}k_{uv}=f_u}
      \prod_{u,v:u \Rightarrow v}
      \frac{p_{uv}^{k_{uv}}}{k_{uv}!} \\
     = & \prod_{u\in V} \left(\sum_{v\in V}
       p_{uv}\right)^{f_u} =1.
\end{split}       
\end{equation}
Moreover, one can easily see similarly to \eqref{multi_kernel_sum}, by the multinomial theorem, that if $\bpi$ is a s.d. of the Markov kernel $P$, then the p.d. $\bvarrho$ defined by \eqref{multinomial} is the s.d. of the induced Markov kernel $R$ defined by \eqref{multinomial_kernel}. Namely, we have the global balance equation 
\begin{equation}
\label{global_balance_traffconfig}
    \sum_{\bff\in\mathcal{F}_k} \bvarrho (\bff) R(\bff,\bg)= \bvarrho (\bg)
\end{equation}
for all $\bg\in \mathcal{F}_k$. (For the proof see \ref{proof}.)

Note that the concepts of traffic configuration and induced Markov kernel on them can be extended to the case of second-order Markov traffic by using the state space $\mathcal{F}(E,\mathbb{N}_0)$.
     
\smallskip

\textbf{Ergodicity of Markov traffic.} Let $\bpi_0$ be a p.d. on $V$. Using $\bpi_0$ as an initial distribution let us define the $n$-th absolute p.d. $\bpi_n$ by the recursion $\bpi_n^\top = \bpi^\top_{n-1}P$, $n\in \mathbb{N}$. Clearly, $\bpi^\top_n = \bpi^\top_0 P^n$, where the product of two Markov kernels $P$ and $Q$ on $V$ is defined as $(PQ)_{uw}:= \sum_{v\in V}p_{uv}q_{vw}$, $u,w\in V$. If $G$ is strongly connected and for the initial distribution $\bpi_0=\bpi$, where $\bpi$ is the unique stationary solution for $P$, see Theorem \ref{statsol}, then $\bpi^\top_n = \bpi^\top P^n =\bpi^\top$ for all $n\in \mathbb{N}$. Thus, in this case, $\bpi_n\to \bpi$ as $n\to \infty$. However, in general, the sequence $(\bpi_n)_{n\in \mathbb{N}}$ does not converge to the stationary distribution for all initial distribution $\bpi_0$ even if the stationary distribution is unique. However, if we consider the average of the $n$-th absolute p.d.'s in time, we have the convergence to the unique stationary distribution. The Markov kernel $P$ is called ergodic if it satisfies this property.

The following result on $G$-compatible Markov kernel is based on Theorem 4.1 in Chapter 3 of \cite{Bremaud1999}. If a road network $G$ is strongly connected then any $G$-compatible Markov kernel $P$ is ergodic and the average Markov kernel $A_n$ converges, i.e.,
\begin{equation*}
   A_n:=(n+1)^{-1}(I+P+\ldots +P^n)
   \to \Pi:= \textbf{1} \bpi^\top
\end{equation*}   
as $n\to\infty$, where $\bpi$ is the unique stationary distribution of $P$. Moreover, the limiting probabilities of the time averages of the absolute p.d.'s satisfy 
\begin{equation}
\label{average_lim}
(n+1)^{-1}(\bpi_0+\bpi_1+\ldots +\bpi_n)
\to \bpi
\end{equation}
as $n\to\infty$ for all initial p.d. $\bpi_0$.

In applications, along absolute p.d.'s, we may also be interested in some functionals of these distributions, e.g., the number of vehicles in a region of the road network. Define the functional
\[
      F(\bpi):=\sum_{v\in V} f_v\pi_v,
\]
of p.d. $\bpi$, where $\bff\in\mathcal{F}(V,\mathbb{R})$. Then, \eqref{average_lim} can be extended that $n^{-1}\sum_{k=1}^n F(\bpi_k)\to F(\bpi)$ as $n\to\infty$, see Theorem 4.1 of \cite{Bremaud1999}.

Instead of time averages, in order to achieve the convergence of $n$-th absolute p.d.'s we need the additional assumption of aperodicity for $G$, see Theorem 2.1 in Chapter 4 of \cite{Bremaud1999}. If $G$ is an aperiodic, strongly connected road network and $P$ is a $G$-compatible Markov kernel on it, then the sequence of Markov kernels $P^n$, $n\in \mathbb{N}$, converges to the limiting Markov kernel $\Pi$. Moreover, the limit of the sequence of $n$-th absolute p.d. $\bpi_n$ is the unique stationary distribution $\bpi$ to the Markov kernel $P$ which is independent of the initial p.d. $\bpi_0$. 

For any functional $F$ we also have that $F(\bpi_n)$ converges to $F(\bpi)$ as $n\to\infty$ on an aperiodic, strongly connected road network. In \cite{Bremaud1999} stronger convergence concepts, e.g., the convergence in variation, are also investigated.

Note that similar results hold for any $G$-compatible Markov kernel $P^\prime$ on $V^\prime =E$.

The above results on Markov kernels on $V$ (or $E$) can be extended to Markov kernels on the state space of traffic configurations. Let $\bvarrho_0$ be an initial p.d. on $\mathcal{F}_k$ and let us define the $n$th absolute p.d. $\bvarrho_n$ on $\mathcal{F}_k$ by the recursion $\bvarrho_n:= R\bvarrho_{n-1}$, $n\in\mathbb{N}$, where $R$ is a Markov kernel on $\mathcal{F}_k$ induced by a $G$-compatible Markov kernel $P$ on $G$, see formula \eqref{multinomial_kernel}. One can prove that the irreducibility and aperiodicity of $P$ imply the same properties for $R$, respectively. 

Our main result on ergodicity of Markov traffic is the following theorem. Note that the $n$th power of $R$ is defined recursively as $R^n \bvarrho:=R(R^{n-1}\bvarrho)$, $n=2,3,\ldots$, by formula \eqref{Rkernel_trans}.

\begin{Thm}\label{Markov_traffic_ergodic}
Let $G$ be a strongly connected and aperiodic road network and $P$ be a $G$-compatible Markov kernel. Then, there is a unique stationary distribution $\bvarrho$ to the Markov traffic described by the Markov kernel $R$ on $\mathcal{F}_k$ induced by $P$ which has the form \eqref{multinomial}. 

Moreover, the Markov traffic is ergodic in the sense that we have
\begin{equation*}
     R^n(\bff,\bg)\to \bvarrho(\bg)
\end{equation*}
as $n\to\infty$ for all $\bff,\bg\in \mathcal{F}_k$ and, for all initial p.d. $\bvarrho_0$ on $\mathcal{F}_k$,
\begin{equation*}
     \bvarrho_n(\bff)\to\bvarrho(\bff)
\end{equation*}
as $n\to\infty$ for all $\bff\in \mathcal{F}_k$.

Furthermore, the p.d. $\bpi$ on $G$ can be unfolded by the limit of state space averages in time as
\begin{equation}
\label{pi_limit}
     \frac{1}{k} \sum_{\bff\in\mathcal{F}_k} f_v \bvarrho_n(\bff) \to \pi_v
\end{equation}
as $n\to\infty$ for all $v\in V$.
\end{Thm}

Note that formula \eqref{pi_limit} follows from the well-kown fact that the expectation vector of a multivariate distribution with parameters $k$ and $\bpi$ is equal to $k\bpi$, see formula (35.6) in \cite{Johnson1997Kotz}.

These convergence results guarantee that the unique stationary distribution of a $G$-compatible Markov kernel can be approximated and thus explored by long run behavior of absolute p.d.'s on the traffic configurations of the road network.

\smallskip

\textbf{Two-dimensional stationary distribution.} Using two-dimensional distributions a Markov traffic can be reparametrized in the following way. Introduce the two-dimensional distribution $Q=(q_{uv})$ on $V\times V$ as $q_{uv}:=\pi_u p_{uv}$, $u,v\in V$. Then, $Q$ is a two-dimensional stationary distribution on $G$ in the following sense: 

\begin{definition}\label{two_stat}
A matrix $Q=(q_{uv})_{u,v\in V}$ is called \textbf{two-dimensional stationary distribution} on $G$ if (i) $q_{uv}\ge 0$ for all $u,v\in V$ and $q_{uv}=0$ for all $u,v\in V$ such that $(u, v)\notin E\cup S$ (i.e., $Q$ is weakly $G$-subordinated); (ii) $\sum_{u,v\in V}q_{uv} = 1$ (i.e., $Q$ is a normalized matrix on $V$); and (iii) $\sum_{v\in V} q_{uv} = \sum_{v\in V} q_{vu}$ for all $u\in V$ (i.e., $Q$ has equidistributed marginals). \end{definition}

A two-dimensional stationary distribution $Q$ on $G$ is called (strictly) positive if $q_{uv}>0$ for all $u,v\in V$ such that ($u\Rightarrow v$) $u\rightarrow v$.

\begin{figure*}
\centering

\begin{tikzpicture}[->,>=stealth',shorten >=1pt,auto,node distance=2.6cm,
                    thick,main node/.style={circle,draw,font=\sffamily\footnotesize\bfseries}]
 
  \node[main node] (3)  {3: 1/7};
  \node[main node] (2) [below left of=3] {2: 2/7};
  \node[main node] (1) [left of=2] {1: 1/7};
  \node[main node] (4) [below right of=3] {4: 2/7};
  \node[main node] (5) [below right of=2] {5: 1/7};  

  \path[every node/.style={font=\sffamily\scriptsize}]
    (1) edge [bend right] node [below] {1/14} (2)
          edge [loop left] node {1/14} (1)
    (2) edge node [left] {1/14} (3) 
        edge [bend right] node[above] {1/14} (1)
        edge [bend right] node[below] {1/14} (4)        
         edge [loop above] node {1/14} (2)
    (3) edge node [right] {1/14} (4)
          edge [loop above] node {1/14} (3)
    (4) edge [bend right] node [above] {1/14} (2)
         edge node [right] {1/14} (5)
          edge [loop right] node {1/7} (4)
    (5) edge node [left] {1/14} (2)
        edge [loop below] node {1/14} (5);  
\end{tikzpicture}
\caption{The two-dimensional stationary distribution (on edges) with its equidistributed marginals (on vertices) on the road network in Fig.~\ref{graph-example-1} for the Markov kernel in Fig.~\ref{pi_graph}. One can easily check that the sums of probabilities written on the edges in and out each vertex are equal, respectively.}
\label{2D_dist_toy}
\end{figure*}
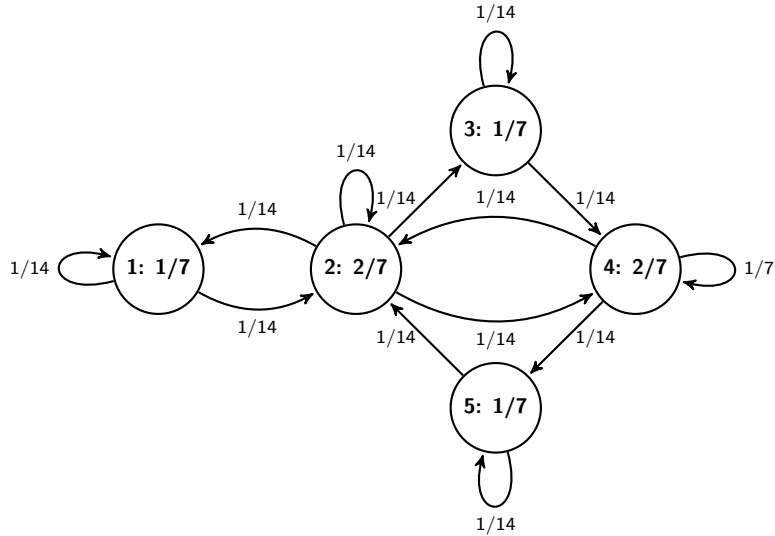

Property (iii) states that the two (row-wise and column-wise) marginal distributions of a two-dimensional stationary distribution on $G$ coincide with each other. Clearly, for a Markov traffic, $Q$ defined above is a positive two-dimensional distribution on $G$. $Q$ can also be considered as a p.d. on the state space $E\cup S$, i.e., if we extend the set $V^\prime$ of vertices of $\text{L}(G)$ as $V^\prime=E\cup S$, on the line digraph. Thus, $Q$ can be interpreted as the distribution of the vehicles on the edges of the road network, i.e., on the line digraph, see formula (11) in \cite{Crisostomietal2011}. The distribution $Q$ can also be visualized on the edges, see, Fig.~\ref{2D_dist_toy} for the toy example and Fig.~\ref{porto_stat_dist} in case of the Porto example discussed later. However, the converse of this statement is not true because there is p.d. on the line digraph which does not satisfy (iii). If $\{X_t\}_{t\in\mathbb{Z}_+}$ is a Markov random walk then the two dimensional distribution of any consecutive pair $(X_t,X_{t+1})$, $t\in\mathbb{Z}_+$, corresponds with $Q$.

Denote by $\mathcal{Q}$ the set of two-dimensional stationary distributions on $G$. One can easily see that $\mathcal{Q}$ is closed with respect to the affine combination. Namely, if $Q_1, Q_2\in\mathcal{Q}$ then $\lambda Q_1 + (1-\lambda)Q_2 \in\mathcal{Q}$ for all $\lambda\in [0,1]$. 

Conversely, for a positive $Q\in\mathcal{Q}$, let us define
\begin{equation}\label{QtoP}
\begin{aligned}
        \pi_u := &\sum_{v\in V} q_{uv} = \sum_{v\in V} q_{vu}, \quad u\in V, \qquad
        \\ p_{uv} := &\frac{q_{uv}}{\pi_u} ,\quad u,v\in V.
\end{aligned}
\end{equation} 

Then, $P=(p_{uv})$ defines a $G$-compatible Markov kernel with stationary distribution $\bpi$ on $G$. Thus, a Markov traffic defined by the quadruple $(G,P,\bpi,k)$ can be introduced by an equivalent way through the triplet $(G,Q,k)$. However, it will turn out later that, from a statistical point of view, the parameter matrix $Q$ can be estimated in a computationally more efficient way than the pair of transition matrix $P$ and its stationary distribution $\bpi$. 

With the help of two-dimensional stationary distribution, we can assign a p.d. to any Markov traffic on the space of traffic configurations which are defined as counting functions on the edges of the road network. Namely, define the traffic configuration $\bh=(h_{uv})_{u\Rightarrow v}$ as a function in $\mathcal{F}(E\cup S,\mathbb{N}_0)$. Then, $h_{uv}$ denotes the number of vehicles on the edge $(u,v)$ where $u,v\in V$ such that $u\Rightarrow v$. Similarly to \eqref{multinomial}, the two-dimensional distribution $\bsig$ on the set of traffic configurations $\bh$ with size $k$ ($k\in\mathbb{N}$), i.e., $\sum_{u\Rightarrow v}h_{uv}=k$, induced by a p.d. $\bpi$ on $G$ can be expressed as a multinomial distribution with parameter $k$ and $Q$ as 
\begin{equation*}
    \bsig(\bh):= k! \prod_{u\Rightarrow v}
    \frac{q_{uv}^{h_{uv}}}{h_{uv}!}
\end{equation*}
for all $\bh\in \mathcal{F}(E\cup S,\mathbb{N}_0)$.

Finally, one can easily see that the concept of two-dimensional stationary distribution can be extended for the second-order Markov traffic as well.

\subsection{Trajectories from public datasets}
\label{public-dataset}

For our experiments, we needed a dataset of real-life traffic trajectory data. In our terminology, a trajectory is a sequence of data that provides information about the path of a vehicle moving from a start to an end point, associating geographic coordinates with timestamps. We required a dataset that satisfies the following criteria:
\begin{enumerate}
\item Contains complete trajectories, i.e., the availability of only the start and end points is not sufficient, intermediate trajectory points must also be available.
\item The trajectory points must be sampled at a high enough frequency, so that the distance between consecutive points should not be too large, (e.g., an average distance of the order of 10 meters is acceptable, but an average distance of the order of 100 meters is definitely not.)
\item The dataset is sufficiently large. It should cover a long enough period of time, preferably uniformly. The number of trajectories per day should be of the order of thousands.
\item Trajectories should cover a relatively small geographic area, e.g., a city or a district.
\item Vehicles should not follow a fixed route, e.g., public transport bus trajectories are not suitable.
\item Publicly available for research purposes.
\end{enumerate}

\begin{figure}[!t]
    \centering
    \includegraphics[width=.49\textwidth]{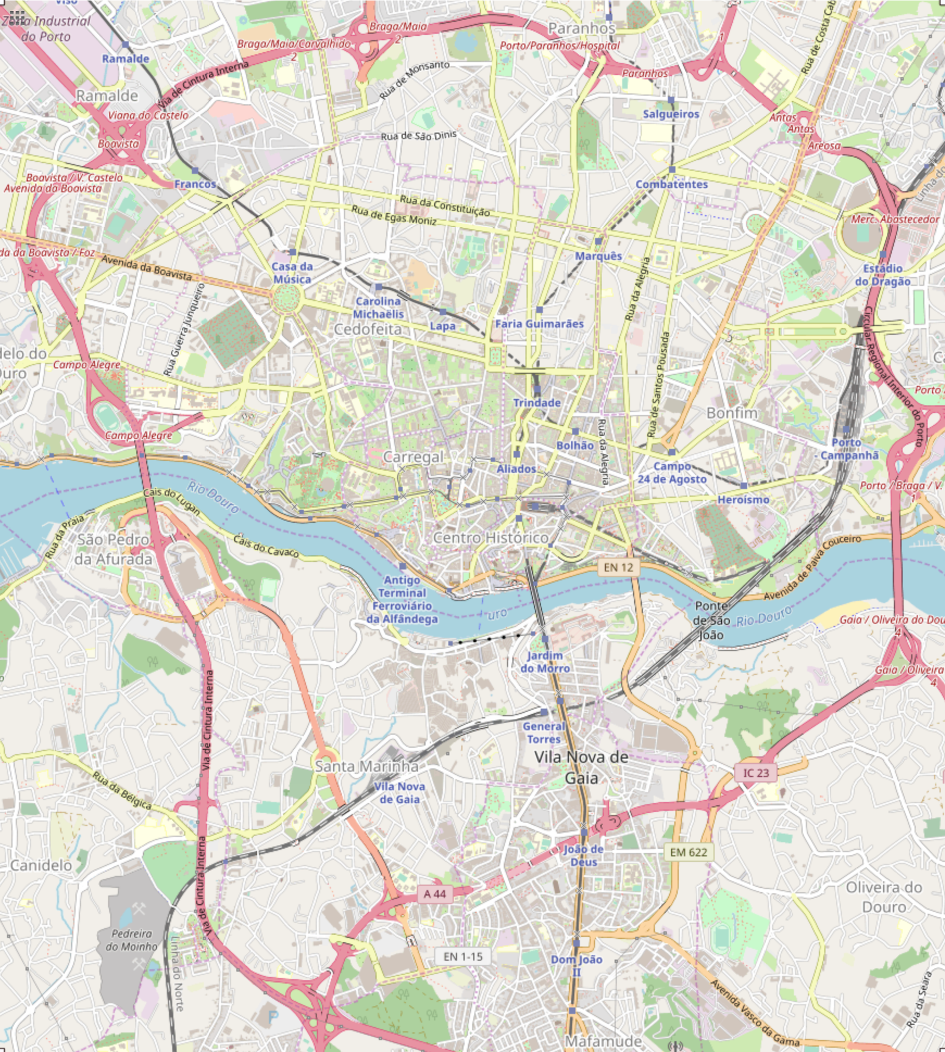}
    \caption{The map of the observed area. The graph created from the OSM data has 33,961 nodes, 53,126 edges, and covers a total of 857.26 km of road. The size of the area is about 43.68 km$^2$. (Image source: openstreetmap.org)}
    \label{porto_map}
\end{figure}

\begin{figure*}[!h]
    \centering
    \subfloat[Histogram of trajectory lengths. The rightmost bar represents trajectories longer than 12,500 meters. The average trajectory length is 3,628.93 meters.]{\includegraphics[width=0.48\textwidth]{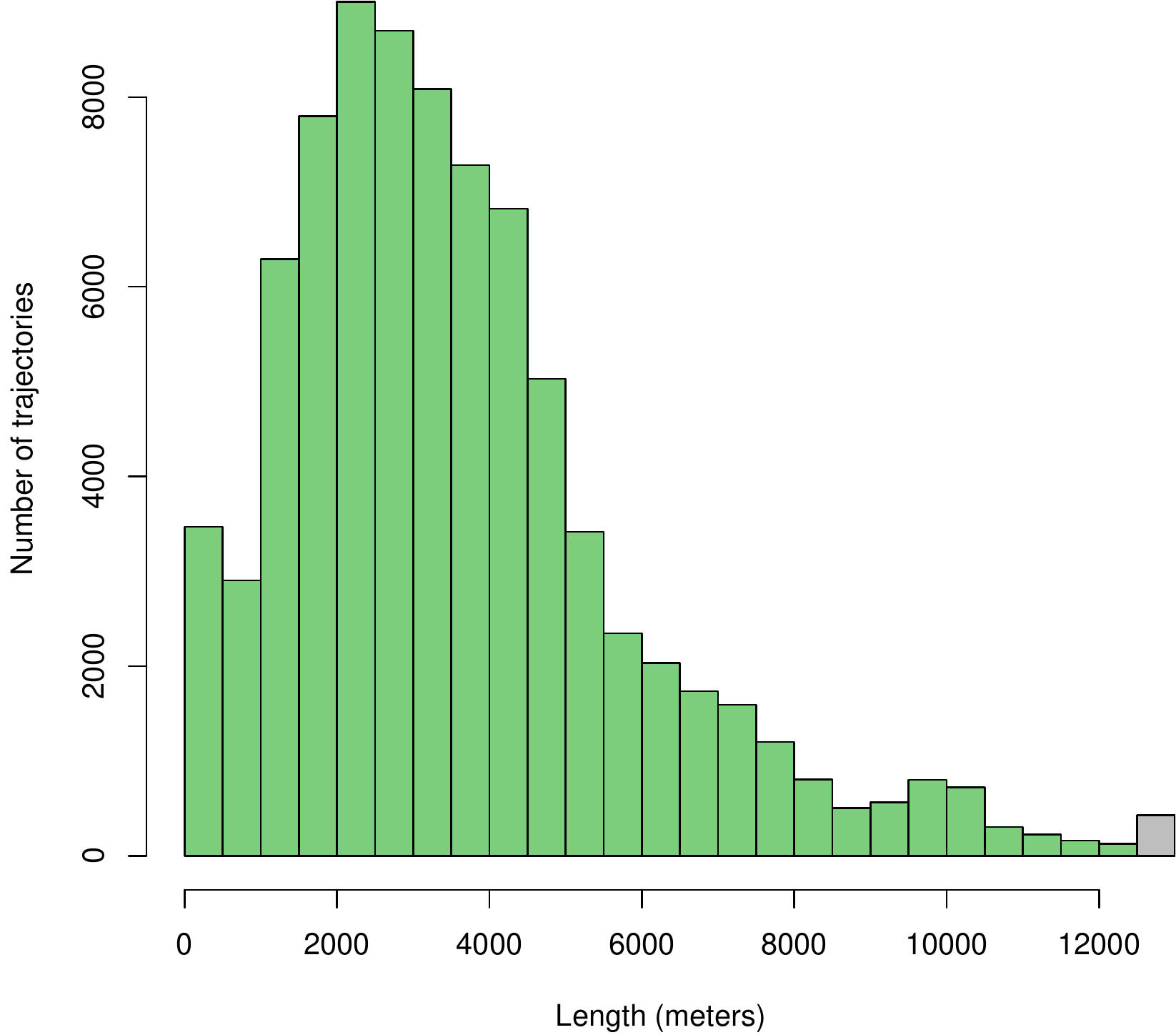}\label{pkdd15_trajectory_length}}
    \hfill
    \subfloat[Histogram of number of sample points per trajectories. The rightmost bar represents trajectories with more than 200 sample points. On the average, a trajectory consists of 40 sample points and takes 10 minutes.]{\includegraphics[width=0.48\textwidth]{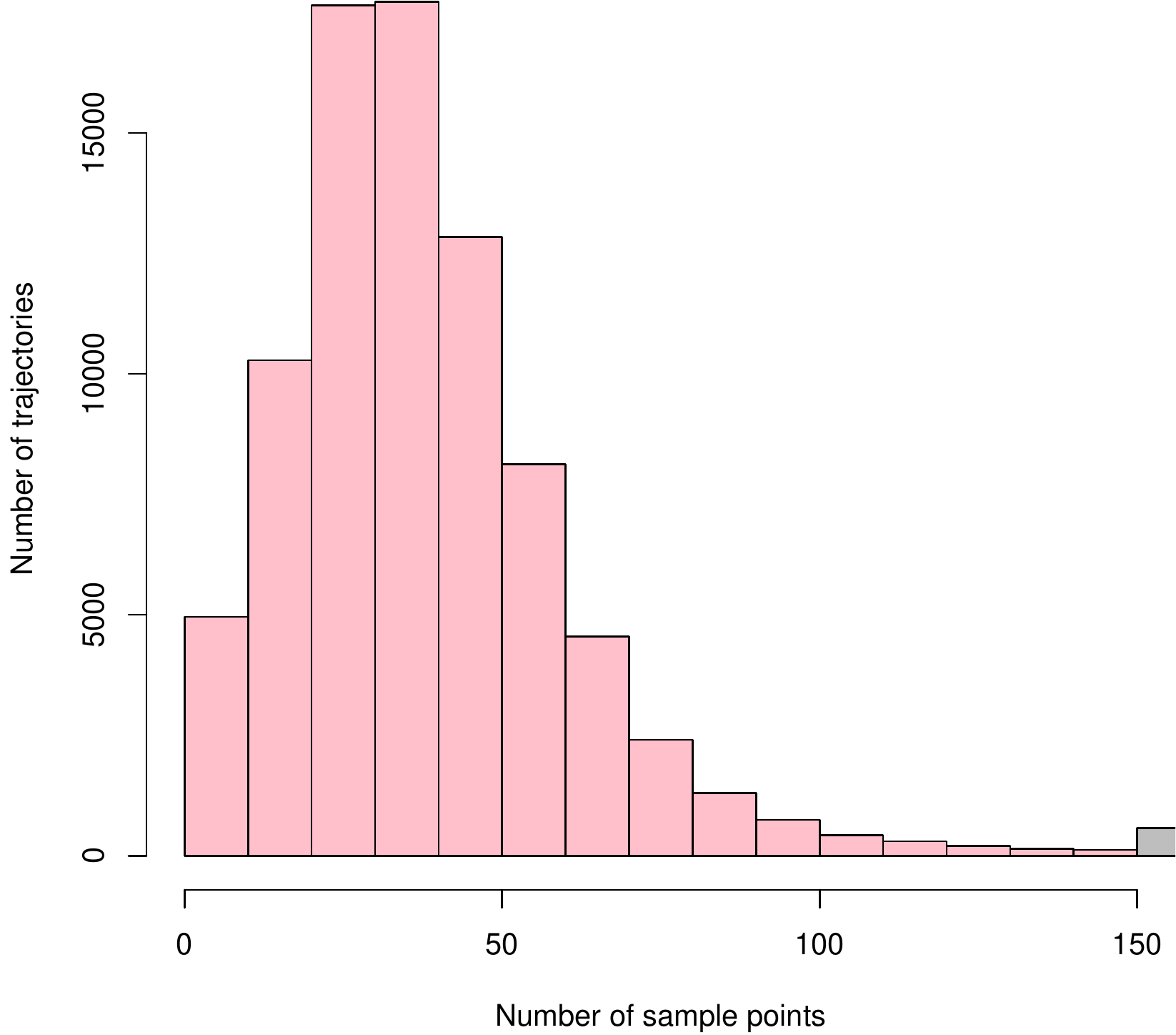}\label{pkdd15_sample_points}}
    \caption{Histograms of lenght of trajectories and sample points.}
    \label{tr_points_hist}
\end{figure*}

\begin{figure*}[!t]
    \centering
    \subfloat[Distribution of all trajectory points shown in a 2D histogram (number of bins: $80\times80$).]{\includegraphics[width=0.32\textwidth]{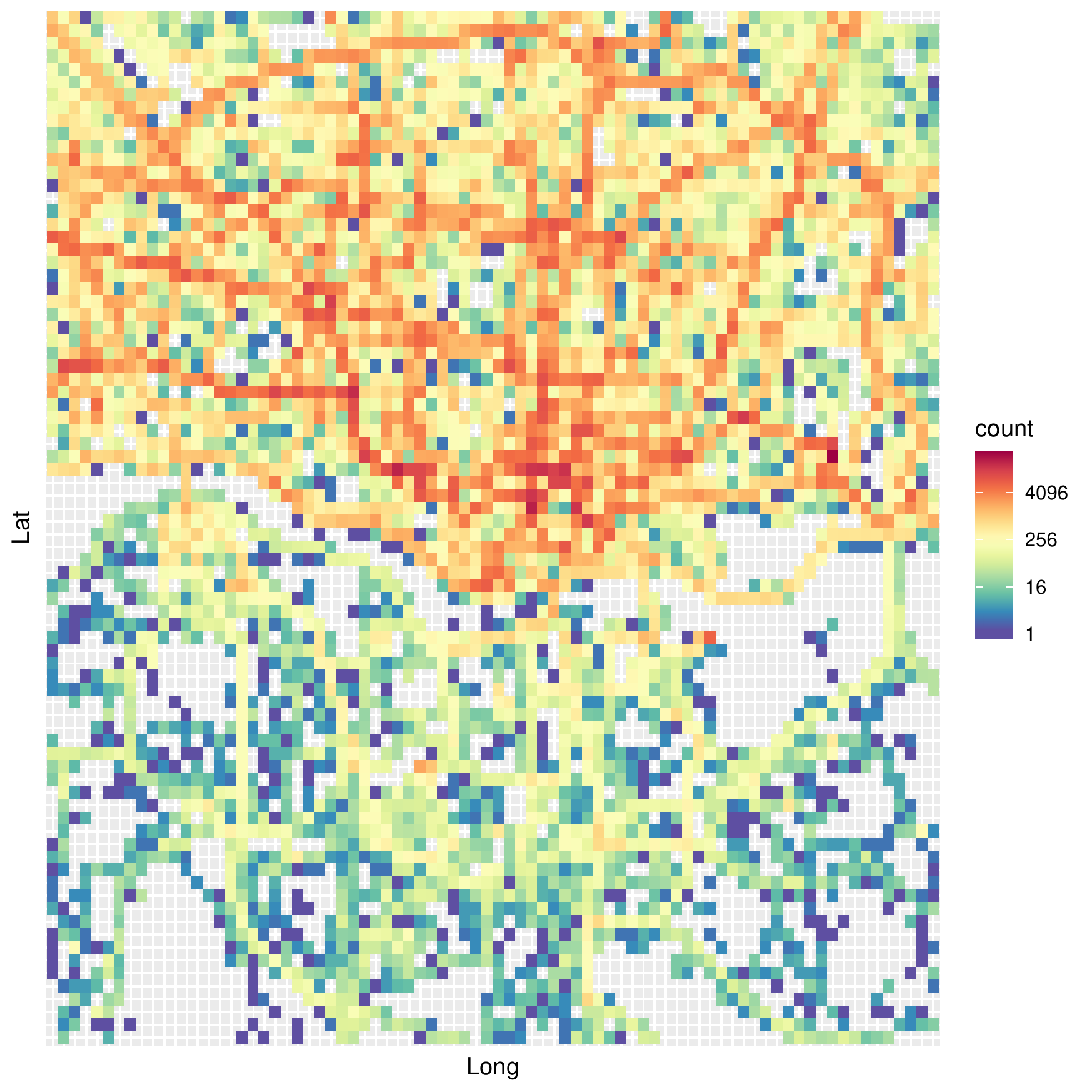}\label{tr_points_all}}
    \hfill
    \subfloat[Difference of trajectory starting and endpoints shown in a 2D histogram (number of bins: $80\times80$). The color of each bin represents the number of trajectory starting points minus the number of trajectory endpoints that fall in that bin.]{\includegraphics[width=0.32\textwidth]{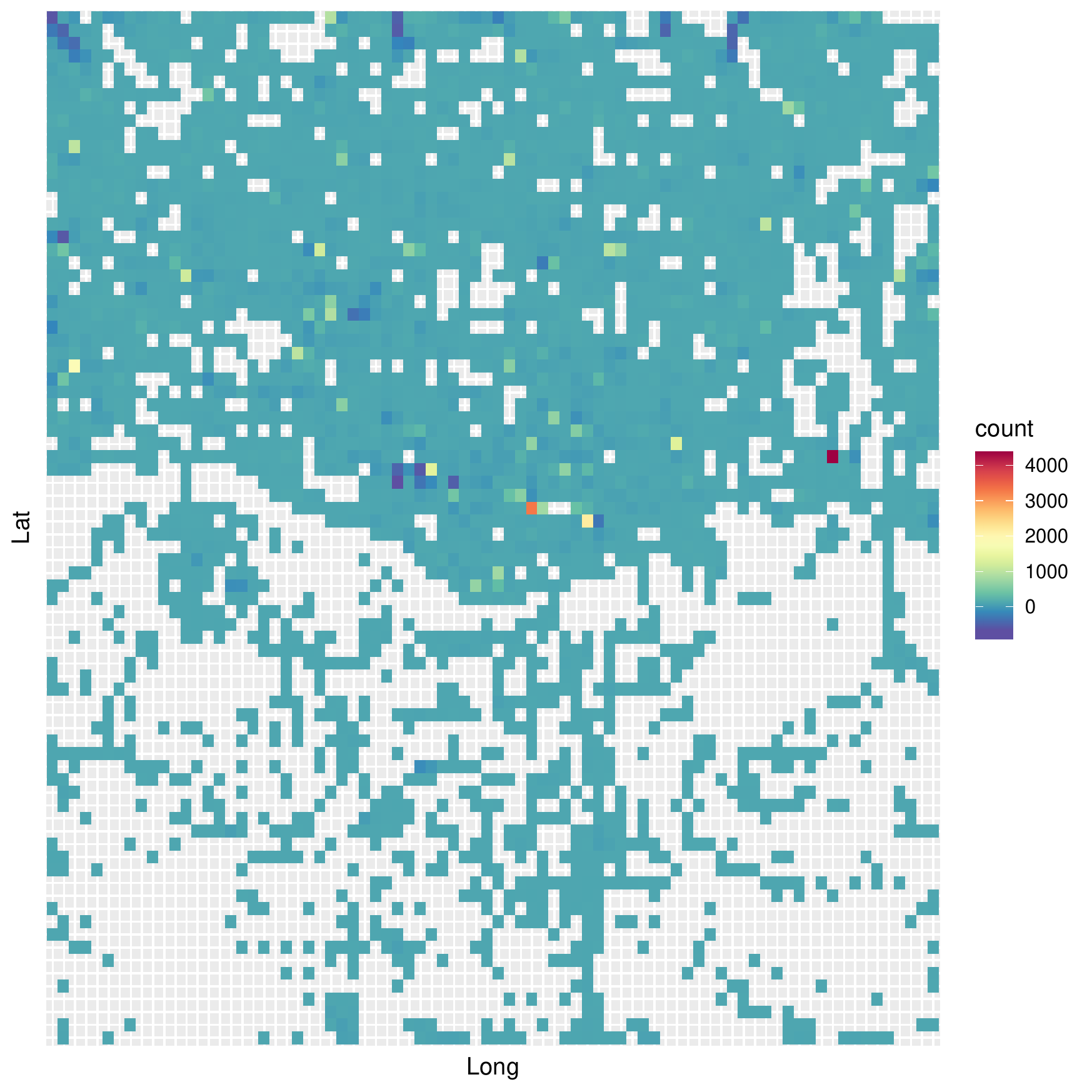}\label{tr_points_start}}
    \hfill
    \subfloat[Histogram of the difference of trajectory starting and endpoints.]{\includegraphics[width=0.32\textwidth]{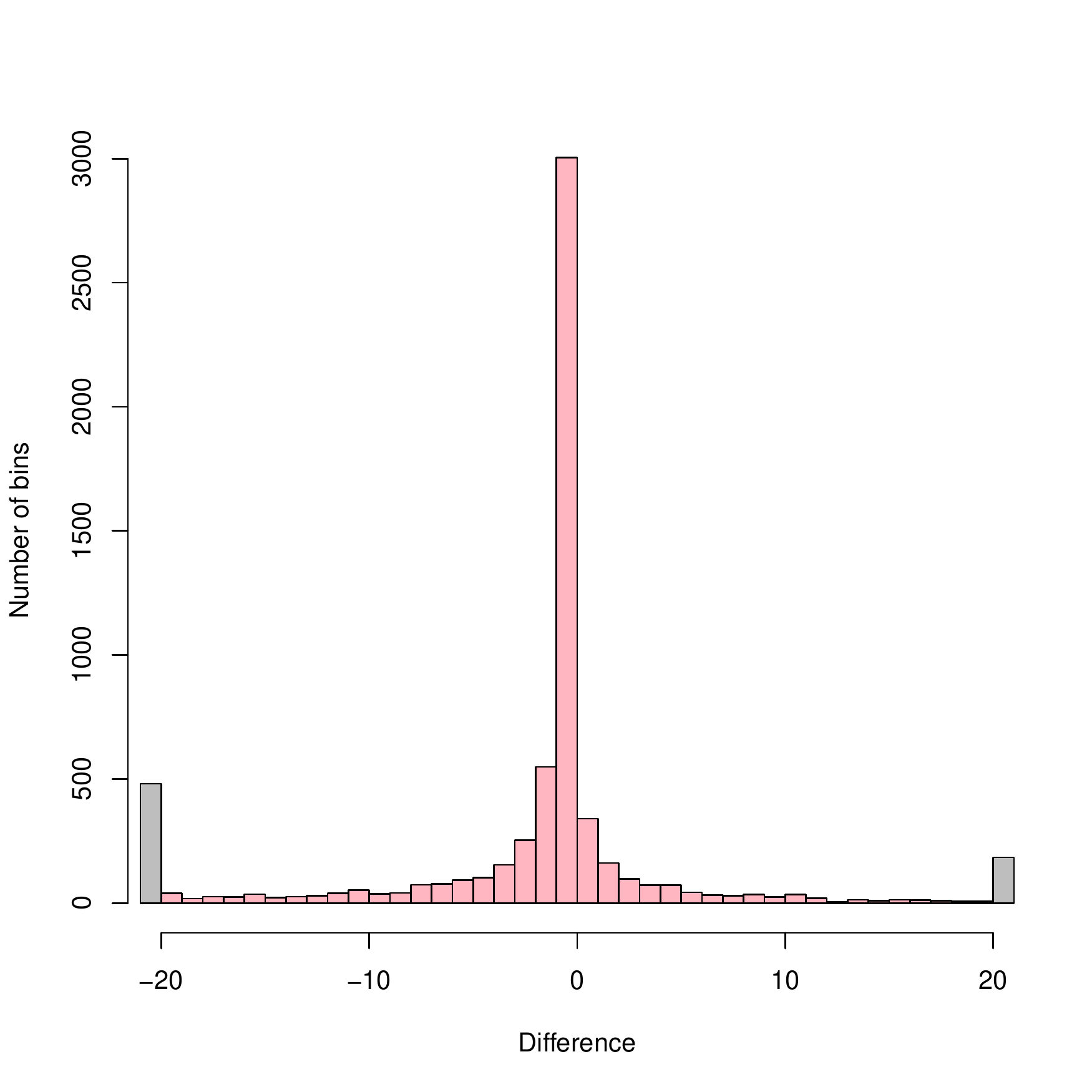}\label{tr_points_end}}
    \caption{Distribution of trajectory points of the filtered dataset.}
    \label{tr_points}
\end{figure*}

These requirements were satisfied by the Taxi Trajectory Prediction (TTP) dataset from Kaggle. The dataset covers a period of one year from July 1, 2013 to June 30, 2014. It is split into a training and a test set, the former contains 1,710,670 trajectories, and the latter contains 320. The trajectories were collected in the city of Porto, Portugal, with a sampling rate of 15 seconds. First, we created a subset of the dataset, filtered to coordinates between W8.6518, W8.5771, N41.1129, N41.1756, see Fig.~\ref{porto_map}. The data samples' features that were not relevant to the research, such as origin of call, identifiers for individual taxi or customers, and type of day (i.e. workday, weekend, holiday) were omitted. The processed format included the time of departure, both as a timestamp and as distinct date attributes, the length of the trajectory, and the points of the trajectory, represented as a list of GPS coordinates. Some data samples contained incomplete trajectories, these were discarded. Because of the properties of the proposed simulation model, the data was filtered to include only those samples that had a time of departure between 8-9 am. As a result, 82,345 trajectories remained. Although the length of trajectories had a wide range (the longest has 2,324 sample points), long trips were rare. Fig.~\ref{pkdd15_trajectory_length} shows the distribution of the length of trajectories. Most routes were around a length of 41 sample points, and routes with over 150 points were less than 1\% of the dataset, see Fig.~\ref{pkdd15_sample_points}. The distribution of the trajectory points (all, difference of start and end points, histogram of the difference) is shown in Fig.~\ref{tr_points}. The descriptive statistics of the dataset is shown in Table \ref{trajectory-lenght-stat}.

\begin{table}[t!]
\centering
\caption{Descriptive statistics of lengths of trajectories. (82,345 total trajectories.)}
\label{trajectory-lenght-stat} 
\begin{tabular}{l|c c}
Name of statistics  & Dist in points & Dist in meters \\ \hline
Mean                & 39.53     & 3,628.93    \\ 
Median              & 35     & 3,176.786   \\ 
Mode                & 34     & 0         \\ 
Standard Deviation  & 31.64  & 2,408.93  \\ 
Kurtosis            & 473.28 & 9.9  \\ 
Skewness            & 12.11   & 1.78  \\ 
Minimum             & 2      & 0         \\ 
Maximum             & 2,324   & 61,055.58  \\ 
\end{tabular}
\label{desc_stat}
\end{table}

\subsection{Building graphs from OpenStreetMap data}
\label{OSM-graph}

OpenStreetMap (OSM) is a community project to build a free map of the world to which anyone can contribute. Data is available under the Open Data Commons Open Database License (ODbL). The representation and storing of map data is based on a simple but powerful model, that uses only three modeling primitives, namely, nodes, ways, and relations\footnote{\url{http://wiki.openstreetmap.org/wiki/Elements}}:
\begin{enumerate*}
\item A node represents a geographical entity with GPS coordinates.
\item A way is an ordered list of at least two nodes.
\item A relation is an ordered list of nodes, ways, and/or relations.
\end{enumerate*}
All of these modeling elements can have associated key-value pairs called tags that describe and refine the meaning of the element to which they belong. Users can export map data at the OSM web site manually, selecting a rectangular region of the map. Alternatively, map data can be extracted via web services.\footnote{\url{http://wiki.openstreetmap.org/wiki/API}} OSM uses two formats for exporting map data, namely OSM XML and PBF. Software libraries for parsing and working with OSM data are available for several programming languages.\footnote{\url{https://wiki.openstreetmap.org/wiki/Frameworks}}

We started our processing by building a graph from the OSM map of Porto, with the same bounding box as the filtered dataset. Specific nodes of the OSM file become the nodes in the graph. Because we only need those nodes that can be reached via vehicles, we had to filter the OSM file and collect only specific types of way nodes. In the OSM file, a way is a sequence of OSM nodes, so naturally, the nodes of ways become nodes in the graph. For every node we store the node's OSM ID, and its coordinates. We also insert an edge into the graph between every nodes in way. The weight of an edge is given by the squared distance between the nodes, which we calculate from the OSM file's data. We used pyosmium library for processing the OSM files and the NetworkX Python library for building the graph.

After building the graph we process the list of trajectories. Because the trajectories are given in GPS coordinates, we first have to translate those coordinates into OSM node IDs. For every coordinate in a trajectory, we search for the closest way node's coordinates in the built graph, so the result nodes have the same domain as the built graph's nodes. Obviously, the original trajectories made up of GPS coordinates does not have the same scaling as the OSM map. The coordinates in the trajectory are sampled in regular, but larger time intervals than the OSM, so they are not aligned. In order to match a trajectory to a way in our graph, we had to perform an interpolation on the result list of node IDs, so we ran a Dijkstra's shortest path algorithm on our graph between every node IDs for every trajectory. Because the OSM database contains errors, it can happen that in real life a route exists between two given places, but in the OSM database, there are no existing routes between those nodes that are representing the given places. In this case, we cut the faulty trajectories into pieces. The result of this process is an aperiodic strongly connected road network augmented by the ideal vertex 0.

\subsection{Statistical inference for Markov traffic using mobile sensors}
\label{inference}

The statistical analysis of a traffic systems described by the Markov traffic model means the estimation of the quadruple $(G,P,\bpi,k)$ or the triplet $(G,Q,k)$ using observed data. (In this subsection only the first-order Markov traffic is considered.) To estimate $G$ we have to explore the road system under study by identifying the set $V$ of vertices and the set $E$ of directed road segments. Fortunately, this exploring has already been done by a few organizations, see, e.g., the Google Maps and the OpenStreetMap. However, we should note that, in case of GPS-based trajectory data, we have to fit the data to the applied map system which is not an evident task at all, see section \ref{OSM-graph}. In the present paper, we propose a method for estimating the two-dimensional stationary distribution $Q$ immediately instead of the pair $(P,\bpi)$ of a transition matrix and its stationary distribution using mobile sensor data which may be gathered by vehicles, passengers etc. In this case, we have trajectories data which consists of the sequences of consecutive vertices, like in the TTP dataset. By \eqref{QtoP}, the estimators for $P$ and $\bpi$ can be easily derived from an estimator of $Q$. Finally, it is supposed that the size $k$ of the traffic is known.

Suppose that, for a Markov traffic, we observed a random sample of trajectories  $\{X^i\}$, $i=1,\ldots,k$, of size $k$ defined by $X_1^i\Rightarrow X_2^i\Rightarrow\ldots\Rightarrow X_{n_i}^i$, $i=1,\ldots,k$, where $n_i$ denotes the length of the $i$-th trajectory. Let $n:=n_1+\ldots +n_k$ be the total sample size. Define the total two-dimensional consecutive empirical frequencies as:
\begin{equation}\label{2freq}
    n_{uv}:= \sum_{i=1}^k n_{uv}^i,
\end{equation}     
$u,v\in V$, where the trajectory-wise two-dimensional consecutive empirical frequencies, $i=1,\ldots,k$, are defined as
\[
    n_{uv}^i := \sum_{j=1}^{n_i-1} I(X_j^i=u,X_{j+1}^i=v),
\] 
$u,v\in V$. Plainly, $n_{uv}^i$ denotes the number of consecutive $(u,v)$ ($u,v\in V$) pairs in the $i$-th trajectory. One can see that since $\{X^i\}$ is a proper Markov random walk we have $n_{uv}^i=0$ for all $(u,v)\notin E\cup S$. Thus, the support of the two-dimensional frequency matrices $N:=(n_{uv})_{u,v\in V}$, $N_i:=(n_{uv}^i)_{u,v\in V}$, $i=1,\ldots,k$, is a subset of $E\cup S$, i.e., they are weakly $G$-subordinated matrices. Clearly, $N= \sum_{i=1}^k N_i$ and we have
\begin{equation}\label{corr_sample_size}
    \sum_{u,v:u\Rightarrow v} n_{uv} = n-k,
\end{equation}
where $n-k$ is the corrected sample size. Introduce
\[
    s_v := \sum_{i=1}^k I(X_1^i=v), \quad
          e_v := \sum_{i=1}^k I(X_{n_i}^i = v),
\] 
$v\in V$, i.e., $s_v$ denotes the number of trajectories which start at vertex $v$ and $e_v$ denotes the number of trajectories which terminate at vertex $v$, respectively. Denote the one-dimensional marginal frequencies of $N$ by $n_{v+}:=\sum_{u\in V} n_{vu}$ and $n_{+v}:=\sum_{u\in V} n_{uv}$, $v\in V$. We obtain that
\begin{equation}\label{start_end}
    n_{v+}+e_v = n_{+v} + s_v = n_v :=\sum_{i=1}^k \sum_{j=1}^{n_i} I(X_j^i=v)
\end{equation} 
for all $v\in V$, where $n_v$ denotes the number of vertex $v$ in all trajectories. Finally,
\begin{equation}\label{start_end_sum}        
    \sum_{v\in V}  s_v = \sum_{v\in V}  e_v = k.
\end{equation} 
Define the vectors $\bs$ and $\be$ on $V$ as $\bs:=(s_v)_{v\in V}$ and $\be:=(e_v)_{v\in V}$, respectively. Then, \eqref{start_end_sum} implies that $\textbf{1}^\top (\be -\bs) = 0$, i.e., the vectors $\be -\bs$ and $\textbf{1}$ are orthogonal.

The traditional maximum likelihood (ML) estimator $\widehat{P}_{\textrm{ML}}$ of the transition matrix $P$ is given by the maximization of the conditional loglikelihood
\[
      \log L = \sum_{u\Rightarrow v} n_{uv}\log p_{uv}
\]
in parameters $p_{uv}$, $u,v\in V$ such that $u\Rightarrow v$, under the constraints $p_{u+}=1$ for all $u\in V$. A solution of this constrained optimization problem is $\widehat{p}^{\textrm{ML}}_{uv}=n_{uv}/n_{u+}$ for all $u,v\in V$ if $n_{u+}>0$ and $\widehat{p}^{\textrm{ML}}_{uv}=\delta_{uv}$ if $n_{u+}=0$ where $\delta$ denotes the Kronecker delta. The maximum likelihood estimator $\widehat{\bpi}_{\textrm{ML}}$ of the stationary distribution $\bpi$ is derived by the solution of the global balance equation $\bpi^\top =\bpi^\top \widehat{P}_{\textrm{ML}}$ in $\bpi$. Thus, the maximum likelihood estimator $\widehat{Q}_{\textrm{ML}}=(\widehat{q}^{\textrm{ML}}_{uv})$ of $Q$ is given by $\widehat{q}^{\textrm{ML}}_{uv}=\widehat{\pi}^{\textrm{ML}}_u \widehat{p}^{\textrm{ML}}_{uv}$, $u,v\in V$. In the sequel, a direct method is proposed for estimating the two-dimensional stationary distribution $Q$.

A na\^ ive estimator for the two-dimensional stationary distribution $Q$ based on the two-dimensional consecutive empirical frequency matrix $N$ is $\widehat{Q}_{\textrm{na\^ ive}}:= (n-k)^{-1}N$. Clearly, $\widehat{Q}_{\textrm{na\^ ive}}$, as a non-negative matrix on $V$, satisfies the properties (i) and (ii) of Definition \ref{two_stat}. However, the problem with this na\^ ive estimator is that its row and column marginals are not necessarily equal, i.e., in general, it does not satisfy the asumption (iii) of Definition \ref{two_stat}. Hence, we have to introduce a new estimator $\widehat{Q}$ which belongs to $\mathcal{Q}$ and is optimal in some sense.

The optimality of the proposed estimator is defined by means of the least squares distance between matrices over $G$. Let $A=(a_{uv})_{u,v\in V}$ and $B=(b_{uv})_{u,v\in V}$ such that $a_{uv}=b_{uv}=0$ for all $u,v\in V$ where $u\nRightarrow v$, i.e., let $A$ and $B$ be weakly $G$-subordinated matrices. The distance between $A$ and $B$ is defined as 
\[
    \| A-B\|_G := \left(\sum_{u,v:u\Rightarrow v} |a_{uv} - b_{uv}|^2\right)^{1/2}.
\]
In fact,  $\| \cdot\|_G$ is the Frobenius norm of the matrices of dimension $|V|\times |V|$ which vanish on the entries outside of $E\cup S$. 

To formulate our error or objective function for estimating the two-dimensional stationary distribution it is convenient to weaken the assumptions of Definition \ref{two_stat} by leaving the normalizing assumption (ii). In the sequel, let $M=(m_{uv})$ denote a non-negative parameter matrix on $G$ which satisfies assumptions (i) and (iii) of Definition \ref{two_stat}, i.e., $M$ is weakly $G$-subordinated and $\sum_{v\in V} m_{uv} = \sum_{v\in V} m_{vu}$ for all $u\in V$. Then, one can easily derive a two-dimensional stationary distribution $Q$ from $M$ by its normalization defining as $Q:=(\textbf{1}^\top M\textbf{1})^{-1}M$.

Based on $k$ number of trajectories, using the Frobenius norm, the optimality criterion is defined as the weighted sum of squared errors (SSE):
\begin{equation}\label{square_error}
    \text{SSE}(M,\bw\,|\,\bN) :=  \sum_{i=1}^k w_i^{-1} 
    \|N_i- w_i M\|_G^2,
\end{equation}
where $M$ is a non-negative parameter matrix satisfying assumptions (i) and (iii) of Definition \ref{two_stat}, $\bw=(w_i)_{i=1,\ldots,k}$ are non-negative unknown weights, i.e., $\sum_{i=1}^k w_i=1$, and $\bN:=(N_i)_{i=1,\ldots,k}$ denotes the data, where $N_i$ is the two-dimensional consecutive empirical frequency matrix for the $i$th trajectory, see \eqref{2freq}. The statistical inference for a Markov traffic means the minimization of the objective function SSE in its parameters $M$ and $\bw$ deriving the weighted least squares (WLS) estimators $\widehat{M}_{\textrm{WLS}}$ and $\widehat{\bw}_{\textrm{WLS}}$. Then, the WLS estimator of $Q$ is defined as $\widehat{Q}_{\textrm{WLS}}:= n_{\textrm{eff}}^{-1}\widehat{M}_{\textrm{WLS}}$ where $n_{\textrm{eff}} := (\textbf{1}^\top \widehat{M}_{\textrm{WLS}}\textbf{1})$ is the so-called effective sample size. Here, $\widehat{Q}$ can be interpreted as the estimated two-dimensional stationary distribution which describes the individual Markov traffic in time. On the other hand, $n_{\textrm{eff}}$ gives the equivalent sample size related to the independent case which may be thought of as the information content of the observed data. Note that $n_{\textrm{eff}}$ is not necessarily an integer and is different from $n$ and $n-k$. Finally, $w_i$ gives the importance of the $i$th trajectory in the sample. One can see that longer trajectory implies higher weight.

To formulate our result on WLS estimation of Markov traffic we need some basic facts on the spectral theory of directed graphs, see \cite{von2007tutorial} for details. The symmetric unnormalized graph Laplacian matrix $L$ of a digraph $G$ is defined as
\[
          L:= D - A-A^\top
\]
where $A$ denotes the adjacency matrix of $G$ and $D:=\diag\{\bd^{+} + \bd^{-}\} $. Note that for the road graph $G$, since there is no loop, we have $l_{vv}= d_{vv}=\deg^{+}(v)+\deg^{-}(v)$ for all $v\in V$. The main theorem of this paper is the following.

\begin{Thm}\label{main_thm}
There is a unique pair $(\widehat{M}_{\textrm{WLS}},\widehat{\bw}_{\textrm{WLS}})$ which minimizes the weighted sum of squared errors SSE defined in \eqref{square_error}. These WLS estimators are derived as
\[
    \widehat{w}^i_{\textrm{WLS}} :=  
    \frac{\Vert N_i\Vert_G}{\sum_{j=1}^k \Vert N_j\Vert_G},
\]
$i=1,\ldots,k$, and
\[
    \widehat{M}_{\textrm{WLS}}:= N+(\textbf{1}\blambda^\top -\blambda\textbf{1}^\top)\circ A,
\] 
where $\blambda\in\mathcal{F}(V,\mathbb{R})$ is called Lagrange vector and defined as a unique solution to the linear equation $L\blambda = \bs-\be$ and $\circ$ denotes the entrywise (Hadamard) product of matrices.
\end{Thm}

Based on the previous theorem, by \eqref{corr_sample_size}, the effective sample size is given as
\begin{equation}\label{eff_sample}
    n_{\textrm{eff}} := (n-k) + (\bd^- - \bd^+)^\top \blambda,
\end{equation}
i.e., $n_{\textrm{eff}}$ depends only on the graph structure of the road network, which is independent of the data, the traffic direction vector $\bs-\be$, and the corrected sample size. However, it does not depend on the data which are inside the trajectories.  

The WLS estimators proposed above can be considered as a kind of composite (or quasi-) likelihood estimators for Markov chains, see \cite{Hjort2008Varin}. The composite likelihood method is widely applied in complex statistical models when the full ML method can be difficult to apply or may not be robust enough. In our method, the objective function is based on pairwise marginal distributions, however, instead of formula (2) in \cite{Hjort2008Varin}, the quasi-likelihood function is a square function, the logarithm of the normal probability density with heteroscedastic variance which depends on the length of trajectories. The latter will be more clear by introducing the mean squared error (MSE) as
\begin{equation}\label{mean_square_error}
    \text{MSE} :=n_{\textrm{eff}}^{-1} \text{SSE} =  \sum_{i=1}^k n_{\textrm{eff}}^i \| (n_{\textrm{eff}}^i)^{-1} 
    N_i- Q\|_G^2,
\end{equation}
where $n_{\textrm{eff}}^i:=w_i n_{\textrm{eff}}$ denotes the effective sample size of the $i$th trajectory, $i=1,\ldots,k$. The parameters of the objective function MSE are the effective sample sizes $\{n_{\textrm{eff}}^i\}$ and the two-dimensional stationary distribution $Q$.
The heuristic explanation of the need to use weights in formulas \eqref{square_error} and \eqref{mean_square_error} is the following. By the Central Limit Theorem, for large $n_i$, the trajectory-wise two-dimensional consecutive empirical frequency matrix $N_i$ can be approximated as $N_i \approx n_i Q + n_i^{1/2}\bxi_i$, where $\bxi_i$ is a normally distributed random matrix on $V$ for all $i=1,\ldots,k$, which is a heteroscedastic equation between the observed $N_i$ and the parameter matrix $Q$. Hence, $\Vert N_i - n_i Q\Vert_G^2 \approx n_i \Vert\bxi_i\Vert_G^2$, where $\Vert\bxi_i\Vert_G^2$, $i=1,\ldots,k$, are independent identically distributed r.v.'s. Thus, we have to normalize the trajectory-wise squared errors proportionally to their lengths, respectively, in order to get balanced error terms.

The estimation theory of finite Markov chains goes back for a long time, see \cite{billingsley1961}. In the traditional ML approach the estimators of the transition and stationary probabilities are derived by corresponding relative frequencies, respectively. However, these estimators have a few problems which imply that they can be applied with limited success for estimating the Markov traffic on a road network. Firstly, they are based on only one long trajectory (or realization). However, in a real traffic dataset there is a large number of relatively short trajectories, i.e., the set $\{n_i,i=1, \ldots,k\}$ are bounded, where $k$ is large or tends to infinity. In our example, for the TTP dataset, the number of trajectories is above 80K with the mean length 40 and maximum length 2K, see Table~\ref{desc_stat}. Secondly, they are asymptotic estimators in the sense that, for finite sample size, the estimated stationary distribution does not satisfy the global balance equation given by the estimated transition probability matrix. The global balance equation holds only asymptotically, i.e., when the sample size tends to infinity. In fact, the inaccuracy in the global balance equation is not too large, however, this little bias can cause significant discrepancy from the \enquote{true} stationary distribution in the simulation. Thirdly, the trajectories are biased during a short time period in the sense that they are starting from some parts of the road network and ending at other parts. For example, in the morning period the vehicles are moving from the residential districts to the business districts of the city and they are moving back in the afternoon period. In other words, the traffic has a definite direction on the road network. To demonstrate this behavior in the case of TTP dataset, Fig.~\ref{tr_points_end} shows the distribution of the elements of the traffic direction vector $\bs-\be$ while Fig.~\ref{tr_points_start} shows their spatial distribution. Neither distributions are concentrated around the zero. The known improvements of the ML estimators, e.g., by using the bootstrap, see \cite{teodorescu2009}, do not solve these problems. However, the WLS estimator of the two-dimensional stationary distribution proposed in this paper is able to handle all of these problems. The estimator $\widehat{Q}_{\textrm{WLS}}$ is taking account of more than one trajectory with their length. It determines uniquely both the transition probability matrix and its stationary distribution by \eqref{QtoP} which satisfy the global balance equation obviously. Finally, by taking account of the traffic direction vector in the estimator, it can correct the bias due to the unbalanced sampling of trajectories on the road network. 

The fundamental statement of Theorem \ref{main_thm}, as one of the main result of this paper, is that the estimator $\widehat{Q}_{\textrm{WLS}}$ (or $\widehat{M}_{\textrm{WLS}}$) consists of two parts: the first part is the na\^ ive estimator for the distribution of the consecutive pairs in trajectories based on the empirical frequencies, while the second part is a correction term ensuring that $\widehat{Q}_{\textrm{WLS}}$ (or $\widehat{M}_{\textrm{WLS}}$) has equidistributed marginals. The second part also depends on two components. The first one is the Laplacian matrix of the road graph which depends only on the graph structure of the road network and independent from the trajectory data. The second one is the traffic direction vector which depends only on the trajectory data. Note that all sufficient statistics, namely the total two-dimensional consecutive empirical frequencies and starting and ending empirical frequencies, can be computed by counting, which is numerically very effective and can be executed even for big data. 

The following propositions and remarks are useful in computing the Lagrange vector $\blambda$ in a numerically effective manner. The first proposition sums up the basic properties of the symmetric unnormalized graph Laplacian matrix. Note that 1)--4) are based on Proposition 1 in \cite{von2007tutorial}.

\begin{Pro}
\label{Laplace_properties}
The matrix $L$ satisfies the following properties:
\begin{enumerate}
\item For all $\balpha\in\mathcal{F}(V,\mathbb{R})$ we have
\begin{equation}\label{Lquadform}
    \balpha^\top L\balpha = \frac{1}{2} \sum_{u,v\in V} 
    (a_{uv} + a_{vu}) (\alpha_u -\alpha_v)^2
\end{equation}
\item $L$ is symmetric and positive semi-definite.
\item The smallest eigenvalue of $L$ is 0, the corresponding eigenvector is the constant one vector $\textbf{1}$.
\item $L$ has $|V|$ non-negative, real-valued eigenvalues $0 =\tau_1\le\tau_2\le\ldots\le\tau_{|V|}$ and corresponding orthonormal eigenvectors $\textbf{1}= \balpha_1,\balpha_2,\ldots,\balpha_{|V|}$ in $\mathcal{F}(V,\mathbb{R})$.
\item If $G$ is strongly connected then $\tau_2>0$, i.e., the multiplicity of the eigenvalue $\tau_1=0$ is 1, and $L$ is invertible on the $|V|-1$-dimensional invariant subspace $\mathcal{S}:= \{\balpha\in\mathcal{F}(V,\mathbb{R})\,|\,\textbf{1}^\top\balpha=0 \}$ and the inverse can be expressed as $L^{-1}_{\mathcal{S}} = \sum_{j=2}^{|V|} \tau_j^{-1} \balpha_j\balpha_j^\top$.
\end{enumerate} 
\end{Pro}

By Proposition \ref{Laplace_properties} the Lagrange vector can be expressed by the eigenvalues and the eigenvectors of $L$ as
\[
    \blambda =L^{-1}_{\mathcal{S}} (\bs-\be) = \sum_{j=2}^{|V|} \tau_j^{-1} (\balpha_j^\top (\bs-\be)) \balpha_j .
\]
This formula can be applied for approximating $\blambda$ by leaving the too large eigenvalues which have negligible inverse in the above finite sum in a large road network. Thus, if $|V|$ is too large then it is enough to store the first few significant eigenvalues and eigenvectors. Remark that these eigenvalues and eigenvectors, being independent from the data, can be computed and stored in advance for a simulation program.

If a sharp bound is needed for the eigenvalues then the application of symmetric normalized Laplacian matrix is more useful. The symmetric normalized adjacency matrix $\widetilde{A}$ and the symmetric normalized graph Laplacian matrix $\widetilde{L}$ of a digraph $G$ are defined as
 \[
    \widetilde{A}:= D^{-1/2}(A+A^\top)D^{-1/2}, \qquad   \widetilde{L}:=  I -  \widetilde{A}.
\]
For a road graph, we have $\widetilde{l}_{vv}=1$ for all $v\in V$. One can see that
\[
    L= D^{1/2}\widetilde{L}D^{1/2}, \qquad   \widetilde{L}:= D^{-1/2} LD^{-1/2}
\]
The next proposition is based on Proposition 3 in \cite{von2007tutorial}, the statements 5)--7) are new as far as we know.

\begin{Pro}\label{Sym_Laplace_properties}
The matrices $\widetilde{A}$ and $\widetilde{L}$ satisfy the following properties:
\begin{enumerate}
\item For all $\balpha\in\mathcal{F}(V,\mathbb{R})$ we have
\begin{equation}\label{Lquadform2}
    \balpha^\top \widetilde{L}\balpha  
    = \frac{1}{2} \sum_{u,v\in V} (a_{uv} + a_{vu}) \left(\frac{\alpha_u}{\sqrt{d_u}} -\frac{\alpha_v}
    {\sqrt{d_v}}\right)^2
\end{equation}
\item $\widetilde{L}$ and $\widetilde{A}$ are symmetric and $\widetilde{L}$ is positive semi-definite.
\item The smallest eigenvalue of $\widetilde{L}$ is 0, the corresponding eigenvector is the vector $D^{1/2}\textbf{1}$.
\item $\widetilde{L}$ has $|V|$ non-negative, real-valued eigenvalues $0 =\widetilde{\tau}_1 \le\widetilde{\tau}_2\le\ldots\le\widetilde{\tau}_{|V|}\le 2$ and corresponding orthonormal eigenvectors $D^{1/2}\textbf{1}= \widetilde{\balpha}_1,\widetilde{\balpha}_2,\ldots, \widetilde{\balpha}_{|V|}$ in $\mathcal{F}(V,\mathbb{R})$.
\item If $G$ is strongly connected then $\widetilde{\tau}_2>0$, i.e., the multiplicity of the eigenvalue $\widetilde{\tau}_1=0$ is 1, and $\widetilde{L}$ is invertible on the $|V|-1$-dimensional invariant subspace $\widetilde{\mathcal{S}}:= \{\widetilde{\balpha}\in\mathcal{F}(V,\mathbb{R})\,|\, \widetilde{\balpha}^\top D^{1/2}\textbf{1} =0\}$ and the inverse can be expressed as $\widetilde{L}^{-1}_{\widetilde{\mathcal{S}}} = \sum_{j=2}^{|V|} \widetilde{\tau}_j^{-1} \widetilde{\balpha}_j \widetilde{\balpha}_j^\top$.
\item $\widetilde{A}$ has $|V|$ real-valued eigenvalues $1 =\widetilde{\varrho}_1 \ge\widetilde{\varrho}_2\ge\ldots\ge\widetilde{\varrho}_{|V|}\ge -1$, where $\widetilde{\varrho}_j=1-\widetilde{\tau}_j$, $j=1,\ldots,|V|$, and corresponding orthonormal eigenvectors $D^{1/2}\textbf{1}= \widetilde{\balpha}_1,\widetilde{\balpha}_2,\ldots, \widetilde{\balpha}_{|V|}$ in $\mathcal{F}(V,\mathbb{R})$.
\item If $G$ is strongly connected and $|V|\ge 3$ then $-1 < \widetilde{\varrho}_{|V|} \le \widetilde{\varrho}_2<1$ and 
\[
\widetilde{L}^{-1}_{\widetilde{\mathcal{S}}} \widetilde{\balpha}= \sum_{k=0}^\infty 
\widetilde{A}^k \widetilde{\balpha} \quad \textrm{for all}\quad \widetilde{\balpha}
    \in \widetilde{\mathcal{S}},
\]
where the infinite series is absolute convergent. Moreover, the convergence is exponentially fast since
\[
\left \Vert\widetilde{L}^{-1}_{\widetilde{\mathcal{S}}} \widetilde{\balpha} - \sum_{k=0}^n 
\widetilde{A}^k \widetilde{\balpha} \right\Vert \le \frac{\kappa^{n+1}}
{1-\kappa} \Vert\widetilde{\balpha} \Vert 
\]
for all $\widetilde{\balpha}\in \widetilde{\mathcal{S}}$, where $\kappa:=\max\{|\widetilde{\varrho}_2|,
|\widetilde{\varrho}_{|V|}|\} <1$.
\end{enumerate} 
\end{Pro}

Instead of computing the inverse of the Laplacian matrices, (e.g., by determining their eigenvalues and eigenvectors, similarly to the Google’s PageRank algorithm, see Chapter 15 in \cite{Langville2006Meyer}), a linear recursion could be computationally more efficient in large-scale problems. For each $\widetilde{\balpha}\in \widetilde{\mathcal{S}}$, $\widetilde{L}^{-1}_{\widetilde{\mathcal{S}}} \widetilde{\balpha}$ can be obtained as the limit of the iteration
\[
    \widetilde{\bbeta}_n = \widetilde{A}\widetilde{\bbeta}_{n-1} +
    \widetilde{\balpha}, \quad n\in\mathbb{N}, \quad \widetilde{\bbeta}_0 = 0,
\]
where $\widetilde{\bbeta}_n \in \widetilde{\mathcal{S}}$ for all $ n\in\mathbb{N}$.

Finally, we remark that $\blambda$ can be computed by the symmetric unnormalized graph Laplacian matrix as follows. By equation \eqref{lambda_eq} in \ref{proof} we have $D^{1/2}\widetilde{L}D^{1/2} \blambda = \bs-\be$ and thus, if $\widetilde{\blambda} :=D^{1/2} \blambda$, we have $\widetilde{L}\widetilde{\blambda} = D^{-1/2}(\bs-\be)$, where $D^{-1/2}(\bs-\be)\in \widetilde{\mathcal{S}}$. Thus $\widetilde{\blambda} = \widetilde{L}^{-1}_{\widetilde{\mathcal{S}}} D^{-1/2}(\bs-\be)$ and $\blambda = D^{-1/2}\widetilde{\blambda}$. By Proposition \ref{Sym_Laplace_properties} we have
\begin{equation*}
\begin{split}
    \blambda= &D^{-1/2} \widetilde{L}^{-1}_{\mathcal{S}} D^{-1/2} (\bs-\be) \\ = &\sum_{j=2}^{|V|} 
    \widetilde{\lambda}_j^{-1}  \left(\widetilde{\balpha}_j^\top D^{-1/2} (\bs-\be)\right) D^{-1/2}\widetilde{\balpha}_j
\end{split}
\end{equation*} 
and $\widetilde{\blambda}$ is the limit of the following iteration
\[
    \widetilde{\blambda}_n = \widetilde{A}\widetilde{\blambda}_{n-1} + D^{-1/2}(\bs-\be), \quad n\in\mathbb{N}, \quad \widetilde{\blambda}_0 = 0,
\]
where $\widetilde{\blambda}_n \in \widetilde{\mathcal{S}}$ for all $n\in\mathbb{N}$.

\section{Results}
\label{results}

To evaluate the performance of the proposed WLS estimation method by comparing it to the traditional ML one discussed in section \ref{inference} a simple simulation study was conducted at different sample sizes for small and medium road network. In the simulations, in order to mimic the real traffic, we tried to keep the length of trajectories low and the number of trajectories high compared to the size of the road network, similarly to the Porto example. The absolute bias of an investigated estimator $\widehat{Q}$ for the two-dimensional stationary distribution $Q$ as a parameter is defined by $\|\widehat{Q} -Q\|_G$. The empirical absolute bias and its standard error (SE) correspond to the mean and standard deviation of absolute biases in 100 replications, respectively. All simulations were carried out in Python using the PyDTMC library developed for analysing discrete time Markov chains\footnote{\url{https://pypi.org/project/PyDTMC/}}. The codes and datasets of our simulation are available upon request.

Table \ref{simulation_small} displays the simulation results for the small road network in Fig.~\ref{graph-example-1} using the Markov kernel of Fig.~\ref{pi_graph}, see the toy example in \ref{example}. The simulation parameters were $k=100,200,500,$ and $1000$ number of trajectories with $n=3,5,$ and $10$ length. The absolute bias and its standard error do not depend on the length $n$ and they are decreasing as $k$ is increasing for both estimation methods. The latter is an expected result. Moreover, while for relatively small $k$ the performance of the WLS and ML methods are similar, in the case of relatively large number of trajectories the ML estimator outperforms the WLS one a little bit. This phenomenon could be due to the asymptotic optimality of the ML estimator because the parameter $k$ is enough large (1000) compared to the size of the road graph (5).

In the second simulation scenario, a strongly connected subgraph, which contains 1000 vertices, of Porto's road network was chosen (exported from the OSM, as well, GPS coordinates W8.6137, W8.5991, N411573, N41.1437). The entries of Markov kernel were generated randomly. The simulation parameters were $k=1000,3000,$ and $5000$ with $n=3,5,$ and $10$. In this scenario, the absolute bias and its standard error are also independent of the length $n$. However, there are significant differences between the performances of the two estimation methods (ML and WLS) related to the parameter $k$. On the one hand, the absolute bias of ML estimator is decreasing as $k$ is increasing while it is constant for WLS estimator. On the other hand, the WLS estimator is better than the ML one in case of $k=1000$ but worse in case of $k=5000$. Since the former parameter setting is closer to the real traffic, this simulation corroborates the superiority of WLS method based on two-dimensional stationary distribution against the traditional maximum likelihood. Finally, in this scenario, the scale of the SE's indicates that the WLS estimator is more stable than the ML one.

\begin{table}[!t]
\centering
\caption{Simulation results, absolute bias and SE (inside parenthesis), for the Markov kernel in Fig.~\ref{pi_graph} on the road network in Fig.~\ref{graph-example-1}. ($k$ - number, $n$ - length of trajectories)}
\label{simulation_small} 
\begin{tabular}{c c | c c}
k & n & ML & WLS  \\ \hline
100 & 3 & 0.034 (0.0103) & 0.034 (0.0109)\\ 
100 & 5 & 0.035 (0.0106) & 0.034 (0.0103)\\ 
100 & 10 & 0.033 (0.0095) & 0.033 (0.0094)\\ 
200 & 3 & 0.024 (0.0066) & 0.026 (0.0066)\\ 
200 & 5 & 0.023 (0.0064) & 0.024 (0.0067)\\ 
200 & 10 & 0.024 (0.0071) & 0.025 (0.0070)\\ 
500 & 3 & 0.015 (0.0046) & 0.017 (0.0049)\\
500 & 5 & 0.015 (0.0041) & 0.017 (0.0049)\\
500 & 10 & 0.016 (0.0047) & 0.017 (0.0049)\\
1000 & 3 & 0.010 (0.0032) & 0.013 (0.0041)\\
1000 & 5 & 0.011 (0.0034) & 0.015 (0.0044)\\
1000 & 10 & 0.010 (0.0030) & 0.014 (0.0040)
\\
\hline
\end{tabular}
\end{table}

\begin{table}[!t]
\centering
\caption{Simulation results, absolute bias and SE (inside parenthesis), for a part of Porto's map with 1000 vertices. ($k$ - number, $n$ - length of trajectories)}
\label{simulation_large} 
\begin{tabular}{c c | c c}
k & n & ML & WLS  \\ \hline
1000 & 3 & 0.166 (0.0559) & 0.025 (0.0007) \\
1000 & 5 & 0.184 (0.1214) & 0.025 (0.0007) \\
1000 & 10 & 0.169 (0.0938) & 0.025 (0.0008) \\
3000 & 3 & 0.064 (0.1725) & 0.023 (0.0005) \\
3000 & 5 & 0.070 (0.1665) & 0.023 (0.0005) \\
3000 & 10 & 0.063 (0.1705) & 0.023 (0.0005) \\
5000 & 3 & 0.016 (0.0150) & 0.023 (0.0004) \\
5000 & 5 & 0.014 (0.0055) & 0.023 (0.0004) \\
5000 & 10 & 0.014 (0.0126) & 0.023 (0.0003) \\
\hline
\end{tabular}
\end{table}

\begin{figure}[!b]
    \centering
    \includegraphics[width=.48\textwidth]{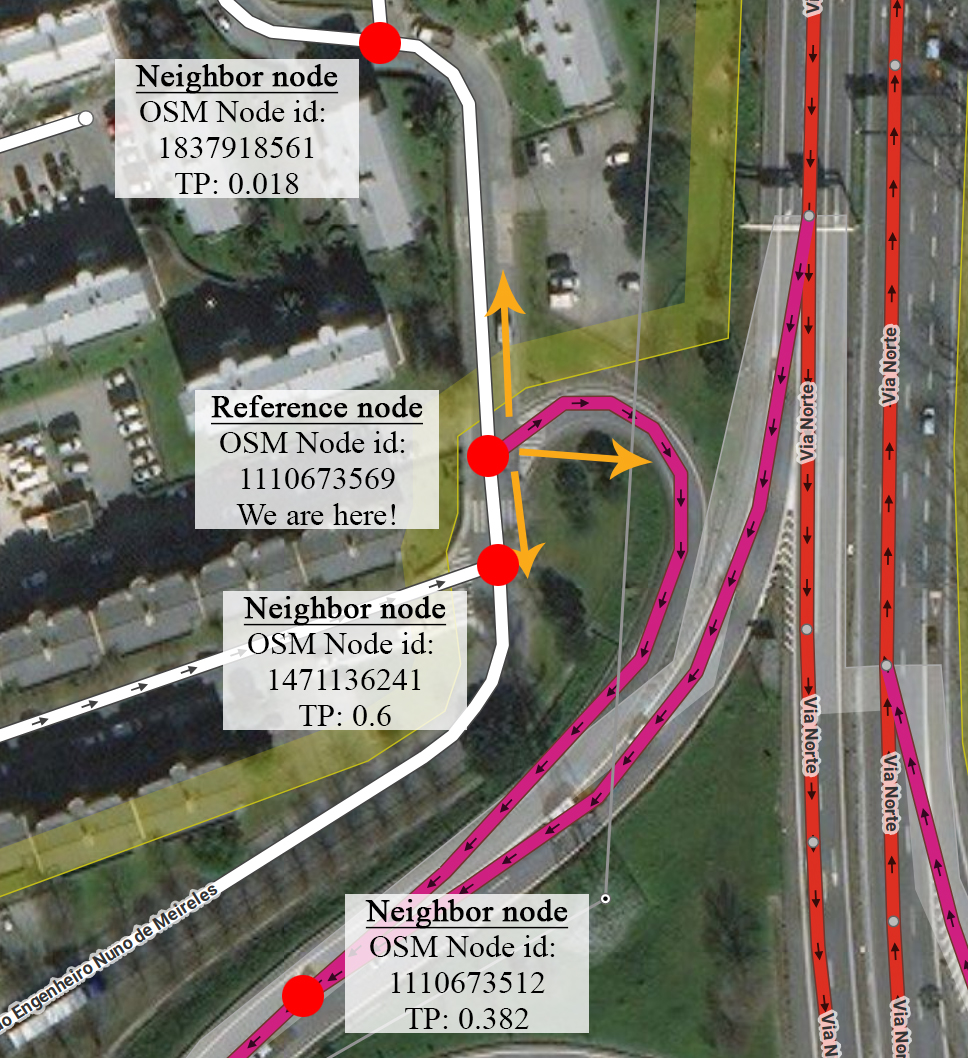}
    \caption{A visual explanation of transitions of intersection 1110673569. TP means transition probability, yellow arrows indicate directions, red dots indicate nodes. (Source of the map: openstreetmap.org, annotated by the authors.)}
    \label{mapexample}
\end{figure}

We have also implemented the model in the OOCWC system. First, we have filtered the TTP dataset as described in section \ref{public-dataset}. Then, we have created the Markov kernel from the filtered dataset following the method described in section \ref{inference} (see also \ref{example} for a toy example). Regarding the RCE, we performed several modifications. First, we extended the operation of the RCE to be able to handle kernel files. This kernel file can be loaded to the RCE software, so all nodes of the simulation graph will have the corresponding transition probability vector from the Markov kernel file. For this, we had to extend the shared memory segment of the RCE.  We should note, however, that not all nodes can be found in the Markov kernel, because it can happen that the dataset does not completely cover the whole map, i.e., not all nodes are part of a trajectory. In this case, we set uniform distribution for the corresponding node. Finally, we had to modify the basic operation of the simulation algorithm. In the original implementation, the cars are moving on the map quite randomly. Now, a car selects the next node based on the transition probability vector of the current node. For this, we use the pseudo-random number generation engine from the Boost Random library that is based on the method presented in paper \cite{Matsumoto}. Let's consider an example. We are at the graph vertex (or intersection) of OSM node ID 1110673569 (with GPS coordinates 41.1752185, -8.6231927). The total transitions of this node (i.e. the total trajectories that cross this intersection) in the dataset is 1,649. The transitions to the neighbor nodes are shown in Table \ref{transitions_example}. Please note that the actual transition probability (TP) is not the same as the ratio of the transitions to the neighbor node and the total transitions of the node which is called frequency (or ML) based transition probabilities. The actual transition probability comes from the Markov kernel of the whole graph. The two kinds of transition probabilities are also compared in Table \ref{transitions_example} where the WLS based transition probabilities have been derived by our method. One can already see in this simple example that the difference between the two methods could be huge. This small example can be observed in Fig.~\ref{mapexample}, as well.

\begin{table}[!t]
\centering
\caption{Transitions of intersection 1110673569. (TP -- transition probability)}
\label{transitions_example} 
\begin{tabular}{c |r r r}
Neighbor node    & \# of transitions & ML TP & WLS TP  \\ \hline
1471136241          & 1449  & 0.879 & 0.6    \\ 
1110673512          & 170  & 0.103   & 0.382   \\ 
1837918561          & 30  & 0.018   & 0.018  \\ 
\hline
Sum & 1649 & 1 & 1
\end{tabular}
\end{table}

\begin{figure*}[!h]
    \centering
    \subfloat[]{\includegraphics[width=0.24\textwidth]{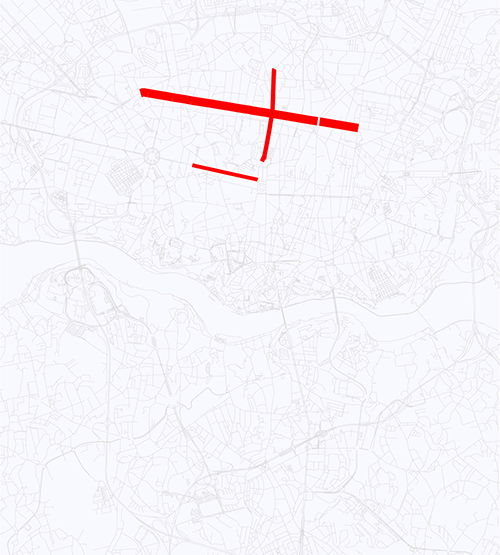}\label{simul_init_5}}
    \hspace{0em}
    \subfloat[]{\includegraphics[width=0.24\textwidth]{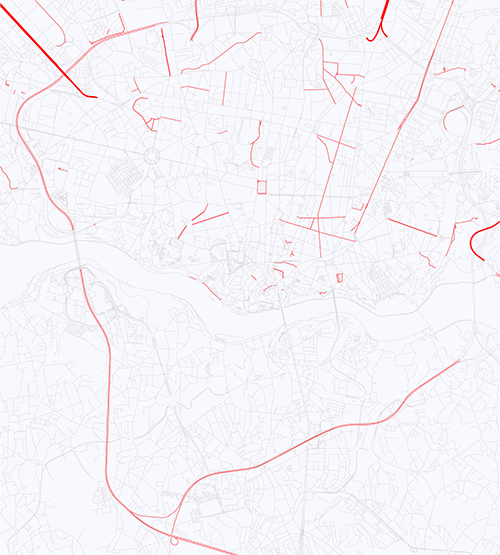}\label{simul_30_5}}
    \hspace{0em}
    \subfloat[]{\includegraphics[width=0.24\textwidth]{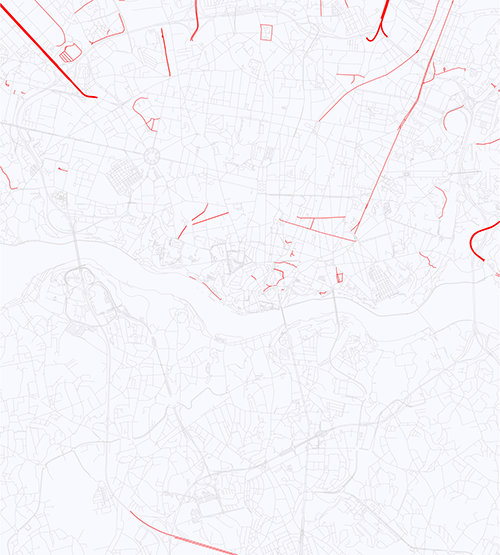}\label{simul_60_5}}
    \hspace{0em}
    \\
    \subfloat[]{\includegraphics[width=0.24\textwidth]{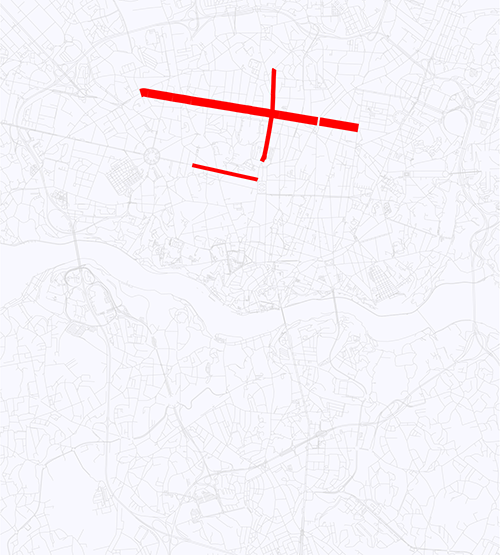}\label{simul_init_10}}
    \hspace{0em}
    \subfloat[]{\includegraphics[width=0.24\textwidth]{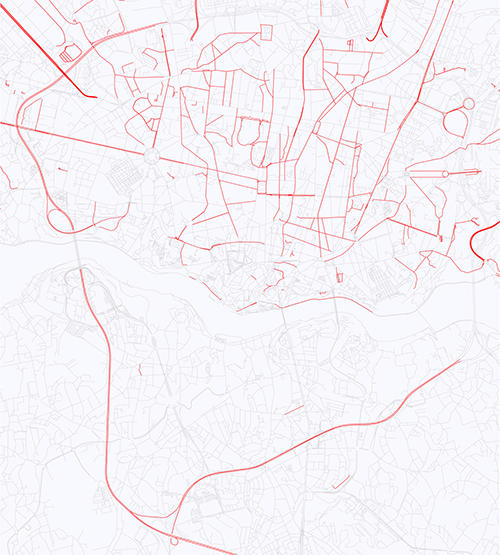}\label{simul_30_10}}
    \hspace{0em}
    \subfloat[]{\includegraphics[width=0.24\textwidth]{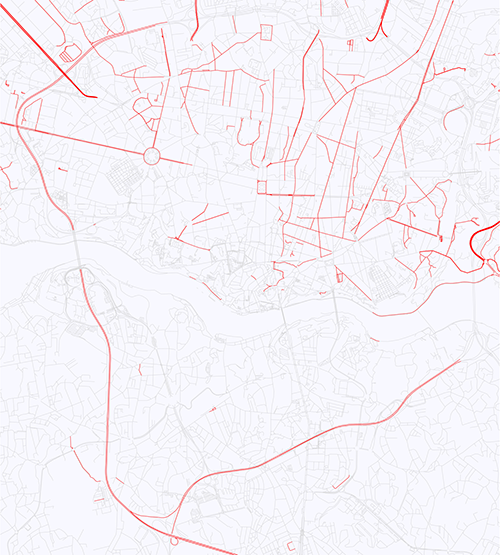}\label{simul_60_10}}
    \hspace{0em}
    \\
    \subfloat[]{\includegraphics[width=0.24\textwidth]{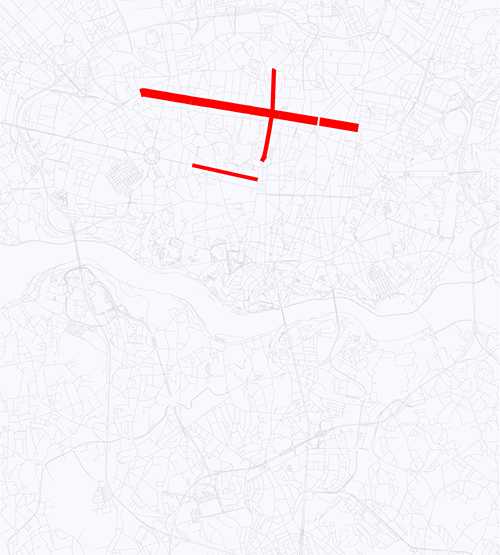}\label{simul_init_20}}
    \hspace{0em}
    \subfloat[]{\includegraphics[width=0.24\textwidth]{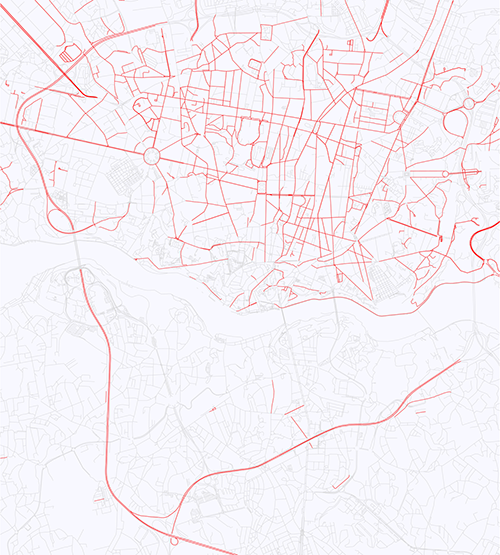}\label{simul_30_20}}
    \hspace{0em}
    \subfloat[]{\includegraphics[width=0.24\textwidth]{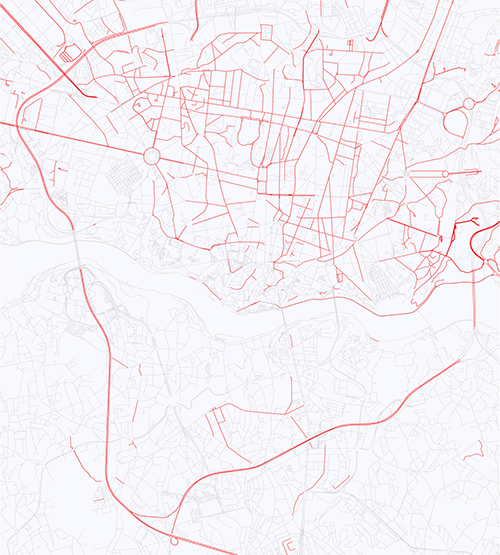}\label{simul_60_20}}
    \hspace{0em}
    \\
    \subfloat[]{\includegraphics[width=0.24\textwidth]{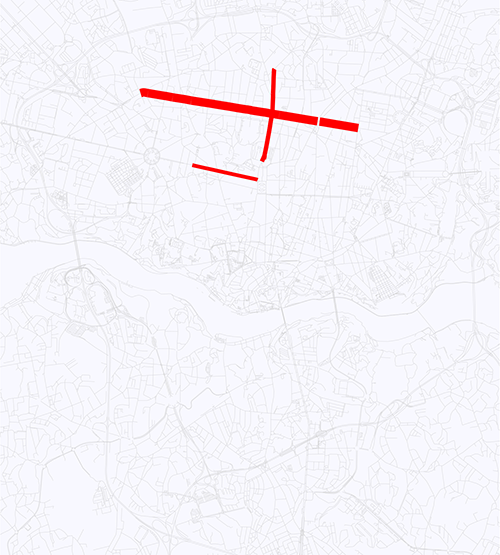}\label{simul_init_50}}
    \hspace{0em}
    \subfloat[]{\includegraphics[width=0.24\textwidth]{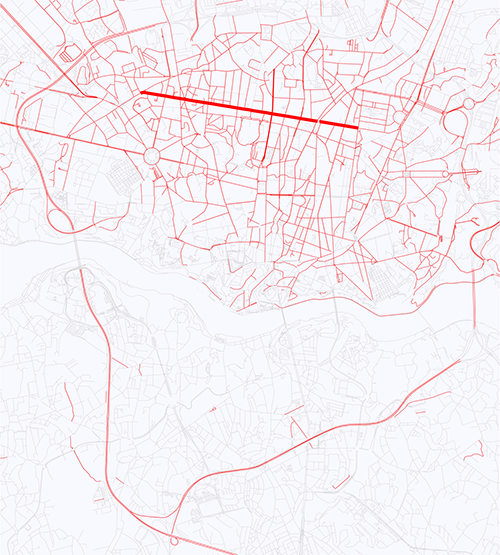}\label{simul_30_50}}
    \hspace{0em}
    \subfloat[]{\includegraphics[width=0.24\textwidth]{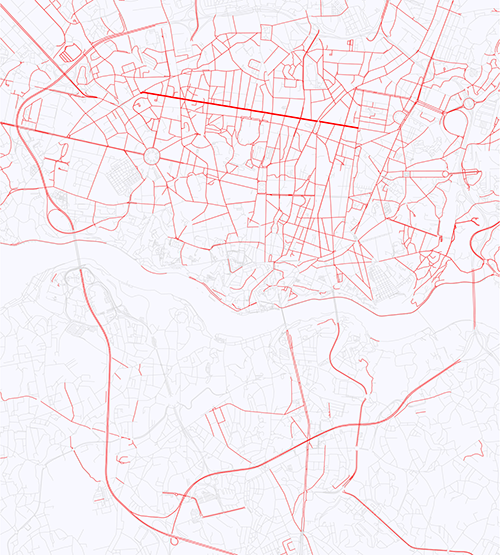}\label{simul_60_50}}
    \caption{The change of the distribution of cars during the simulation (5,000 (\textit{a-c}), 10,000 (\textit{d-f}), 20,000 (\textit{g-i}) and 50,000 (\textit{j-l}) cars, initial state, after 30 mins, after 60 mins, respectively). The thickness of the street is proportionate with the number of cars on the street.}
    \label{cardist}
\end{figure*}

The initialization phase of the simulation adds traffic units to the map. Each unit is placed to an OSM node, i.e., on a vertex of the simulation graph. There exist two ways to do this, one is following a prescribed distribution (e.g.~uniform), the other is following measured data. In our test case, we initialized simulations with fictional measured data. We put units only to the streets Rua de Antero de Quental, Rua da Constitui\c{c}\~ao and Rua da Boavista (25.6\%, 51.4\%, 23\% of the cars, respectively), i.e., the simulation starts from the traffic configuration which is concentrated on three nodes of the road graph. In addition, we can set the number of the simulation units. We run simulations with $k=5,000, 10,000, 20,000, 30,000$ and $50,000$ units. The simulation starts when all simulation units are added to the map. Fig.~\ref{cardist} shows the change of the distribution of cars during the simulation.

\begin{figure}[!b]
    \centering
    \includegraphics[width=.49\textwidth]{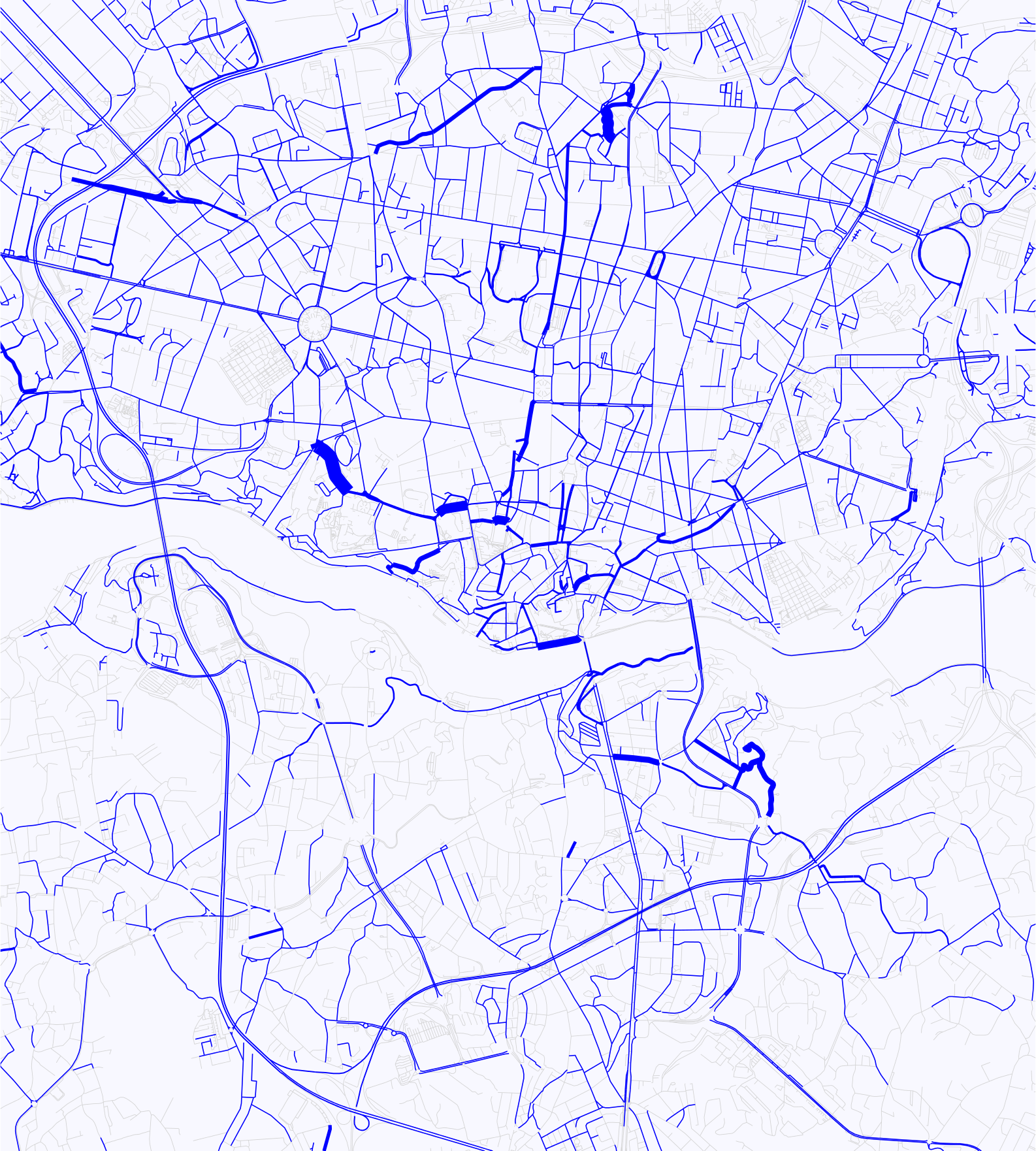}
    \caption{The two-dimensional stationary distribution of cars in Porto based on the TTP dataset.}
    \label{porto_stat_dist}
\end{figure}

The RCE produces a logfile that contains the position of every simulation unit in every simulation step. From this file, we create a new file that contains the number of cars by streets in every minute, so we can observe the change of distribution of the cars. In addition, we calculated the stationary distribution of cars for streets in the city of Porto, see Fig.~\ref{porto_stat_dist}. This latter one tells us, what is the probability that a car is on a given street. It is worth noting the similarity between this figure and Fig.~\ref{tr_points_all}. The ticker line on Fig.~\ref{porto_stat_dist} corresponds to increasingly hot color on Fig.~\ref{tr_points_all}.

\begin{figure}[!t]
    \centering
    \includegraphics[width=.49\textwidth]{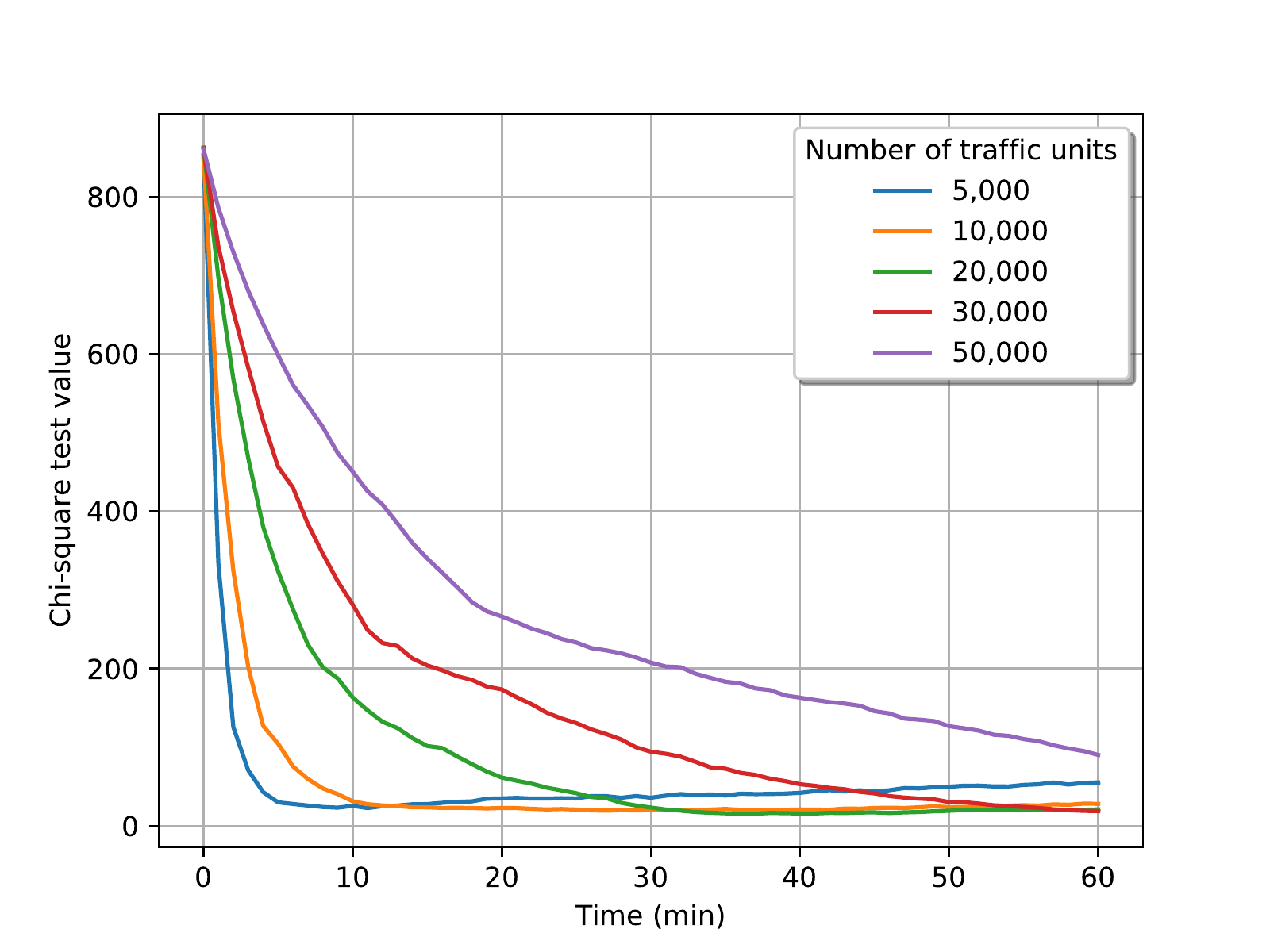}
    \caption{Chi-square test results.}
    \label{DIST_chi_sq}
\end{figure}

To obtain a quantitative measure that describes the \enquote{goodness} of our simulation algorithm, we applied the Pearson's chi-squared test. We expect, by Theorem \ref{Markov_traffic_ergodic}, that during the simulation, independently from the initial distribution, within a certain time period, the distribution of the cars become close to the previously calculated stationary distribution. Fig.~\ref{DIST_chi_sq} shows the test results. We can observe that in the first few minutes the test statistic is significantly high, meaning that the distribution of the cars is still far from the steady-state. However, after a time period that depends on the number of cars, the test statistic becomes low, meaning that the distribution became steady. One can observe that it takes more time to reach the steady-state with more traffic units, which is reasonable. Another notable trend is the case of 5,000 cars, where the line is elevating after reaching the steady-state. This can be caused by the low number of cars. The number of individual streets (named or unnamed, e.g.~motorway junctions) is 2,194. 5,000 cars are simply not enough to reach and hold a steady-state in this type of simulation.

Finally, we should note some implementation details and possible drawbacks of this model that may have an impact on the model's overall performance. 

In small and medium graphs, the proposed algorithm and its implementation performs as it is expected. But in the case of our Porto example, where the graph has 33,961 nodes and 53,126 edges and the TP matrix is very sparse, numerical problems may occur. One problem can occur when we calculate the $\widehat{Q}_{\textrm{WLS}}$ estimator and then TP matrix using the method described in section \ref{inference}. For matrices with this size (34,000 x 34,000), we cannot solve the linear equation of the Lagrange vector in Theorem \ref{main_thm} always numerically, thus we could only use the least square solution for a numerically stable calculation. In some cases, this causes impossible numbers to present in the TP matrix, e.g.~for a node, the TP vector is [1.17489, -0.174894], which is obviously impossible. It is interesting to note that the sum of these \enquote{malfunctional} TP vectors are 1 all the time, and mostly occurs if the node has a low number of transitions (less than 20). In such cases, we use the frequency based TP. Another numerical problem can occur when we calculate the stationary distribution $\bpi$, namely, negative values may present in the results. We need to handle this problem when we calculate the Pearson's chi-squared test. We chose to shift every value of $\bpi$ until we get a sum of 1 for $\bpi$.

Some minor issues can occur with the map database and the differences between the Porto dataset and the OSM data. In some cases, we could not calculate a route between two consecutive trajectory points using OSM data. This can happen because of the imperfection of the OSM data or false GPS measurement. We handle this case by splitting the trajectory into pieces.

Another minor issue arises in the calculation of the Pearson's chi-squared test. Since the OSM Porto map and the trajectory dataset do not cover each other perfectly, we only know the stationary distribution $\bpi$ for a subgraph of the whole map. During the simulation the units can traverse the whole map graph, so, it can happen that a traffic unit reaches an edge which is not part of the subgraph where we know the stationary distribution $\bpi$.
During the calculation of the Pearson's chi-squared test, we consider only those cars that are present on the road network, where the stationary distribution $\bpi$ is known.

\section{Conclusions}
\label{conclusion}

In this paper, we have described our traffic simulation model that is called \enquote{Markov traffic} based on tools from graph theory and Markov modelling. The aim was to provide a simulation method that is able to keep the distribution of the cars on the map in a steady-state on a large scale road network. We have proven that, under general assumptions, the stationary distribution is unique for any Markov transition mechanism on a wide class of road networks. An explicit formula has also been derived for the stationary distribution.

We have shown that the stationary distribution, with the related transition mechanism, can be explored from observed data based on sample trajectories. We have provided a statistical method and proved its optimality with which we can create the Markov kernel necessary to obtain a Markov traffic on a given road graph. Using this kernel, we can initiate traffic simulations that provide a stationary distribution of the cars on the map.

To provide an example for creating this kernel file, we have used a publicly available dataset, namely the Taxi Trajectory Prediction dataset. Our simulation uses OpenStreetMap, a free map database.

To test our theories, we have implemented the proposed model in our simulation program (RCE). We have run simulations and it has been proved to provide a stationary distribution on the map graph of Porto, Portugal. The whole project (including the RCE) is available for download.\footnote{see \url{https://github.com/rbesenczi/Crowd-sourced-Traffic-Simulator/blob/master/justine/install.txt}} Some simulation video is available at the YouTube channel of the first author.\footnote{see \url{http://bit.ly/2FRpPxL}}

Future work will focus on the possible applications of our simulation approach, e.g., modelling the pollution or energy consumption in a city due to multi-modal traffic with gasoline, diesel, electric and plug-in hybrid vehicles. 

\section*{Acknowledgement}

This publication is supported by the EFOP-3.6.1-16-2016-00022 project. The project is co-financed by the European Union and the European Social Fund. The authors would like to thank all actual and former members of the Smart City group of the University of Debrecen. Special thanks to Louis Mattia for a close reading of the manuscript. We are especially grateful to all of the participants of the OOCWC competitions and the students of the BSc courses of \enquote{High Level Programming Languages} at the University of Debrecen. Finally we would also like to give special thanks to M{\'a}rton Vona and Bal{\'a}zs K{\'o}ti for designing the graphical elements of the OOCWC.

\bibliography{msmatf}

\appendix
\section{A Toy Example}
\label{example}

In order to demonstrate the main concepts and methods of this paper we present a simple toy example.\footnote{We implemented this example in Python, see: \url{https://github.com/rbesenczi/Crowd-sourced-Traffic-Simulator/blob/master/model-sources/Markovkernel/example_graph.py}} Consider the road network $G=(V,E)$ on Fig.~\ref{graph-example-1} where $V := \{1, 2, 3, 4, 5\}$ and $E := \{(1, 2), (2, 1), (2, 3), (2, 4), (3, 4),\\ (4, 2), (4, 5), (5, 2)\}$. Then $|V| = 5$ and $|E| = 8$. The adjacency matrix $A_G$ of $G$, where we denote the vertices as well, can be derived as:
\[ A_G := \begin{array}{c|ccccc}
 & 1 & 2 & 3 & 4 & 5 \\
 \hline
1 & 0 & 1 & 0 & 0 & 0 \\
2 & 1 & 0 & 1 & 1 & 0 \\
3 & 0 & 0 & 0 & 1 & 0 \\
4 & 0 & 1 & 0 & 0 & 1 \\
5 & 0 & 1 & 0 & 0 & 0
\end{array}.\] 

Clearly, $G$ is a strongly connected digraph. Since $1 \rightarrow 2 \rightarrow 1$ and $2 \rightarrow 3 \rightarrow 4 \rightarrow 2$ are cycles of length 2 and 3, respectively, we have $per(G) = 1$ and thus $G$ is aperiodic. The first power $k$ that $A^k_G > 0$ is $k = 4$ and
\[ A^4_G := \left[ \begin{array}{ccccc}
2 & 2 & 2 & 2 & 1 \\
2 & 5 & 2 & 4 & 2 \\
1 & 2 & 1 & 2 & 1 \\
2 & 4 & 2 & 3 & 2 \\
2 & 2 & 2 & 2 & 1
\end{array} \right]. \]
The entries of this matrix are the number of directed walks of length 4 between the pairs of vertices.

One can see that the in- and outdegree of vertices are given as:
\[ 
\bd^{−} = \bd^{+} = \left[ \begin{array}{ccccc}
1 & 3 & 1 & 2 & 1 \\
\end{array} \right]^T. 
\]

Define the Markov kernel $P$ on the road network $G$ as:
\[
    P := \begin{array}{c|ccccc} &1&2&3&4&5 \\ \hline 1&1/2&1/2&0&0&0 \\ 2&1/4&1/4&1/4&1/4&0 \\ 
    3&0&0&1/2&1/2&0 \\ 4 &0&1/4&0&1/2&1/4 \\ 5 &0&1/2&0&0&1/2\end{array}     
\]
Clearly, $P$ is weakly $G$-subordinated. Fig.~\ref{pi_graph} displays the Markov kernel $P$ denoting the transition probabilities on the edges and its stationary distribution $\bpi$ denoting on the vertices. Note that $\bpi = (1/7,2/7, 1/7,2/7,1/7)^\top$.

The symmetric unnormalized graph Laplacian matrix $L$ and the symmetric normalized graph Laplacian matrix $\widetilde{L}$ of the road network $G$ are given as:
\begin{equation*}
L = \begin{bmatrix} 2 &-2 & 0 & 0 & 0 \\  
     -2 & 6 &-1 &-2 &-1 \\ 
     0 &-1 & 2 &-1 & 0 \\
    0 &-2 &-1 & 4 &-1 \\
     0 &-1 & 0 &-1 & 2 \end{bmatrix}
\end{equation*}
and
\begin{equation*}
\widetilde{L} = \begin{bmatrix} 1  & -0.577 & 0 &  0  & 0 \\
     -0.577 & 1 & -0.288  & -0.41 & -0.288 \\
     0  &  -0.288 & 1 &  -0.35 & 0  \\
     0  &  -0.41 & -0.35 & 1  &   -0.35  \\
     0  &  -0.288 &  0 &  -0.35 & 1 \end{bmatrix}.
\end{equation*}
The eigenvalues and eigenvectors of the symmetric unnormalized graph Laplacian matrix $L$ are given as:
\begin{equation*}
    S = \begin{bmatrix} 0 & 0 & 0 & 0 & 0 \\ 
    0 & 1.72 &0&0&0 \\ 0&0&  2 &0&0 \\ 
    0&0&0&  4.46 &0\\
    0&0&0&0&7.82 \end{bmatrix}
\end{equation*} 
and
\begin{equation*}
    O = \begin{bmatrix} 0.447 & -0.82 & 0 &  -0.18 &  0.29 \\
    0.447 & -0.11 &  0 &   0.23 & -0.86 \\
    0.447 &  0.36 &   0.7 & -0.4 & 0.07 \\
    0.447 &  0.22 &  0 &   0.76 & 0.41 \\
    0.447 & 0.36 &  -0.71 &  -0.4 & 0.077 \end{bmatrix}
\end{equation*}
where $S$ contains the eigenvalues in its diagonal and $O$ is the orthonormal matrix of eigenvectors in its columns. The multiplicity of the smallest eigenvalue 0 is 1 which shows that the road network is strongly connected. The inverse of $L$ on the subspace $\mathcal{S}$ which is the generalized inverse or Moore-Penrose inverse of $L$, can be derived as
\[
\footnotesize
    L_{\mathcal{S}}^{-1} = O S^{-1} O^\top = \begin{bmatrix}  0.41 &  0.01 & -0.15 & -0.12 & -0.15 \\
    0.01 &  0.11 & -0.05 & -0.02 & -0.05 \\
    -0.15 & -0.05 &  0.36 & -0.02 & -0.136 \\
     -0.12 & -0.02 & -0.02 & 0.18 & -0.02 \\
    -0.15 & -0.05 & -0.136 & -0.02 &  0.36 \end{bmatrix}
\]
where $S^{-1}$ is the generalized inverse of $S$. The eigenvalues and eigenvectors of the symmetric normalized graph Laplacian matrix $\widetilde{L}$ are given as:
\begin{equation*}
    \widetilde{S} = \begin{bmatrix} 0 & 0 & 0 & 0 & 0 \\ 0 & 0.77 &0&0&0 \\ 0&0&  1 &0&0 \\ 0&0&0&  1.5 &0\\
    0&0&0&0& 1.73 \end{bmatrix}
\end{equation*}
and
\begin{equation*}
\widetilde{O} =\begin{bmatrix} 0.35 & 0.73 & 0 & 0 & -0.58\\ 
    0.61  & 0.29 & 0 &  0 & 0.74 \\
    0.35  &  -0.31 &   0.71 & -0.5 & -0.17  \\
    0.5  &  -0.44 &  0  & 0.71  & -0.24 \\
    0.35 &  -0.31 & -0.71 & -0.5 & -0.17  \end{bmatrix}.
\end{equation*}
Clearly, the eigenvalues of $\widetilde{L}$ are in $[0,2]$ and the smallest one is 0 with multiplicity 1. In this example, the speed of convergence in Proposition \ref{Sym_Laplace_properties} is $\kappa = \max \{0.23,0.73\} = 0.73$.

Let the following trajectories be observed in the road network $G$:
\[
\begin{array}{ccc} \textrm{Trajectory}&\textrm{Length}&\textrm{Count} \\ \hline 1\rightarrow 2 \rightarrow 3 \rightarrow 4 &4&150 \\ 
1\rightarrow 2 \rightarrow 4 \rightarrow 5&4&100 \\ 3\rightarrow 4\rightarrow 5 &3&200 \\
5\rightarrow 2\rightarrow 1&3&250 \\ 5\rightarrow 2\rightarrow 3 &3&50 \\ 
3\rightarrow 4\rightarrow 2\rightarrow 1 & 4&100 \\ 5\rightarrow 2\rightarrow 4&3&50 \\
4\rightarrow 2 \rightarrow 1 &3&100 \\ \hline && 1000
\end{array}  
\] 
Then, the total sample size is $n=3350$, the number of trajectories is $k=1000$ and the two-dimensional consecutive frequency matrix $N$ is given by
\[
    N = \begin{bmatrix}  0&250&0&0&0 \\ 450&0&200&150&0 \\ 0&0&0&450&0 \\
    0&200&0&0&300 \\ 0&350&0&0&0 \end{bmatrix}.
\]
The statistics for the starting and ending points of the trajectories are:
\[
    \begin{array}{c|cc|c} \textrm{Node} &\textrm{Start} & \textrm{End} & \textrm{Diff} \\ \hline 1&250&450&-200 \\ 2&0&0&0 \\ 3&300&50&250 \\
    4&100&200&-100 \\ 5&350&300&50 \\ \hline \textrm{Sum} & 1000&1000&0\end{array} 
\] 
The Lagrange multiplicators are given as:
\[
    \blambda = \begin{bmatrix} -116.66 &  -16.66 &  116.66 & 0& 16.66 \end{bmatrix}^\top.
\]        
Thus, the correction matrix $R$ is given by
\[
    R = \begin{bmatrix}   0 &  100 & 0 & 0 & 0 \\ -100 & 0 & 133.33 &  16.66 &  0 \\
     0  &  0 & 0 & -116.66 &  0  \\ 0 &  -16.66 & 0 & 0 &  16.66 \\
     0  & -33.33 & 0 & 0 & 0  \end{bmatrix}
\]   
Since $\bd^{-} = \bd^{+}$ we have $n_{\textrm{eff}} = (n-k) = 2350$, and
\begin{equation*}
\begin{split}
    N+R = &\begin{bmatrix}   0 &  350 & 0 & 0 & 0 \\ 350 & 0 & 333.33 &  166.66 &  0 \\
     0  &  0 & 0 & 333.33 &  0  \\ 0 &  183.33 & 0 & 0 &  316.66 \\
    0  & 316.66 & 0 & 0 & 0  \end{bmatrix} \\
    \widehat{Q}_{\textrm{WLS}}  = & \begin{bmatrix}   0 &  0.149 & 0 & 0 & 0 \\ 0.149 & 0 & 0.142 &  0.07 &  0 \\
    0  &  0 & 0 & 0.142 &  0  \\ 0 &  0.078 & 0 & 0 &  0.135 \\
    0  & 0.135 & 0 & 0 & 0  \end{bmatrix}   
\end{split}
\end{equation*}   
The stationary distribution is given as
\[
    \widehat{\bpi}_{\textrm{WLS}} = \begin{bmatrix} 0.149 &  0.362 &  0.142 & 0.213 &  0.135 \end{bmatrix}^\top
\]
and one can easily check that $\widehat{\bpi}_{\textrm{WLS}}$ is indeed the stationary distribution of the estimated Markov kernel:
\[
\widehat{P}_{\textrm{WLS}} =  \begin{bmatrix} 0 & 1 & 0 & 0 & 0 \\ 0.41 & 0 &  0.39 & 0.2 & 0 \\
0 & 0 & 0 & 1 & 0 \\ 0 & 0.37 & 0 & 0 & 0.63 \\
0 & 1 & 0 & 0 & 0 \end{bmatrix}
\]
Note that the estimated Markov kernel using the standard maximum likelihood estimator is given as
\[
\widehat{P}_{\textrm{ML}} =  
\begin{bmatrix} 0 & 1 & 0 & 0 & 0 \\ 
     0.5625 & 0 &  0.25 & 0.1875 & 0 \\
     0 & 0 & 0 & 1 & 0 \\ 0 & 0.4 & 0 & 0 & 0.6 \\
     0 & 1 & 0 & 0 & 0 \end{bmatrix}
\]
which has stationary distribution
\[
    \widehat{\bpi}_{\textrm{ML}} = \begin{bmatrix} 
    0.224 &  0.398 & 0.1 & 0.174 & 0.104 \end{bmatrix}^\top.
\]

\section{Proofs}
\label{proof}

\textit{Proof of formula \eqref{global_balance_traffconfig}.} For all $\bg\in\mathcal{F}_k$ we have by formulas \eqref{multinomial} and \eqref{multinomial_kernel} and the multinomial theorem that
\begin{equation*}
\begin{split}
    \sum_{\bff\in\mathcal{F}_k} & \bvarrho (\bff) R(\bff,\bg)= \\ = & k!\sum_{\bff\in\mathcal{F}_k} \prod_{u\in V} \pi_u^{f_u} \sum_{K\in\mathcal{M} (\bff,\bg)}\prod_{u,v:u \Rightarrow v}
      \frac{p_{uv}^{k_{uv}}}{k_{uv}!} \\
    = & k!\sum_{\bff\in\mathcal{F}_N} 
    \sum_{K\in\mathcal{M} (\bff,\bg)}\prod_{u,v:u \Rightarrow v}
      \frac{(\pi_u p_{uv})^{k_{uv}}}{k_{uv}!} \\
     = & k!\sum_{\sum\limits_{u\in V}k_{uv}=g_v} \prod_{u,v:u 
     \Rightarrow v} \frac{(\pi_u p_{uv})^{k_{uv}}}{k_{uv}!} \\
     = & k! \prod_{v\in V} (g_v!)^{-1} \left(\sum_{u\in V} \pi_u
       p_{uv}\right)^{g_v} \\ 
       = & k! \prod_{v\in V} \frac{\pi_v^{g_v}}{g_v!}=
       \bvarrho (\bg).
\end{split}       
\end{equation*}

\textit{Proof of Theorem~\ref{main_thm}.}  
By the bias-variance decomposition, see Section 3.2 in \cite{bishop2006}, we have
\begin{equation}\label{bias-variance}
\begin{split}
    \text{SSE}(M,\bw\,|\,\bN) = & \sum_{i=1}^k w_i^{-1} \Vert N_i - w_{i} M\Vert_G^2 \\ = & \sum_{i=1}^k w_i^{-1} \Vert N_i -w_i N\Vert_G^2 + \Vert N - M \Vert_G^2 . 
\end{split}
\end{equation}
Clearly, if $\langle\cdot,\cdot\rangle_G$ denotes the inner product induced by the norm $\Vert\cdot\Vert_G$,
\begin{equation*}
\begin{aligned}
    \Vert N_i -w_i M \Vert_G^2 = &\Vert N_i - w_i N + w_i N - w_i M\Vert_G^2 \\
    = &\Vert N_i - w_i N \Vert_G^2 + \Vert w_i (N - M)\Vert_G^2 \\ 
    & + 2 \langle N_i - w_i N,
    w_i (N - M) \rangle_G.
\end{aligned}
\end{equation*}
Thus, we have \eqref{bias-variance} since
\begin{equation*}
\begin{split}
    \sum_{i=1}^k & w_i^{-1}\Vert w_i (N - M)\Vert_G^2 \\ 
    &= \sum_{i=1}^k  w_i \Vert N - M \Vert_G^2  
    = \Vert N - M \Vert_G^2 
\end{split}          
\end{equation*}
and
\begin{equation*}
\begin{split}
    \sum_{i=1}^k & w_{i}^{-1}\langle N_i - w_i N, 
    w_i (N - M) \rangle_G \\ 
    &= \sum_{i=1}^k \langle N_i - w_i N, N - M\rangle_G \\ 
    &= \langle {\sum_{i=1}^k} N_i -
    {\sum_{i=1}^k} w_i N, N - M\rangle_G =0
\end{split}          
\end{equation*}
where we applied that $\sum_{i=1}^k N_i = N$ and $\sum_{i=1}^k w_i =1$.

Minimizing SSE in its parameters $M$ and $\bw$ is equivalent by minimizing the right hand side of \eqref{bias-variance} partially: minimizing the first term with respect to $\bw$; and minimizing the second term with respect to $M$. Since
\begin{equation*}
    \Vert N_i -w_iN\Vert_G^2 = \Vert N_i\Vert_G^2 + w_i^2\Vert N\Vert_G^2 - 2w_i\langle N_i,N\rangle_G
\end{equation*}
we have to minimize
\begin{equation*}
\begin{split}
    \sum_{i=1}^k &w_i^{-1} \Vert N_i -w_iN\Vert_G^2 \\ &= \sum_{i=1}^k w_i^{-1}
    \Vert N_i\Vert_G^2 + \sum_{i=1}^k w_i \Vert N\Vert_G^2 - 2\sum_{i=1}^k\langle N_i,N\rangle_G \\
    &= \sum_{i=1}^k w_i^{-1} \Vert N_i\Vert_G^2 -  \Vert N\Vert_G^2 
\end{split}          
\end{equation*}
under the constraint $\sum_{i=1}^k w_i =1$. The Lagrange function is given as:
\begin{equation*}
    \ell = \sum_{i=1}^k w_i^{-1} \Vert N_i\Vert_G^2 -\Vert N\Vert_G^2 +\mu \left(\sum_{i=1}^k w_i -1\right)
\end{equation*}
where $\mu$ is the Lagrange multiplicator. For the first-order condition we have, for all $i=1,\ldots,k$,
\begin{equation*}
    \frac{\partial\ell}{\partial w_i} = - w_i^{-2}\Vert N_i\Vert_G^2 + \mu = 0.
\end{equation*}
Thus, $w_i=\mu^{-1/2}\Vert N_i\Vert_G$, $i=1,\ldots,k$, and hence $\mu^{1/2}=\sum_{i=1}^k \Vert N_i\Vert_G$
and $\widehat{w}_i=\Vert N_i\Vert_G/\sum_{j=1}^k \Vert N_j\Vert_G$, $i=1,\ldots,k$, minimizes the first term of \eqref{bias-variance}. 

To minimize the second term in \eqref{bias-variance} define the estimator $\widehat{M}=(\widehat{m}_{uv})$ as a correction to the two-dimensional consecutive frequency matrix $N$ in the following way: $\widehat{M} := N+R$ where $R:=(r_{uv})_{u,v\in V}$ is a real matrix on $V$ such that (i) $r_{uv}=0$ if $(u,v)\notin E\cup S$ (i.e., $u\nRightarrow v$); (ii) $\widehat{m}_{v+}= \widehat{m}_{+v}$ for all $v\in V$; and (iii) $\Vert N -\widehat{M}\Vert_G^2= \sum_{u\Rightarrow v} r_{uv}^2$ is minimal. Property (i) means that there is correction on the set $E\cup S$ only. Property (ii) means that $\widehat{M}$ is an unnormalized two-dimensional stationary distribution on $G$. Finally, property (iii) implies that $\widehat{M}$ is optimal in the least square sense. By equation \eqref{start_end} one can see that (ii) implies 
\begin{equation}\label{r_constraint}
    r_{v+}-r_{+v}-e_v+s_v =0, \qquad v\in V.
\end{equation} 
Thus, the correction matrix $R$ is given as a solution to the constrained optimization problem defined by
\[
    \frac{1}{2}\sum_{u,v: u\Rightarrow v} r_{uv}^2 \to \min
\]
under the constraints \eqref{r_constraint}.

The Lagrange function of this constrained optimization problem is given as:
\[
    \ell := \frac{1}{2}\sum_{u,v:u\Rightarrow v} r_{uv}^2 + \sum_{v\in V} \lambda_v (r_{v+}-r_{+v}-e_v+s_v) 
\] 
where $\lambda_v$, $v\in V$, are the Lagrange multipliers. Taking the gradient of the Lagrange function we have the first-order condition for extrema as
\[
    \frac{\partial\ell}{\partial r_{uv}} =r_{uv} + \lambda_u - \lambda_v = 0, 
    \quad \textrm{for all}
    \quad u,v\in V: u\Rightarrow v.
\] 
Thus, we have $r_{uv}=\lambda_v-\lambda_u$ for all $u\Rightarrow v$ which implies $r_{vv}=0$ for all $v\in V$ and, for all $u,v\in V$, 
\begin{equation*}
\begin{split}
    r_{u+} = &\sum_{v:u\rightarrow v} (\lambda_v-\lambda_u) = \sum_{v:u\rightarrow v} \lambda_v
    - \deg^+(u) \lambda_u \\ = &\sum_{v\in V} a_{uv} \lambda_v - \deg^+(u) \lambda_u, \\
    r_{+v} = &\sum_{u:u\rightarrow v} (\lambda_v-\lambda_u)
    = \deg^-(v) \lambda_v -  
    \sum_{u:u\rightarrow v}\lambda_u \\
    = & \deg^-(v) \lambda_v - \sum_{u\in V} a_{uv} \lambda_u.
\end{split}
\end{equation*}
Thus, by \eqref{r_constraint}, we have the linear equation for the vector $\blambda\in\mathcal{F}(V,\mathbb{R})$ defined by
$\blambda:=(\lambda_v)_{v\in V}$:
\begin{equation*}\label{lambda_eq}
    \bs-\be = (r_{+v} - r_{v+})_{v\in V} = (D-A-A^\top)\blambda = L\blambda
\end{equation*} 
By equation \eqref{start_end_sum} $\bs-\be\in\mathcal{S}$ which implies that $\blambda = L^{-1}_{\mathcal{S}} (\bs-\be) + c\textbf{1}$, where $c\in\mathbb{R}$ is arbitrary. One can easily see that $R$ is not dependent on $c$ thus we can suppose that $c=0$ and we have $R=(\textbf{1} \blambda^\top - \blambda\textbf{1}^\top)\circ A$ and we finished the proof. \qed

The formula \eqref{eff_sample} for the effective sample size follows from 
\begin{equation*}
\begin{split}
    n_{\textit{eff}} :=& \textbf{1}^\top \widehat{M}\textbf{1} = \sum_{u,v:u\Rightarrow v} \widehat{m}_{uv} = 
    \sum_{u,v:u\Rightarrow v} (n_{uv} + r_{uv}) \\
    = & n-k + (\bd^- - \bd^+)^\top \blambda 
\end{split}
\end{equation*} 
since
\begin{equation*}
\begin{split}
    \sum_{u,v:u\Rightarrow v} r_{uv} = & \sum_{u,v:u\rightarrow v} (\lambda_v-\lambda_u)
    \\ = &\sum_{v\in V} \deg^-(v)\lambda_v - \sum_{u\in V} \deg^+(u)\lambda_u \\
    = &(\bd^{-} - \bd^{+})^\top \blambda.
\end{split}
\end{equation*} 

\textit{Proof of Proposition~\ref{Laplace_properties}.} The entries of $L$ can be expressed as
\[
    l_{uv} = \begin{cases} \sum_{w: w\neq u} (a_{uw} + a_{wu}) \quad &\textrm{if}\quad u=v,\\
    - (a_{uv} + a_{vu})\quad &\textrm{if}\quad u\neq v.\end{cases}
\]
Thus, we have
\begin{equation*}
\begin{split}
    \sum_{u,v\in V} l_{uv}\alpha_u\alpha_v =& \sum_{u,w: u\neq w} (a_{uw} + a_{wu})\alpha_u^2 \\
     & - \sum_{u,v: u\neq v}  (a_{uv} + a_{vu})\alpha_u \alpha_v  \\
    = &\frac{1}{2}\sum_{u,v: u\neq v} (a_{uv} + a_{vu})(\alpha_u^2-2\alpha_u\alpha_v +\alpha_v^2)
\end{split}
\end{equation*}
which implies \eqref{Lquadform}. The parts 2), 3), and 4) can be proved similarly to Proposition 1 in \cite{von2007tutorial}. The strict positivity of the second eigenvalue in 5) follows from the strong connectedness of $G$ by Proposition 2 in \cite{von2007tutorial}. Finally, the remaining part of 5) is a direct consequence of the spectral decomposition of $L$ which implies $L = \sum_{j=1}^{|V|} \tau_j \balpha_j \balpha_j^\top$ and the fact that $\balpha_2,\ldots,\balpha_{|V|}$ form a basis of $\mathcal{S}$. \qed

\textit{Proof of Proposition~\ref{Sym_Laplace_properties}.} The entries of $\widetilde{L}$ can be expressed as
\[
    \widetilde{l}_{uv} = \begin{cases} 1-\frac{2 a_{uu}}{d_u} \quad &\textrm{if}\quad u=v,\\
     - \frac{a_{uv} + a_{vu}}{\sqrt{d_u d_v}}\quad &\textrm{if}\quad u\neq v.\end{cases}
\]
Thus, we have
\begin{equation*}
\begin{split}
    \sum_{u,v\in V} \widetilde{l}_{uv}&\alpha_u\alpha_v =\\
    =& \sum_{u\in V} \left(1-\frac{2 a_{uu}}{d_u}\right)
    \alpha_u^2 - \sum_{u,v: u\neq v}  \frac{a_{uv} + a_{vu}}{\sqrt{d_u d_v}}\alpha_u \alpha_v  \\
    = &\frac{1}{2}\sum_{u,v: u\neq v} (a_{uv} + a_{vu})\left(\frac{\alpha_u^2}{d_u}-\frac{2\alpha_u\alpha_v}
    {\sqrt{d_u d_v}} +\frac{\alpha_v^2}{d_v}\right)
\end{split}
\end{equation*}
which implies \eqref{Lquadform2}. The proof of 2), 3), 4) and 5) is similar to the Proposition \ref{Laplace_properties}. We have to prove only $\widetilde{\tau}_{j}\le 2$ for all $j$. Since $(a-b)^2\le 2(a^2 +b^2)$ for all $a,b\in\mathbb{R}$, and $=$ if and only if $a=-b$ this follows from
\[
    \balpha^\top \widetilde{L}\balpha \le \sum_{u,v\in V} (a_{uv} + a_{vu}) \left(\frac{\alpha_u^2}{d_u}
    + \frac{\alpha_v^2}{d_v}\right) \le 2 \sum_{v\in V} \alpha_v^2,
\]
where $=$ is in the first inequality if and only if $\frac{\alpha_u}{\sqrt{d_u}} = - \frac{\alpha_v} {\sqrt{d_v}}$ for all $u,v\in V$. However, this is impossible if $|V|\ge 3$. \qed

\end{document}